\documentclass[twocolumn,twocolappendix]{aastex631}

\usepackage{amsmath}
\usepackage{multirow}

\hypersetup{linkcolor=cyan,citecolor=cyan,filecolor=cyan,urlcolor=cyan}

\begin{document}

\title{ICM-SHOX III. The case of MACS J0018.5+1626, a radio relic that looks like a radio halo?}

 \author[0000-0001-7058-8418]{P. Dom\'inguez-Fern\'andez}
\affiliation{Center for Astrophysics $\vert$ Harvard \& Smithsonian, 60 Garden Street, Cambridge, MA 02138, USA}

\author[0000-0003-3175-2347]{J. A. ZuHone}
\affiliation{Center for Astrophysics $\vert$ Harvard \& Smithsonian, 60 Garden Street, Cambridge, MA 02138, USA}

\author[0000-0002-1616-5649]{E. M. Silich}
\affiliation{Cahill Center for Astronomy and Astrophysics, California Institute of Technology, Pasadena, CA 91125, USA}

\author[0000-0001-6411-3686]{E. Bellomi}
\affiliation{Center for Astrophysics $\vert$ Harvard \& Smithsonian, 60 Garden Street, Cambridge, MA 02138, USA}

\author[0000-0002-8213-3784]{J. Sayers}
\affiliation{Cahill Center for Astronomy and Astrophysics, California Institute of Technology, Pasadena, CA 91125, USA}

\author[0000-0003-3816-5372]{T. Mroczkowski}
\affiliation{Institute of Space Sciences (ICE-CSIC), Carrer de Can Magrans, s/n, 08193 Cerdanyola del Vallès, Barcelona, Spain}

\author[0000-0002-9325-1567]{A. Botteon}
\affiliation{Istituto di Radio Astronomia, INAF, Via Gobetti 101, 40121 Bologna, Italy}

\author[0000-0002-0587-1660]{R. J. van Weeren}
\affiliation{Leiden Observatory, Leiden University, PO Box 9513, NL-2300 RA Leiden, The Netherlands}

\author[0000-0001-6950-1629]{L. Hernquist}
\affiliation{Center for Astrophysics $\vert$ Harvard \& Smithsonian, 60 Garden Street, Cambridge, MA 02138, USA}

\author[0000-0003-4195-8613]{G. Brunetti}
%\altaffiliation{AASTeX v6+ programmer}
\affiliation{Istituto di Radio Astronomia, INAF, Via Gobetti 101, 40121 Bologna, Italy}

\author[0000-0002-4421-0267]{J. Golec}
\affiliation{Department of Astronomy, University of Massachusetts Amherst, Amherst, MA 01003, USA}

\author[0000-0002-1030-8012]{S. Gupta}
\affiliation{Department of Astrophysical Sciences, Princeton University, 4 Ivy Lane, Princeton, NJ 08544, USA}



\begin{abstract}

We present the first detailed numerical modeling of the radio emission from MACS J0018.5+1626 as part of the Improved Constraints on Mergers with SZ, Hydrodynamical simulations, Optical, and X-ray (ICM-SHOX) project. By matching X-ray, thermal and kinetic Sunyaev–Zel'dovich, optical and lensing observables to simulations, the ICM-SHOX pipeline indicates that MACS J0018.5+1626 is undergoing a binary merger close to pericenter passage and is observed along a line of sight nearly aligned with the merger axis. We perform three-dimensional magnetohydrodynamic simulations of binary cluster mergers coupled to tracer particles and a Fokker–Planck solver to model the radio emission. Exploring variations in the most likely initial conditions within the ICM-SHOX parameter space, such as the relative cluster velocity and impact parameter, we find that the resulting merger configuration consistently produces two merger-driven shocks with typical average Mach numbers $\mathcal{M}_s \sim 2$--$3$ with corresponding standard deviations of $\sigma_{\mathcal{M}} \sim 0.5$--$1.5$. Within this framework, we examine the cluster conditions under which standard diffusive shock acceleration can reproduce LOFAR observations. In particular, we discuss the possibility that the apparent radio halo seen by LOFAR arises from the superposition of two radio relics viewed nearly face-on.

\end{abstract}



%%%%%%%%%%%%%%%%%%%%%%%%%%%%%%%%%%%%%%%%%%%
\section{Introduction} \label{sec:intro}
%%%%%%%%%%%%%%%%%%%%%%%%%%%%%%%%%%%%%%%%%%%

%

Observations of the intracluster medium (ICM) provide key insights into galaxy cluster mergers and their imprints, such as shocks, turbulence, and bulk motions. Through these diagnostics, we can probe both the thermal and non-thermal energy budget of the ICM and, ultimately, gain a deeper understanding of structure formation \citep[see][for a review]{2012ARA&A..50..353K}. However, to obtain a complete physical picture of galaxy cluster mergers, it is essential to establish a coherent connection between the non-thermal components of the ICM (e.g., radio observations), the thermal components (e.g., X-ray and thermal Sunyaev–Zel'dovich observations), and the underlying mass distribution (e.g., gravitational lensing observations). The main focus of this work is on radio observations, which trace the non-thermal component of the ICM through the interaction between cosmic rays (CR) and magnetic fields.

Diffuse radio emission on Mpc scales is commonly observed in galaxy clusters.
The typical Mpc-scale diffuse radio emission comes in the form of radio halos, radio mini-halos, and radio relics, among other features \citep[see][for a review]{2019SSRv..215...16V}. Radio halos are typically located in the cluster center and have a relatively round morphology tracing the X-ray emission \citep[e.g.,][]{2014IJMPD..2330007B}. Radio relics, on the other hand, are typically classified as diffuse radio emission located in the cluster periphery, characterized by elongated morphologies and high degrees of polarization \citep[e.g.,][]{2012SSRv..166..187B,2019SSRv..215...16V}. Such elongated structures are commonly interpreted as merger-driven shocks \citep[][]{2003ApJ...593..599R} propagating toward the cluster outskirts.
However, growing observational evidence indicates that radio relics display a wide range of complex morphologies that do not always conform to this simplified picture \citep[see e.g.,][for one of the most complex morphologies]{Kamlesh_Abell2256}. Moreover, the interpretation and classification of diffuse radio emission are challenged by projection effects, which can alter the apparent morphology or structure of the emission.

A key first step toward understanding and modeling diffuse radio emission in the ICM is to characterize the global properties of cluster mergers, including the masses of the cluster components, gas density profiles, impact parameters, initial relative velocities, merger stage, and viewing angle. However, a major and recurring limitation is the availability of deep, multi-probe datasets required to adequately constrain these parameters.
In this work, we address this challenge by using the results of Improved Constraints on Mergers with SZ, Hydrodynamical simulations, Optical, and X-ray  \citep[ICM-SHOX,][]{2024ApJ...968...74S,2024EPJWC.29300050S} pipeline as our basis to model the radio emission  of the galaxy cluster MACS J0018.5+1626. ICM-SHOX is a powerful tool for inferring multiple galaxy cluster merger parameters by comparing multi-probe observables to mock observables generated from idealized hydrodynamical simulations. The multi-probe observables include projected total mass maps from strong gravitational lensing models, X-ray surface brightness and temperature maps, gas density and velocity information derived from the Sunyaev–Zel'dovich effect, and galaxy velocities inferred from spectroscopic redshifts of cluster members. One of the aims of this work is to demonstrate the potential of combining multi-probe data for individual galaxy clusters as a powerful approach to link the thermal and non-thermal components of the ICM.

Modeling radio emission in galaxy clusters requires accounting for particle acceleration. The mechanisms most commonly evoked to explain radio observations are the diffusive shock acceleration mechanism \citep[DSA e.g.,][]{1983RPPh...46..973D,1987PhR...154....1B} and turbulent (re)acceleration \citep[see][for a review]{2014IJMPD..2330007B}. Nevertheless, these mechanisms are not fully understood for the ICM. In terms of radio relics or radio shocks, the standard DSA of a thermal population of electrons seems not to be enough to explain the radio luminosity in various clusters hosting radio relics \citep[][]{2020A&A...634A..64B}. A proposed alternative is a population of mildly relativistic or ``fossil'' ($10 \lesssim \gamma \lesssim 10^4$) electrons that participate in the DSA mechanism rather than the thermal population \citep[e.g.,][]{2012ApJ...756...97K,Pinzke2013}. However, the origin of such fossil cosmic-ray electrons (CRe) remains an open question. For example, one alternative could be CRe in jets launched by AGN \citep[][]{2012ARA&A..50..455F,2014ApJ...785....1B,2017NatAs...1E...5V} that could be dispersed throughout the cluster environment up to several hundred kpc or even Mpc by shocks and/or sloshing motions \citep[][]{2021ApJ...914...73Z,2021A&A...653A..23V,2022MNRAS.510.4000F,2024ApJ...977..221D}.

In the present paper, we model the synchrotron emission of MACS J0018.5+1626, the first cluster explored as part of the ICM-SHOX project \citep{2024ApJ...968...74S,2024EPJWC.29300050S}. 
We use a hybrid particle–fluid numerical framework in which we couple the magnetohydrodynamic (MHD) code AREPO \citep[][]{2010MNRAS.401..791S,2011MNRAS.418.1392P} to a Fokker–Planck solver \citep{Donnert_2013}. This approach employs Lagrangian particles embedded in the large-scale MHD flow, each carrying its own particle momentum spectrum. MACS J0018.5+1626 is covered by the LOFAR Two-metre Sky Survey \citep[LoTSS;][]{2019A&A...622A...1S,2022A&A...659A...1S} in the 120–168~MHz frequency range, and we use these observations as a benchmark for comparison with our simulated radio emission. The LOFAR data reveal diffuse radio emission at the cluster center that can be classified as a radio halo. However, results from the ICM-SHOX pipeline suggest that MACS~J0018.5+1626 is undergoing a binary merger close to pericenter passage and is observed nearly along the merger axis. This configuration is consistent with two axial shocks viewed almost face-on and motivates a reassessment of the origin and nature of the observed diffuse radio emission. To test if this interpretation is plausible, we consider a simplified scenario in which the radio emission is powered by these two shocks via the DSA mechanism.

The paper is structured as follows. In Sec.~\ref{sec:methods}, we describe our methods including our numerical set-up and initial conditions in Sec.~\ref{sec:num_set-up_MHD}. In Sec.~\ref{sec:fkp} we describe the Fokker-Planck solver and the modeling of the radio emission. In Sec.~\ref{sec:results} we present the results of our simulations of MACS J0018.5+1626. We discuss our results and compare them to LOFAR observations at 144 MHz. We summarize our work in Sec.~\ref{sec:conclusions}.

%%%%%%%%%%%%%%%%%
\section{Methods}\label{sec:methods}
%%%%%%%%%%%%%%%%%

%%%%%%%%%%%%%%%%%%%%%%%%%%%%%%%%%%%%%%%%%%%%%%%%%%%%%%
\subsection{MHD Simulations}\label{sec:num_set-up_MHD}
%%%%%%%%%%%%%%%%%%%%%%%%%%%%%%%%%%%%%%%%%%%%%%%%%%%%%%

We carry out 3D non-radiative MHD simulations of idealized binary mergers of galaxy clusters using the moving-mesh code AREPO \citep[][]{2010MNRAS.401..791S,2011MNRAS.418.1392P} which employs a finite-volume Godunov method on an unstructured moving
Voronoi mesh to solve the ideal MHD equations, and a Tree-PM solver to compute the
self-gravity from gas and dark matter. The MHD Riemann problems at cell interfaces
are solved using a Harten-Lax-van Leer-Discontinuities (HLLD) Riemann solver \citep[see][]{2011MNRAS.418.1392P}. The condition on $\nabla \cdot \vec{B}$ is controlled using the Powell 8-wave scheme \citep[][]{1999JCoPh.154..284P} employed in \cite{2013MNRAS.432..176P, 2018MNRAS.480.5113M}. We keep track of the shocked regions/cells with the AREPO shock-finder throughout the simulations, where the minimum threshold for the Mach number is $\mathcal{M}_{\rm th}$=1.3 \citep[see details in][]{2015MNRAS.446.3992S}. This threshold is commonly selected to filter out weaker shocks in various MHD simulations \citep[see e.g.,][]{2003ApJ...593..599R, 2011MNRAS.418..960V,2015MNRAS.446.3992S,wittor17,dominguezfernandez2020morphology,dominguez2021}.

In this work we closely follow the numerical setup for idealized galaxy cluster mergers in AREPO used in previous works \citep{ZuHone2019,2021ApJ...914...73Z,2021Galax...9...91Z,2023arXiv231009422B,2024ApJ...977..221D,2025arXiv251212754B}. We choose the simulation domain to be a cubic box of size $L = 40$ Mpc with periodic boundary conditions for hydrodynamics, though the gravitational field has isolated boundary conditions. Each simulation initially has $\sim 1.15 \times 10^7$ gas cells and $10^7$ DM particles. The gas cells are initialized with the same mass but are allowed to undergo mesh refinement and derefinement during the simulation. The reference gas cell mass for our runs is $3.58 \times 10^7~\mathrm{M}_{\odot}$. We perform a total of 6 MHD simulations, while we run the Fokker-Planck solver (see Sec.~\ref{sec:fkp}) with different models for each MHD simulation. We present a total of 30 different simulations with modeled radio emission. Finally, we output the simulation data every $\Delta{t}=0.02$ Gyr and run the simulations well past the pericenter passage, $t_f = 1.9$ Gyr. 

%%%%%%%%%%%%%%%%%%%%%%%%%%%%%
\subsubsection{Initial conditions: MACS J0018.5+1626}
%%%%%%%%%%%%%%%%%%%%%%%%%%%%%

We initialize our simulations with two spherically symmetric galaxy clusters in hydrostatic and virial equilibrium following the method described in \cite{2011ApJ...728...54Z}. The gas in each cluster is fully ionized, with an assumed mean molecular weight $\mu = 0.6$, ideal, with a constant adiabatic index $\gamma_0 = 5/3$, and, magnetized, with an initial plasma beta $\beta_p = P_{th}/P_B = 100$ and a variation of $\beta_p =200$. The initial magnetic field is tangled. We generate a divergence-free Gaussian random magnetic field
following the method of \cite{2019ApJ...883..118B}. The initial magnetic field power spectrum follows a Kolmogorov spectrum with a minimum wavenumber $k_0=2\pi / \lambda_0$ and a maximum wavenumber modeled with an exponential cutoff at $k_1=2\pi / \lambda_1$. The corresponding scales are $\lambda_0=500$ kpc and $\lambda_1=10$ kpc.

% ref to 1994sse..book.....K

In this work, we present the radio emission modeling of the merging cluster MACS J0018.5+1626. This cluster is located at $z_{\rm obs}=0.5456$ \citep[][]{2007ApJ...661L..33E}, has an X-ray-derived mass of ${\rm M}_{500}=1.65 \times 10^{15}~{\rm M}_{\odot}$ \citep[][]{2010MNRAS.406.1759M,2019ApJ...880...45S}, and a strong gravitational lensing-derived mass of ${\rm M}_{500}=1.1 \times 10^{15}~{\rm M}_{\odot}$ \citep[see more details in Sec. 3.2 of][]{2024ApJ...968...74S}.
This is the first cluster explored as part of the ICM-SHOX project \citep{2024ApJ...968...74S,2024EPJWC.29300050S}.  In \cite{2024ApJ...968...74S}, the authors successfully constrained the global properties of MACS J0018.5+1626: component masses, gas profiles, viewing angle, epoch, initial impact parameter, and initial relative velocity. We set up our simulations according to the best-match initial condition parameters obtained with the ICM-SHOX pipeline in \cite{2024ApJ...968...74S}. In the following, we briefly describe the initial conditions and refer the reader to \cite{2024ApJ...968...74S} for a more detailed description:

\begin{itemize}
    \item[1)] The clusters are initially separated by a distance of 3 Mpc.  The initial relative velocity, $v_i$, is defined along the merger axis ($X$-axis) while the initial impact parameter, $b$, is defined along an orthogonal axis ($Y$-axis).
    
    \item[2)] The total mass of each cluster follows a truncated Navarro–Frenk–White (NFW) profile \citep[][]{2009JCAP...01..015B}. We use the concentration-mass relation model from \citet{2019ApJ...871..168D} with cosmological parameters from \citet{2020A&A...641A...6P}.

    \item[3)] The electron number density follows the modified $\beta$-model profile from
\cite{2006ApJ...640..691V}. The total gas mass is normalized to ${\rm M}_{500c}$ through the cosmic gas fraction, $f_{\rm gas} = 0.115$.
\end{itemize}

We summarize the parameters used for the initial conditions and profiles of MACS J0018.5+1626 in Tab.~\ref{tab:ics}. While these initial conditions define our nominal runs, we also include other runs with different initial relative velocities and impact parameters. Specifically, $v_i=3000$ km~s$^{-1}$ and $b=100$ kpc, which are also favored parameters by the ICM-SHOX pipeline to reproduce the observations of MACS J0018.5+1626. We summarize all the runs in Tab.~\ref{tab:runs}.
Finally, we ensure that the initial condition is free of spurious gas density and pressure fluctuations by following the method from \cite{2021ApJ...914...73Z}, where a mesh relaxation step for $\sim 100$ timesteps is performed. 

\begin{deluxetable}{lc}
\tablecaption{Initial conditions\label{tab:ics}}
\tablehead{
\colhead{Parameter} & \colhead{Value}
}
\startdata
\multicolumn{2}{c}{\textbf{Initial conditions}} \\
\hline
$M_{500}$                & $1.15\times10^{15}\,M_\odot$ \\
$z_{\rm obs}$            & 0.546 \\
$R = M_{200,1}:M_{200,2}$& 1.5 \\
$b$                      & 250\,kpc \\
$v_i$                    & 2400\,km\,s$^{-1}$ \\
\hline
\multicolumn{2}{c}{\textbf{Modified $\beta$ profile}} \\
\hline
$r_{c,1}$, $r_{c,2}$     & $0.3\,r_{2500,1}$, $0.5\,r_{2500,2}$ \\
$\alpha_1$, $\alpha_2$   & 0.3, 0.1 \\
$\beta_1$, $\beta_2$     & 0.67, 0.67 \\
$\gamma_1$, $\gamma_2$   & 3, 3 \\
$\epsilon_1$, $\epsilon_2$ & 3, 3 \\
$r_{s,1}$, $r_{s,2}$     & $1.1\,r_{200,1}$, $1.1\,r_{200,2}$ \\
\enddata
\tablecomments{
Initial parameters used in our simulations, as reported in \citet{2024ApJ...968...74S}. Subscript~1 refers to the main cluster, and subscript~2 to the secondary cluster. These values correspond to our nominal setup. Variations from this configuration include runs with $v_i = 3000$~km~s$^{-1}$ and $b = 100$~kpc (see Table~\ref{tab:runs}).
}
\end{deluxetable}

%%%%%%%%%%%%%%%%%%%%%%
\subsubsection{Tracer particles} \label{sec:num_setup_tracers}
%%%%%%%%%%%%%%%%%%%%%%

We use tracer particles to represent CRe in our simulations. Each tracer is advected
with the fluid flow, and therefore, keeps track of the fluid's Lagrangian evolution.  We use the ``velocity field'' tracer particles implemented in AREPO \citep[see][]{2013MNRAS.435.1426G}. The numerical scheme works by moving each tracer particle according to the linearly interpolated velocity field of its parent gas cell.

We assign the tracers to the MHD simulation by placing the tracers covering only the two galaxy clusters region up to $r_{\rm max}=5$ Mpc. 
We have a total of $N_{\rm tr} = 10^7$ tracer particles. We assign a constant mass of $m_{\rm tr} = 3.58 \times 10^7 \, \rm M_{\odot}$ (the same as the reference gas cell mass). Given this constant mass, throughout the simulation we can assign each of the tracers a volume
\begin{equation}
    V_{\rm tr,i} = \frac{m_{\rm tr, i}}{\rho_{\rm cell}},
\end{equation}
where $\rho_{\rm cell}$ is the gas density at the parent cell at the current simulation time.

We associate a cosmic-ray particle momentum distribution to each tracer, and evolve each distribution independently (see Sec.~\ref{sec:fkp}). This evolution depends on the state of the fluid variables that the tracer encounters as a function of time. Therefore, 
we interpolate the relevant fluid variables onto each tracer from the gas cells. We describe this method in App.~\ref{app:interpolation}. The final step is to compute the synchrotron emissivity to model the radio emission in galaxy clusters (see Sec.~\ref{sec:sync}). Combining tracers with MHD grids is a relatively common method to model radio emission in different MHD setups and with different numerical codes
\citep[see e.g.,][for the numerical implementation of this method in different astrophysical simulation codes]{Donnert_2013, 2013ApJ...762...78Z, 2015ApJ...801..146Z, wittor17, 2018ApJ...865..144V, 2019MNRAS.488.2235W, 2023MNRAS.519..548B}.

In this work, we limited ourselves to analyze the shock acceleration of CRe. Therefore, we selected only those tracers that have encountered a shocked cell(s). 

\begin{deluxetable*}{lccccc}
\tablecaption{Simulation runs and shock-acceleration efficiency parameters\label{tab:runs}}
\tablehead{
\colhead{\textbf{Run ID}} &
\multicolumn{3}{c}{\textbf{Merger parameters}} &
\multicolumn{2}{c}{\textbf{Shock acceleration efficiency}} \\
\cline{2-4}\cline{5-6}
& \colhead{$v_i$ [km s$^{-1}$]} &
\colhead{$\beta_p$} &
\colhead{$b$ [kpc]} &
\colhead{$\eta_0$} &
\colhead{Model}
}
\startdata
% -------------------------------
v2400\_beta200\_b250\_DSAconst\_eta0.1   & 2400 & 200 & 250 & 0.1   & Constant \\
v2400\_beta200\_b250\_DSAconst\_eta0.01  &      &     &     & 0.01  & Constant \\
v2400\_beta200\_b250\_DSAconst\_eta0.001 &      &     &     & 0.001 & Constant \\
\hline
v2400\_beta200\_b250\_DSAK07\_eta1       & 2400 & 200 & 250 & 0.1   & \citealt{Kang_2007} \\
v2400\_beta200\_b250\_DSAK07\_eta0.01    &      &     &     & 0.01  & \citealt{Kang_2007} \\
v2400\_beta200\_b250\_DSAK07\_eta0.001   &      &     &     & 0.001 & \citealt{Kang_2007} \\
\hline
v2400\_beta100\_b250\_DSAconst\_eta0.1   & 2400 & 100 & 250 & 0.1   & Constant \\
v2400\_beta100\_b250\_DSAconst\_eta0.01  &      &     &     & 0.01  & Constant \\
v2400\_beta100\_b250\_DSAconst\_eta0.001 &      &     &     & 0.001 & Constant \\
\hline
v2400\_beta100\_b250\_DSAK07\_eta1       & 2400 & 100 & 250 & 0.1   & \citealt{Kang_2007} \\
v2400\_beta100\_b250\_DSAK07\_eta0.01    &      &     &     & 0.01  & \citealt{Kang_2007} \\
v2400\_beta100\_b250\_DSAK07\_eta0.001   &      &     &     & 0.001 & \citealt{Kang_2007} \\
\hline
v2400\_beta100\_b100\_DSAK07\_eta1       & 2400 & 100 & 100 & 0.1   & \citealt{Kang_2007} \\
v2400\_beta100\_b100\_DSAK07\_eta0.01    &      &     &     & 0.01  & \citealt{Kang_2007} \\
v2400\_beta100\_b100\_DSAK07\_eta0.001   &      &     &     & 0.001 & \citealt{Kang_2007} \\
\hline
% -------------------------------
v3000\_beta200\_b250\_DSAconst\_eta0.1   & 3000 & 200 & 250 & 0.1   & Constant \\
v3000\_beta200\_b250\_DSAconst\_eta0.01  &      &     &     & 0.01  & Constant \\
v3000\_beta200\_b250\_DSAconst\_eta0.001 &      &     &     & 0.001 & Constant \\
\hline
v3000\_beta200\_b250\_DSAK07\_eta1       & 3000 & 200 & 250 & 0.1   & \citealt{Kang_2007} \\
v3000\_beta200\_b250\_DSAK07\_eta0.01    &      &     &     & 0.01  & \citealt{Kang_2007} \\
v3000\_beta200\_b250\_DSAK07\_eta0.001   &      &     &     & 0.001 & \citealt{Kang_2007} \\
\hline
v3000\_beta100\_b250\_DSAconst\_eta0.1   & 3000 & 100 & 250 & 0.1   & Constant \\
v3000\_beta100\_b250\_DSAconst\_eta0.01  &      &     &     & 0.01  & Constant \\
v3000\_beta100\_b250\_DSAconst\_eta0.001 &      &     &     & 0.001 & Constant \\
\hline
v3000\_beta100\_b250\_DSAK07\_eta1       & 3000 & 100 & 250 & 0.1   & \citealt{Kang_2007} \\
v3000\_beta100\_b250\_DSAK07\_eta0.01    &      &     &     & 0.01  & \citealt{Kang_2007} \\
v3000\_beta100\_b250\_DSAK07\_eta0.001   &      &     &     & 0.001 & \citealt{Kang_2007} \\
\hline
v3000\_beta100\_b100\_DSAK07\_eta1       & 3000 & 100 & 100 & 0.1   & \citealt{Kang_2007} \\
v3000\_beta100\_b100\_DSAK07\_eta0.01    &      &     &     & 0.01  & \citealt{Kang_2007} \\
v3000\_beta100\_b100\_DSAK07\_eta0.001   &      &     &     & 0.001 & \citealt{Kang_2007} \\
\enddata
\tablecomments{
Simulation IDs for runs with different merger initial parameters (see also Table~\ref{tab:ics}) and electron shock-acceleration efficiencies from the thermal pool. The case $v_i = 3000$~km~s$^{-1}$ corresponds to the highest initial relative velocity constrained by the ICM-SHOX pipeline.
}
\end{deluxetable*}

%%%%%%%%%%%%%%%%%%%%%%%%%%%%%%
\subsection{Fokker-Planck Solver}\label{sec:fkp}
%%%%%%%%%%%%%%%%%%%%%%%%%%%%%%

The Fokker-Planck equation describes the evolution of the distribution of charged particles due to being accelerated by interactions with plasma waves, energy gains/losses, and direct injections/removals. Due to its complexity, it is common to make certain assumptions\footnote{Some of this material can be found in \citet{1995ApJ...446..699P}}, such as:

\begin{itemize}
    \item[1)] \textit{The distribution of particle momenta is isotropic}. This is true if the rate of pitch angle scattering is much faster than the change in energy rate or rate of particle escape for example. This means that we can integrate the Fokker-Planck equation over a timescale sufficiently large compared to the pitch angle diffusion, but shorter than the timescale for energy diffusion, effectively leaving out the pitch angle variable.
    \item[2)] \textit{No dependence on the spatial variable}. This dependence can be eliminated by using a volume-averaged distribution.
    \item[3)] \textit{Spatial diffusion}. The scattering mean free path of particles must be much smaller than the size of the turbulent region of interest. Particles leaving this region will no longer be accelerated. The loss of these particles can then be modeled with an escape term $T$.
\end{itemize}

Under these assumptions the evolution of the electron spectrum can be described by the following Fokker-Planck equation \citep[see e.g.,][]{1970JCoPh...6....1C}:
\begin{equation}\label{eq:FP}
    \frac{\partial f}{\partial t} = \frac{1}{p^2}\frac{\partial}{\partial p} \left[ p^2 D_{pp}(x) \frac{\partial f}{\partial p} + p^2 H(p) f \right]
    - \frac{f}{T(p)} + q(p,t),
\end{equation}
where  $f(p,t) 4\pi p^2dp$ is the number of particles in the interval $p$ and $p+dp$ at time $t$, $D_{pp}$ is the turbulent re-acceleration coefficient, 
$T(p)$ describes the spatial loss of particles outside of the turbulent region, $q(p,t)$ is a source term representing the the rate of particle injection into the acceleration region, and the function $H(p)$ is a generalized cooling function defined as:
\begin{equation}
    H(p) = \sum_{i} \left\lvert \frac{dp}{dt} \right\rvert - \frac{1}{3} (\nabla \cdot v) p,
\end{equation}
where energy losses and adiabatic expansion are considered. The total number of particles is defined as
%and energy density are
%
\begin{equation}
    N_{tot} = \int 4 \pi p^2 f(p) dp.
\end{equation}
Defining $n(p) = 4\pi p^2 f(p)$ and $Q(p,t)=4\pi p^2 q(p,t)$, we can re-write Eq.~\ref{eq:FP} in the following form:
\begin{equation}\label{eq:FP-n}
    \frac{\partial n}{\partial t} = \frac{\partial}{\partial p} \left[ D_{pp} \frac{\partial n}{\partial p} + \left( H - 2\frac{D_{pp}}{p} \right) n \right]
    - \frac{n}{T} + Q.
\end{equation}
We solve Eq.~\ref{eq:FP-n} following the procedure in \citet{Donnert_2013}. In the following, we briefly describe the numerical method to solve the Fokker-Planck equation:
\begin{itemize}
    \item\textit{Finite difference scheme}: Following the flux conservative difference scheme proposed by \citet{1970JCoPh...6....1C}, we discretize Eq.~\ref{eq:FP-n} as 
\begin{equation}\label{eq:discretized}
 \frac{n_{m}^{n+1}-n_{m}^{n}}{\Delta t}   =
  \frac{F_{m+1/2}^{n+1} - F_{m-1/2}^{n+1}}{\Delta p_m} - \frac{n_{m}^{n+1}}{T_m} + Q_{m}
\end{equation}
for $m=0,...,M-1$, where the $m$ and $n$ indices correspond to momentum and time, respectively.
%, $A_m$ is the phase-space factor 
The particle flux, $F$, arises from re-writing Eq.~\ref{eq:FP-n} as:
\begin{equation}
    F(p,t) = C(p,t) \frac{\partial n}{\partial p} - B(p,t)n,
\end{equation}
where
\begin{equation}
    C(p,t) = D_{pp}(p,t), 
\end{equation}
and
\begin{equation}
    B(p,t) = H(p,t) - 2\,\frac{D_{pp}(p,t)}{p}.
\end{equation}
Given a particular choice of $F_{m}^{n+1}$ Eq.~\ref{eq:discretized} results in a tridiagonal system of linear
equations, which can be written as
\begin{equation}
\begin{split}
-a_m n_{m-1}^{n+1} + b_m n_m^{n+1} - c_m n_{m+1}^{n+1} &= \Delta t Q_m + n_m^n, \\
a_0 &= c_{M-1} = 0.
\end{split}
\end{equation}
This tridiagonal system  \citep[see Eqs.~26 and 27 in][]{Donnert_2013} is then solved using a tridiagonal matrix solver \citep[e.g.,][]{1992nrfa.book.....P}. Finally, applying a back substitution algorithm, the updated energy spectrum, $n_{m}^{n+1} = d_{m}^{n+1}n_{m+1}^{n+1} + e_{m}^{n+1}$, can be obtained, where\footnote{We note that this part was omitted from the original paper in \citet{Donnert_2013}}
\begin{equation}
d_{m}^{n+1} = \frac{a_m}{b_m - c_m d_{m-1}^{n+1}}, 
\end{equation}
and
\begin{equation}
e_{m}^{n+1} = \frac{n_m^n + c_m e_{n-1}^{m+1}}{b_m - c_m d_{m-1}^{n+1}},
\end{equation}
 \item\textit{Mesh}: We use a logarithmic momentum grid $p_i=p_{\rm min} \, 10^{i \Delta \log p}$ with $M$ points where the momentum grid-stepping is $\Delta \log p = \log_{10}(\frac{p_{\rm max}}{p_{\rm min}})(M - 1)^{-1}$. 
 
 \item\textit{Energy losses}: We take into account inverse Compton (IC), synchrotron, Coulomb scattering, and Bremsstrahlung energy losses and adiabatic expansion \citep[see Eqs.~2 and 5 in][]{Donnert_2013}. We show this in App.~\ref{app:shock_energy}.
\end{itemize}

Our numerical method is the same as in \citet{Donnert_2013}. However, our updated version of the code differs in the following:  
1) we do not use the compression algorithm based on interpolation by cubic Hermite polynomials \citep[see Section 3.5,][]{Donnert_2013}; 2) we have extended the input/output of the code to read general HDF5 files; 3) we have included an emission module to directly output the synchrotron emissivity at each tracer particle (see also Sect.~\ref{sec:sync}). This has been included for practical purposes of not storing all the particle spectra if desired; 4) we have included the redshift dependence in the inverse Compton energy losses; 5) we have included different shock efficiency models (see Sec.~\ref{sec:DSA}). The precise details will be detailed in a forthcoming paper (Dom\'inguez-Fern\'andez in prep.), which will also introduce another turbulent acceleration model. For the present modeling of MACS J0018.5+1626, no spatial diffusion, escape ($T \rightarrow \infty$) or turbulent acceleration terms are considered. Finally, we considered $M=128$ points for the momentum grid, and a fixed time-step for the Fokker-Planck solver with $dt_{\rm FP}=0.686$ Myr for all simulations in this work. This ensures that the Fokker-Planck solver timestep is smaller than the simulation timestep and the particle distribution evolves several cycles before the next simulation timestep \citep[see more details in][]{Donnert_2013}. For reference, the typical electron cooling timescales due to synchrotron losses are of the order of 50-100 Myr in the ICM.

The simulations in this work are non-cosmological. Therefore, we assign a redshift to each simulation time\footnote{We assume a $\Lambda$CDM model with the WMAP 3 year cosmological parameters \citep[][]{2007ApJS..170..377S}}. To do so, we select a time close to pericenter passage, $t_{\rm peri}$, as the time corresponding to the observed redshift of MACS J0018.5+1626, $z_{\rm obs}=0.5456$. Specifically, we select the time at which the radio power reaches its maximum (see Sec.~\ref{sec:radio_power}). This is a reasonable assumption given that \citet{2024ApJ...968...74S} finds that MACS J0018.5+1626 is a system very close to the pericenter passage. We note that $t_{\rm peri}$ can differ depending on the initial conditions mainly due to the assumed initial relative velocity. Therefore, the redshifts assigned will be slightly different for each simulation in this work (see Tab.~\ref{tab:runs}); however, these differences are negligible and do not lead to significant variations in our results.

%%%%%%%%%%%%%%%%%%%%%%%%%%%%%%%%%%%%%
\subsubsection{The distribution function}
%%%%%%%%%%%%%%%%%%%%%%%%%%%%%%%%%%%%%

\begin{figure*}
    \centering
    \includegraphics[width=1\columnwidth]{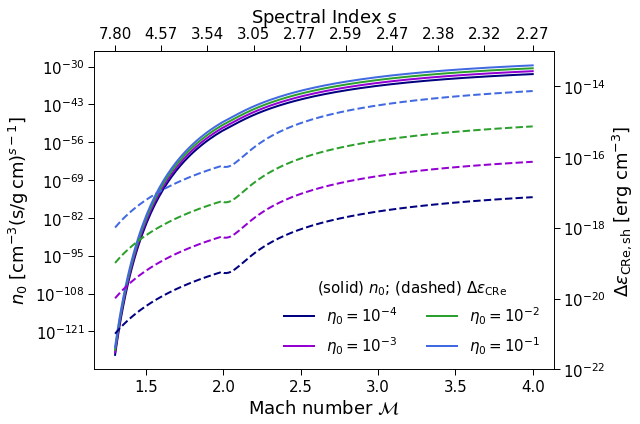} 
    \includegraphics[width=1\columnwidth]{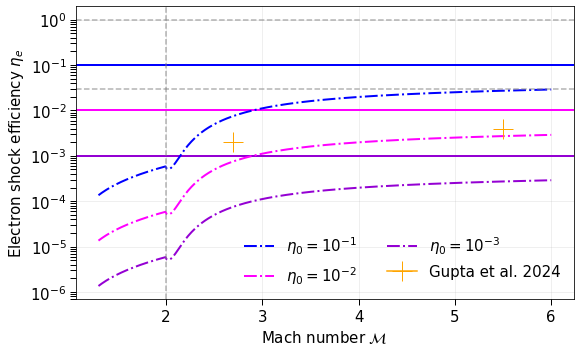}
    \caption{\textit{Left panel}: CRe normalization, $n_0$, (left axis) and CRe energy density injected from the shock, $\Delta \varepsilon_{\text{CRe},sh}$, (right axis) as a function of the Mach number. We considered the $\eta(\mathcal{M})$ model proposed by \citet{Kang_2007}. We assume a fixed density of $8.7\times10^{-28}$ g~cm$^{-3}$ and a fixed temperature of $10^7$ K for Eq.~\ref{eq:norm}. \textit{Right panel}: Shock efficiency models considered for this work: constant (solid lines) and the \citet{Kang_2007} model (dot-dashed lines; see Eq.~\ref{eq:Kang}).}
    \label{app:CRe_vs_Mach}
\end{figure*}

For a given initial momentum distribution function,
\begin{equation}\label{eq:part_injection}
n(p,t=0) = n_0 \, p^{-s} ,   
\end{equation}
 where $n_0$ is the normalization factor in units of [$\mathrm{cm}^{-3} (\mathrm{s}/\mathrm{g}\, \mathrm{cm})^{s -1}$], the total CR number density and CR energy density are derived from
\begin{equation}
\begin{split}
    n_{\rm CRe} &= \int_{p_{0}}^{\infty} n_0 \, p^{-s} dp \\
    &= \frac{n_0}{s - 1} p_{0}^{1-s},
\end{split}
\end{equation}
 and,
 \begin{small}
\begin{equation}\label{eq:energy_integral}
\begin{split}
   \varepsilon_{\rm CRe}  &= \int_{p_{0}}^{\infty} n_0 p^{-s} E(p) dp \\
   &= \frac{n_0 \, m_e c^2}{s - 1} \\
   & \times \left[ -p^{1-s} {}^{}_2F_1\left(-\frac{1}{2}, \frac{1-s}{2}; \frac{3-s}{2};-\frac{p^2}{m_e^2 c^2} \right) \right]_{p_0}^{\infty},
\end{split}
\end{equation}
\end{small}
where $\varepsilon_{\rm CRe}$ is in units of [$\mathrm{erg} \, \mathrm{cm}^{-3}$], $p_0$ is a minimum momentum, 
\begin{equation}
    E(p)=\left[\sqrt{1+(p/m_e c)^2} - 1\right]m_ec^2
\end{equation}
is the kinetic energy of an electron with momentum $p$, and $^{}_2F_1(a,b;c;x)$ is the hypergeometric function.

%%%%%%%%%%%%%%%%%%%%%%%%%%%%%%%%%%%%%%%%%
\subsubsection{Shock primaries injection}
\label{sec:DSA}
%%%%%%%%%%%%%%%%%%%%%%%%%%%%%%%%%%%%%%%%%%

In this section, we briefly describe the code implementation of the shock primaries injection via DSA \citep[see][for more details]{Donnert_2013}, $Q(p)$, in the code which is defined as:
\begin{equation}
    Q(p) = \frac{n_{0} \cdot p^{-s}}{\Delta t}, 
\end{equation}
where $\Delta t$ is the time step and $s$ is the spectral index computed as:
\begin{equation}
    s = 2 \frac{\mathcal{M}^2 + 1}{\mathcal{M}^2 - 1}.
\end{equation}
If the shock's Mach number $\mathcal{M}$ is less than the threshold $\mathcal{M}_{\text{thr}} = 1.3$, no injection occurs. The same threshold is adopted as in the shock finder (see Sec.~\ref{sec:num_set-up_MHD}), consistent with the methodology of \citet{Donnert_2013}. We note that the minimum Mach number above which DSA operates efficiently in shocks in the ICM is uncertain. Recent studies suggest that DSA may be effective only for shocks with sonic Mach numbers exceeding a critical value, $\mathcal{M}_{\rm cr} \sim 2.3$ \citep[see, e.g.,][]{2019ApJ...876...79K,2021ApJ...915...18H}. However, this issue remains an open question. The normalization, $n_{0}$,  ensures that the CRe energy spectrum integrates to the desired shock's energy density. In the following, we describe the steps followed to compute this normalization factor.

A natural expectation is that only suprathermal particles with a gyroradius greater than the shock thickness\footnote{The shock thickness is of the order of the upstream ion Larmor radii $R_L = v_{th,i}/\omega_{ci}$, where $v_{th,i}$ is the ion thermal speed and $\omega_{ci}$ is the ion cyclotron frequency. This can be 
expressed as $R_L = \sqrt{\beta_p}\,d_i$, where 
 $d_i = c/\omega_{pi}$ is the ion skin depth. For typical ICM conditions, $\beta_p \sim 20$--$100$ and $n_e \sim 10^{-4}$--$10^{-3}$ cm$^{-3}$, giving $d_i \sim 10^3$--$10^4$ km and hence $R_L \sim 10^4$--$10^5$ km \citep[see e.g.][]{2009A&ARv..17..409T}.
} 
are expected to cross the shock transition layer. In reality, \citealt{2015ApJ...798L..28C} and \citealt{2025ApJ...994L..34G} showed that injection into DSA depends on proton and electron speeds, rather than their gyroradii. 
Specifically, particles with momentum $p\gtrsim 5 \, p_{\rm th,p}$ exhibit non-thermal tails in their distribution,
where $p_{\rm th,p}=\sqrt{2 m_p k_B T_{dw}}$ and $T_{dw}$ is the downstream temperature. The ratio of CRe  number to cosmic-ray protons (CRp) number $K_{e/p}$ determines the injection of electrons into DSA. In strong supernova remnant shocks, observational constraints suggest $K_{e/p}\sim10^{-2}$ at $\sim$GeV energies, corresponding to electrons contributing $\sim1\%$ of the Galactic CR flux \citep[][]{2008ApJ...680L..41R,2002cra..book.....S}. 
However, the value of $K_{e/p}$ for weak shocks in the ICM is rather uncertain. Particle-in-cell (PIC) simulations have shown that $K_{e/p}$ depends on the local magnetic geometry of shocks. 
For example, quasi-parallel shocks are efficient at accelerating both electrons and ions \citep[][]{2014ApJ...783...91C,2015PhRvL.114h5003P} and
quasi-perpendicular shocks can accelerate electrons \citep[][]{2014ApJ...797...47G}. However, whether a strictly perpendicular shock can efficiently accelerate particles (electrons and ions) remains an open question. In reality, mixed regions of quasi-parallel and quasi-perpendicular shocks should facilitate the acceleration of both electrons and protons, making it difficult to estimate the value of $K_{e/p}$. \citet{2014MNRAS.437.2291V} showed that a nominal value of $K_{e/p}=0.01$ is the most conservative assumption in an analysis based on observations of radio relics. The authors showed that this value minimizes the acceleration of CRp while still being in agreement with the observed upper limits of hadronic gamma-ray emission.
Given these uncertainties, we follow a simple model \citep[][]{Kang_2010}
in which the downstream electrons above a certain injection or transition\footnote{The transitional momentum above which the downstream particle distribution transitions into a power-law.}
momentum, $p_{\rm inj}=Q_{\rm inj} p_{\rm th,p}$, are assumed to be injected to the CR population, where $Q_{\rm inj}$
is a parameter that depends on the
shock Mach number and turbulent magnetic field amplitude in
the thermal leakage injection model.

We determine the injection or transition momentum $p_{\rm inj}$ for the definition of the injection function $Q(p)$ by determining the momentum at which the Maxwellian thermal distribution
\begin{equation}
n_{\text{th}} (p) = n_{\text{dw}} \, \frac{4 \pi p^2}{ \left( 2 \pi m_e k_B T_{dw} \right)^{3/2}} \exp\left(-\frac{p^2}{2 m_e k_B T_{dw}}\right)
\end{equation}
transitions into the power-law given by DSA, %$n_{\text{th}}(p_{\text{inj}}) = n_{\text{CRe}}(p_{\text{inj}})$, where
\begin{equation}
n_{\mathrm{CRe}}(p) = n_0 \, p^{-s}.
\end{equation}
We determine the value of $p_{\rm inj}$ by imposing continuity in the distribution function, $n_{\text{th}}(p_{\text{inj}}) = n_{\text{CRe}}(p_{\text{inj}})$, and using a bisection root-finding method. We define a weighted difference,
\begin{equation}
\delta = \frac{n_{\text{th}}(p_{\text{inj}}) - n_{\text{CRe}}(p_{\text{inj}})}{n_{\text{th}}(p_{\text{inj}}) + n_{\text{CRe}}(p_{\text{inj}})},
\end{equation}
and iterate until it converges, $|\delta| < \delta_0$, to an accuracy factor, $\delta_0$, defined by the user\footnote{In our case, we use $\delta_0 \sim 10^{-3}$.}. We compute the normalization factor by evaluating the CR energy density integral (see Eq.~\ref{eq:energy_integral}) in units of [erg $\mathrm{cm}^{-3}$]. The integral is evaluated at $p_0 = p_{\textrm{inj}}$. Finally, we obtain the normalization factor:
\begin{equation}\label{eq:norm}
    n_0 = \frac{\Delta \varepsilon_{\mathrm{CRe, sh}} }{\int_{p_{\textrm{inj}}}^{\infty} p^{-s} E(p) dp},
\end{equation}
in units of [$\textrm{cm}^{-3}\, (\mathrm{s}/\mathrm{g}\, \mathrm{cm})^{s-1}$].
We use a trapezoidal rule in the code to compute the integral defined in Eq.~\ref{eq:energy_integral}.

We compute the energy density of CRs, $\Delta \varepsilon_{\text{CR},sh}$, by assuming that the shock-dissipated energy $E_{\rm diss}$ that is converted into the acceleration of CR:
\begin{equation}
   E_{\text{CR}} = \eta(\mathcal{M}) \, E_{\mathrm{diss}}.
\end{equation}
We refer the reader to App.~\ref{app:shock_energy} for the details on the computation of the shock-dissipated energy.
The shock efficiency function $\eta(\mathcal{M})$ describes what fraction of the shock's
energy is injected to the particles as a function of the strength of the shock. 

The shock efficiency function in the ICM is poorly constrained. For this reason, we explore different models. In the first model we assume a simple constant efficiency, $\eta(\mathcal{M}) = \eta_0$. In the second model, we assume the model from \citealt{Kang_2007},
\begin{equation}\label{eq:Kang}
    \eta(\mathcal{M}) = 
\begin{cases} 
1.96 \times 10^{-3} \cdot (\mathcal{M}^2 - 1) & \mathcal{M} \leq 2, \\
\sum_{i=0}^{4} c_i \cdot \frac{(\mathcal{M} - 1)^i}{\mathcal{M}^4} & \mathcal{M} > 2,
\end{cases}
\end{equation}
where $c_i$ are empirically derived coefficients: $c_0 = 5.46$, $c_1 = -9.78$, $c_2 = 4.17$, $c_3 = -0.334$, $c_4 = 0.570$. This model is motivated by results from cosmological simulations in \citealt{Kang_2007}. The authors estimated the CR acceleration
efficiency $\eta(\mathcal{M})$ as a function of the shock Mach number assuming Bohm diffusion and incorporating
self-consistent treatments of thermal leakage injection and Alfv\'en wave propagation
\citep[][]{2007APh....28..232K}. 

\begin{figure*}
    \centering
    \includegraphics[width=0.88\columnwidth]{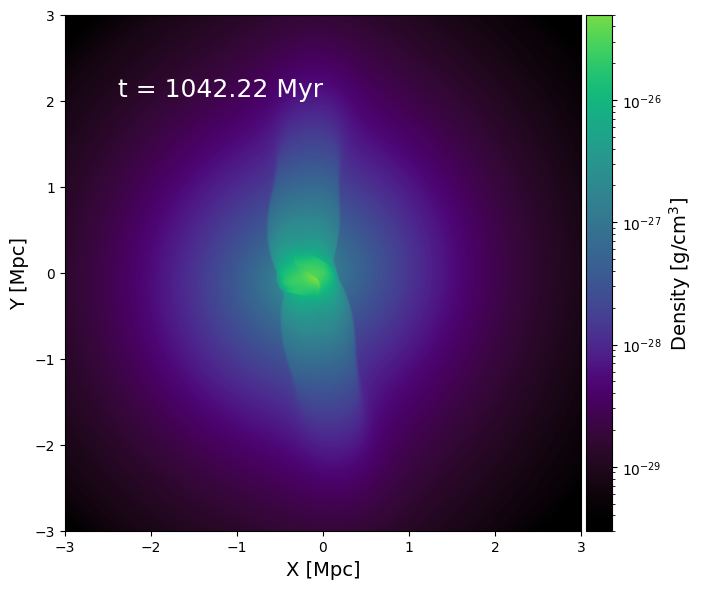}
    \includegraphics[width=0.88\columnwidth]{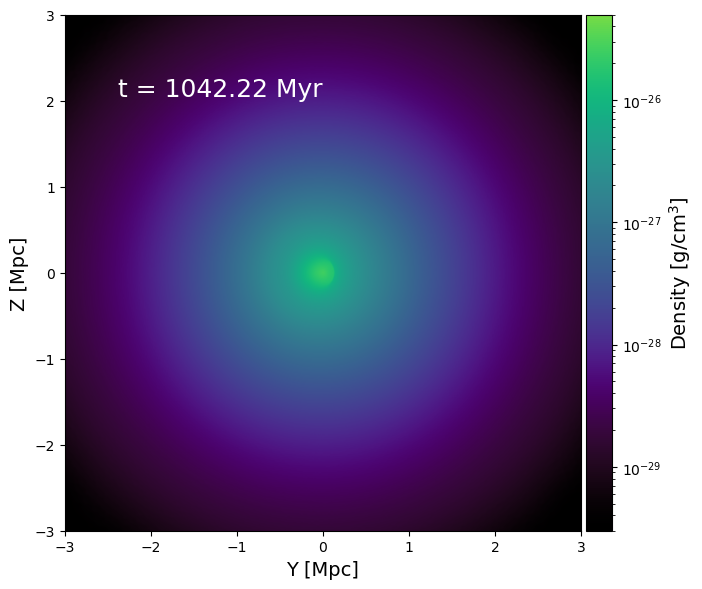} \\
    \includegraphics[width=0.88\columnwidth]{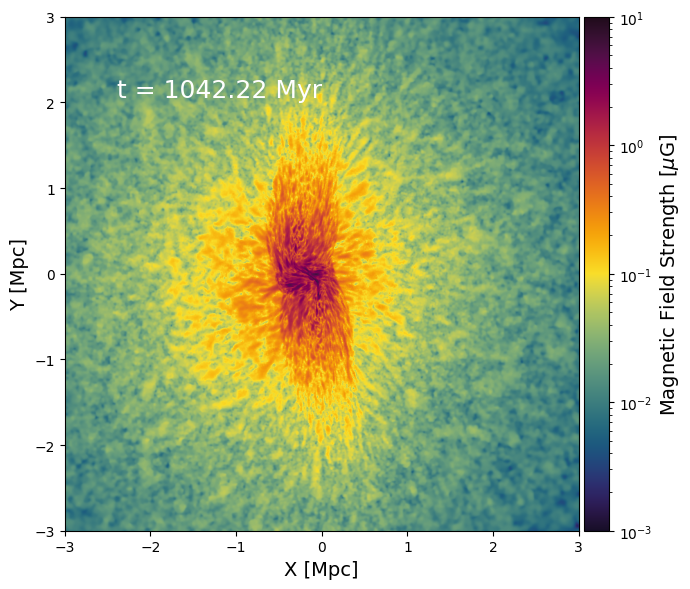}
    \includegraphics[width=0.88\columnwidth]{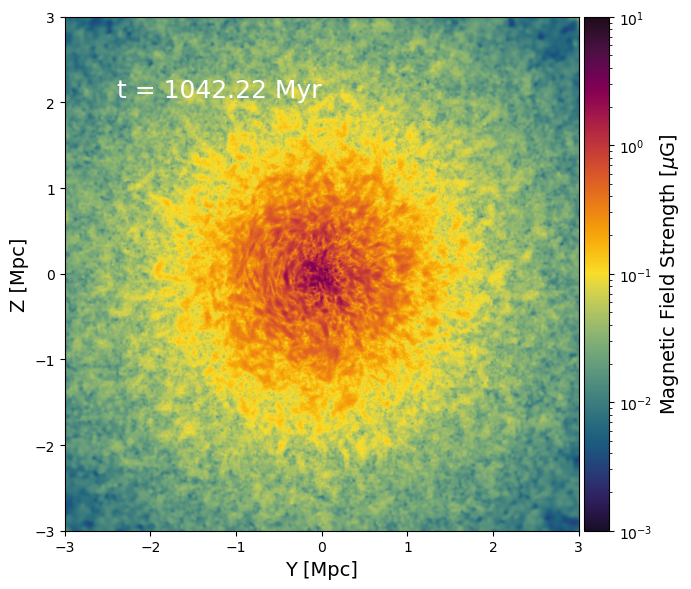} \\
    \includegraphics[width=0.88\columnwidth]{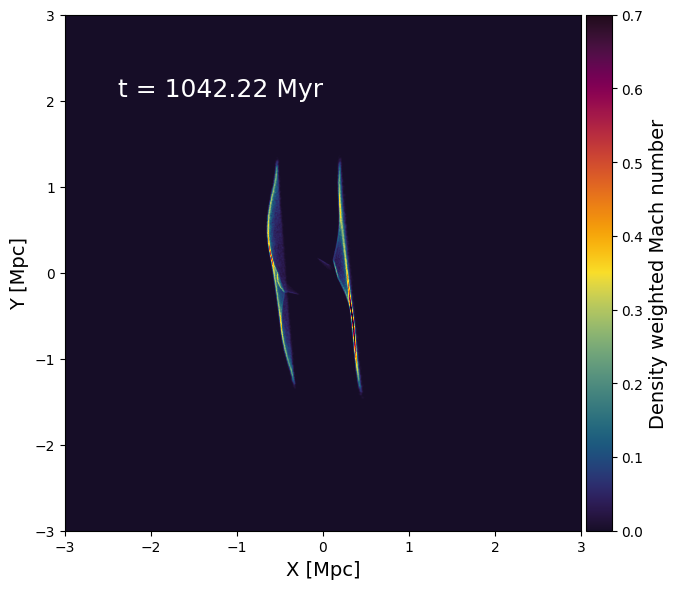}
    \includegraphics[width=0.88\columnwidth]{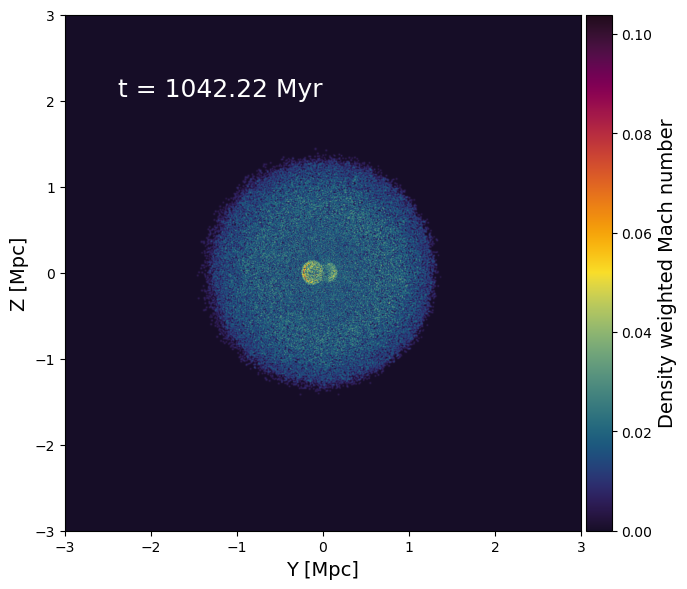} \\
    \caption{\textit{Upper and middle panels}: Density and magnetic field strength slice plots of the run initialized with $v_i=2400$ km~s$^{-1}$, $b=250$ kpc, $\beta_p=200$. \textit{Lower panels}: Density-weighted Mach number projection maps. For all rows, the left panels show a slice or projection normal to the $Z$-axis (perpendicular to the merger axis), and the right panels show a slice or projection normal the $X$-axis (along the merger axis).}
    \label{fig:proj_maps}
\end{figure*}

The injection energy density is effectively channeled into accelerating both protons and electrons. The electron shock efficiency is related to the proton shock efficiency via the CR electron-to-proton ratio, $\eta_e \simeq K_{e/p} \, \eta$. However, as discussed before, $K_{e/p}$ is highly uncertain. 
Hence, the electron shock acceleration efficiency that we considered is $\eta_e(\mathcal{M}) = \eta_0 \, \eta(\mathcal{M})$, where $\eta_0$ is a free parameter that we vary accounting for the combined uncertainties in $K_{e/p}$ and $\eta$. We show a summary of all the runs and shock acceleration models in Tab.~\ref{tab:runs}. 

Finally, the angle between the upstream background magnetic field and the normal of the shock, ${\theta_{Bn}}$, or what is commonly referred to as obliquity, also plays a role in the acceleration of particles at shocks \citep[][]{2014ApJ...797...47G,2019ApJ...876...79K}. However, in this work we assume that the acceleration efficiency does not depend on shock obliquity for simplicity. Under these considerations, only the Mach number dependency in the $\eta({\mathcal{M}})$ function plays a role in our DSA models. 
In Fig.~\ref{app:CRe_vs_Mach}, we show how the normalization of the momentum distribution and the CRe energy density injected vary considering the first model of shock acceleration efficiency and its dependence on the Mach number.
%

%%%%%%%%%%%%%%%%%%%%%%%%%%%%%%%%%%%
\subsubsection{Synchrotron emission}
\label{sec:sync}
%%%%%%%%%%%%%%%%%%%%%%%%%%%%%%%%%%%

The synchrotron emissivity of a tracer particle in a local magnetic field $\mathbf{B}$ in the direction $\mathbf{\hat n_{\rm los}}$ specifying the unit vector along the line of sight (LOS), per unit solid angle, volume and frequency is given by
\begin{small}
\begin{equation}
    \mathcal{J}_{\rm syn}(\nu,\mathbf{\hat n_{\rm los}},\mathbf{B})
    = \int N(E',\mathbf{\hat n_{los}})
    \mathcal{P}(\nu',E',\phi',\mathbf{B}) dE' d\Omega',
\end{equation}
\end{small}
where $\mathcal{P}(\nu,E',\phi')$ is the synchrotron power per unit frequency and unit solid angle emitted by a single particle that has energy $E'$, and $\phi'$ is the angle that the velocity vector of the particle makes with the direction $\mathbf{\hat n_{\rm los}}$. We have assumed that the distribution of electrons can be considered homogeneous and isotropic, $N(E',\mathbf{\hat n_{\rm los}}) = (1/4\pi) N(E')$. Following \citet{1965ARA&A...3..297G}, the synchrotron emissivity in units of [erg cm$^{-3}$ s$^{-1}$ Hz$^{-1}$ str$^{-1}$] is
\begin{equation}\label{eq:17}
     \mathcal{J}_{\rm syn}(\nu,\mathbf{\hat n_{\rm los}},\mathbf{B})
     = \frac{\sqrt{3} e^3}{4\pi m_e c^2} 
     | \mathbf{B} \times \mathbf{\hat n_{\rm los}} | \int N(E') F(\xi) \, dE',
\end{equation}
where $\nu=\nu_{\rm obs} (1+z)$, $\mathbf{\hat n_{\rm los}}$ is the unit vector in the direction of the LOS in the comoving frame and $F(\xi)$ is a Bessel function integral given by
\begin{equation}\label{eq:Bessel}
    F(\xi) = \xi \int_{\xi}^{\infty} K_{5/3}(e')\,de',
\end{equation}
where
\begin{equation}\label{J_pol}
\begin{split}
    \xi &= \frac{\nu}{\nu_c} \\
    &= \frac{4\pi m_{e}^3c^5 \nu}{3eE'^2| \mathbf{B} \times \mathbf{\hat n_{\rm los}}|} \\
    &= 1.596 \times 10^{-10} \frac{\nu_{\rm [GHz]}}{E_{[{\rm erg}]}^2 B_{\rm [G]}\sin \alpha_s},
\end{split}
\end{equation}
where $\nu_{c}$ is defined as the critical frequency at which the emission peaks and $\alpha_s$ is the angle between $\mathbf{B}$ and $\mathbf{\hat n_{\rm los}}$. 

Note that strictly speaking only those particles with a pitch angle coinciding with the angle between $\mathbf{B}$ and $\mathbf{\hat n_{\rm los}}$ contribute to the emission along the LOS in Eq. (\ref{eq:17}). Nevertheless, in this work, we assume an isotropic distribution of pitch angles, i.e., $\langle \sin \alpha_s \rangle = \pi/4$.

Once the synchrotron emissivity is assigned back to the tracers, we produce particle projection maps\footnote{We use \url{https://yt-project.org} for the analysis of our simulations.}. The specific intensity maps can then be obtained by integrating along a LOS as
\begin{equation}\label{eq:intensity}
    I_{\nu}(X,Y) = \int \mathcal{J}_{\rm syn}(\nu,X,Y,Z)dZ,
\end{equation}
in units of [erg cm$^{-2}$ s$^{-1}$ Hz$^{-1}$ str$^{-1}$]\footnote{We note that in the case of off-axis projections (see Sec.~\ref{sec:view_angle}) we will refer to $Z$ as the LOS axis, and correspondingly, that X and Y will be the image plane coordinates, instead of the $X$- and $Y$-axes of the simulation box.}. The observed radio flux density is estimated assuming a Gaussian beam with $\theta$ width. The surface brightness maps in units of [mJy/beam] are obtained computing 
\begin{equation}
    S_{\nu} \approx I_{\nu, \rm obs} \theta^2
\end{equation} 
where $\theta^2= \pi \theta_1 \theta_2/(4\ln(2))$ is the beam area of the telescope and where the observed intensity,
\begin{equation}
    I_{\nu, \rm obs} = I_{\nu} (1+z)^{-3},
\end{equation}
has been taken into account.
We compute the radio spectral index maps between two frequencies $\nu_1$ and $\nu_2$ with 
\begin{equation}\label{eq:spectral_index}
    - \alpha(X,Y) =
    \frac{\log \left[I_{\nu_{2}}(X,Y)/I_{\nu_{1}}(X,Y) \right]}{\log(\nu_{2} / \nu_{1})}.
\end{equation}
Finally, the integrated spectra (or net flux) can be obtained by integrating the specific intensity $I_{\nu}$ over the area covered by the source in the plane of the sky, that is
\begin{equation}\label{eq:flux}
   F_{\nu} = \int I_{\nu}(X,Y)dXdY, 
\end{equation}
in units of [erg s$^{-1}$ Hz$^{-1}$ str$^{-1}$].

In this work, we compute the synchrotron emissivity and derived quantities at six frequencies: 0.05, 0.144, 0.4, 0.75, 1.5, and 6.5 GHz. Throughout the manuscript, we primarily focus on the LOFAR observing frequency of 144 MHz. However, the additional frequencies are used to perform the radio spectral index analysis (see Sec.~\ref{sec:spectra}).

%%%%%%%%%%%%%%%%%%%%%%%%%%%%%%%%%%%%%%%%%%%%
\section{Results}\label{sec:results}
%%%%%%%%%%%%%%%%%%%%%%%%%%%%%%%%%%%%%%%%%%%%

%%%%%%%%%%%%%%%%%%%%%%%%%%%%%%%
\subsection{MHD}
%%%%%%%%%%%%%%%%%%%%%%%%%%%%%%%%

\begin{figure}
    \centering
    \includegraphics[width=0.9\columnwidth]{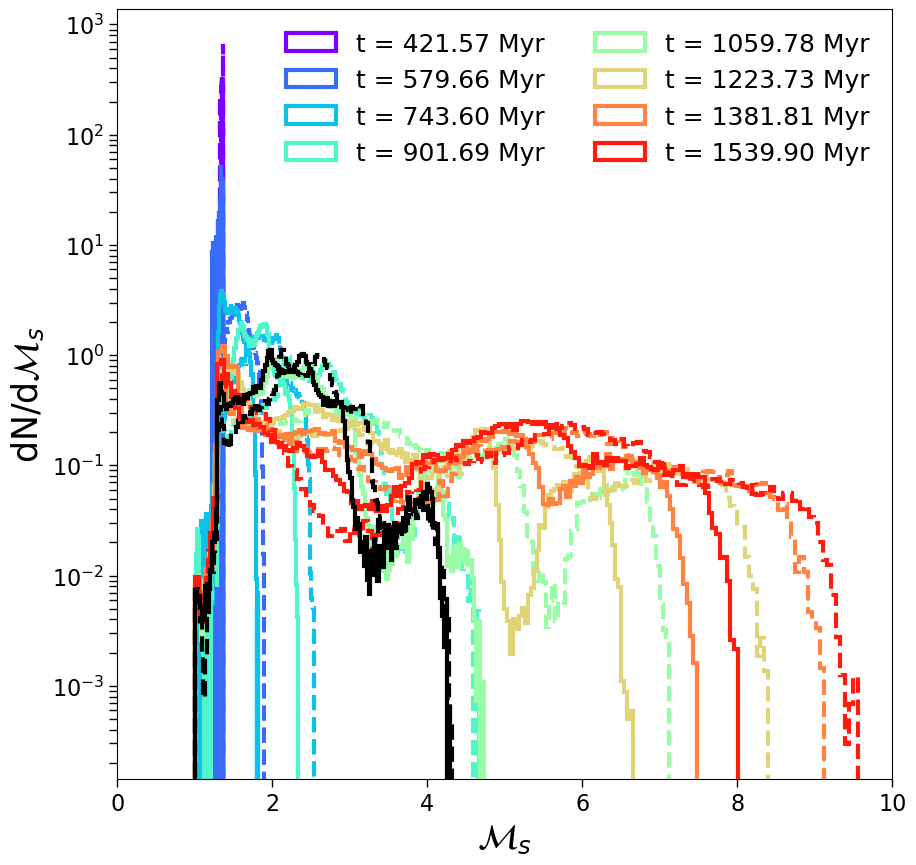}
    \includegraphics[width=0.9\columnwidth]{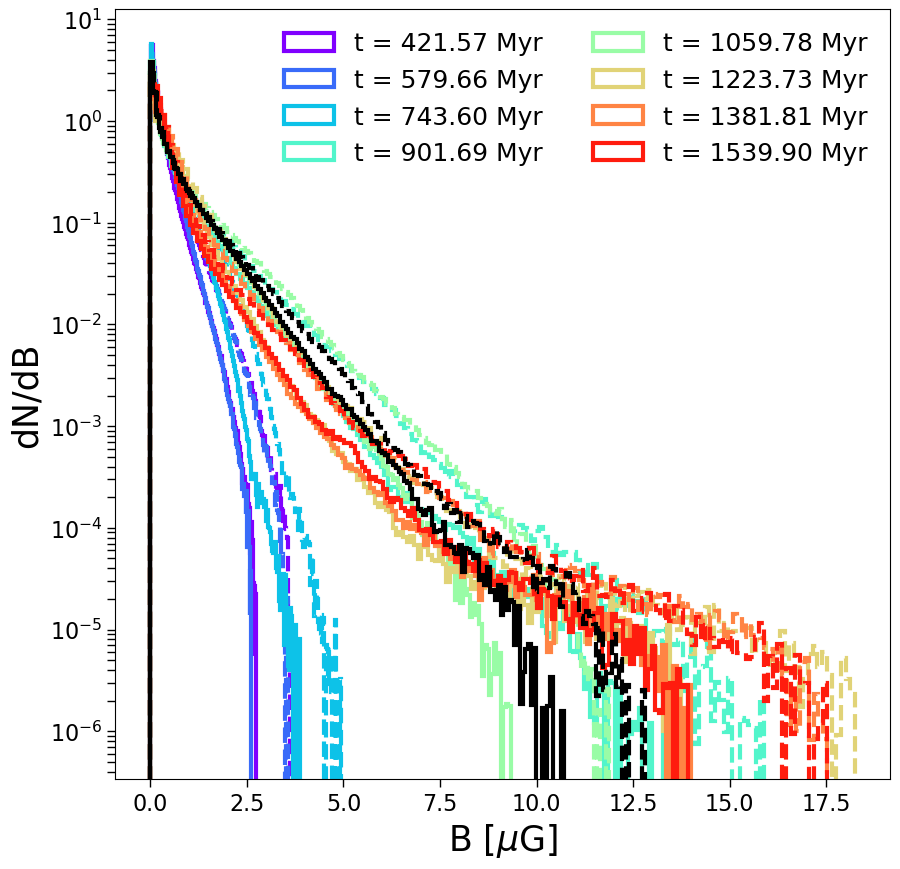}
    \caption{Fluid variables. \textit{Upper panel}: Time evolution of the mass-weighted Mach number distribution ($v_i=2400$ and $v_i=3000$ cases). The solid lines correspond to the run initialized with $v_i=2400$ km~s$^{-1}$, $b=250$ kpc, $\beta_p=200$. The dashed lines correspond to the run initialized with $v_i=3000$ km~s$^{-1}$, $b=250$ kpc, $\beta_p=200$. The black lines correspond to the time at the pericenter passage, $t=1042.22$ Myr and $t=884.13$ Myr, respectively. \textit{Lower panel}: Evolution of the mass-weighted magnetic field strength distribution ($\beta_p=200$ and $\beta_p=100$ cases). The solid lines correspond to the same run initialized with $v_i=2400$ km~s$^{-1}$, $b=250$ kpc, $\beta_p=200$ (same as solid lines on the upper panel). The dashed lines correspond to the run initialized with $v_i=2400$ km~s$^{-1}$, $b=250$ kpc, $\beta_p=100$.}
    \label{fig:distributions}
\end{figure}

\begin{figure}
    \centering
    \includegraphics[width=0.9\columnwidth]{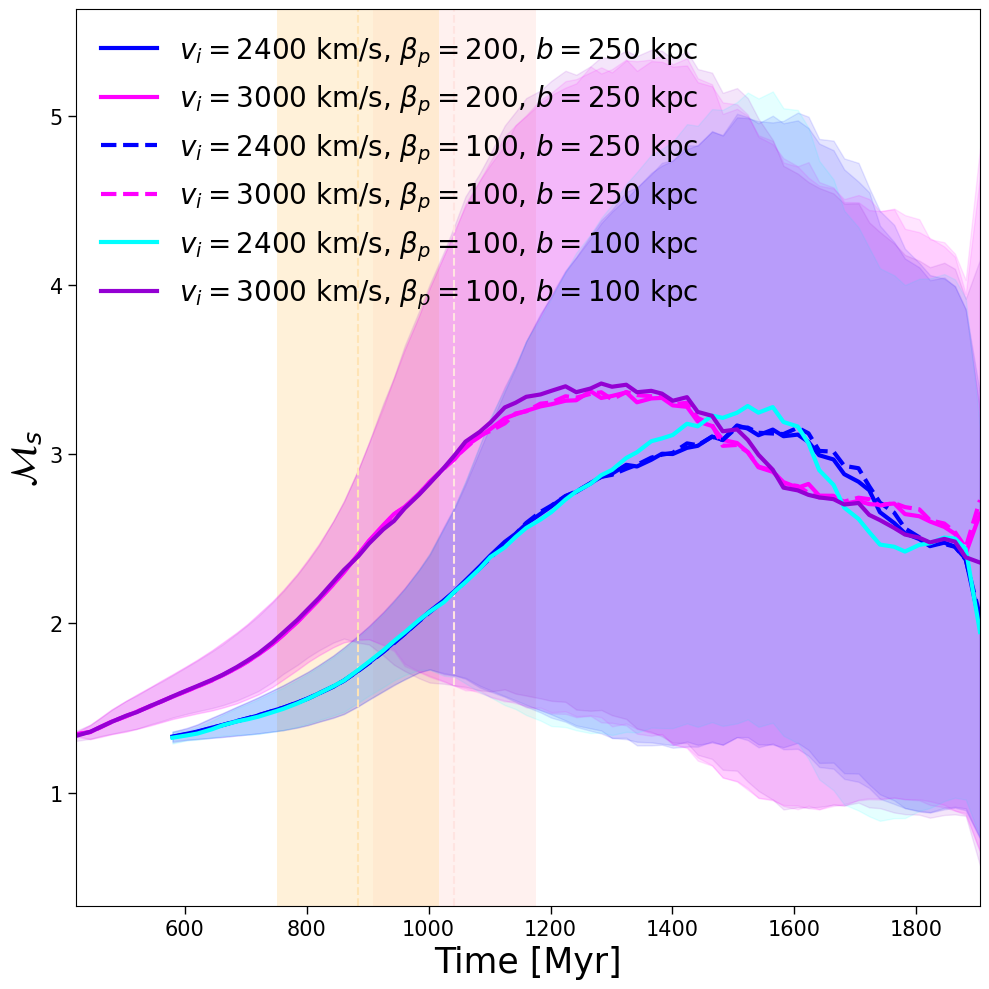}
    \caption{Tracer variables. Shock Mach number detected at the tracer particles as a function of time for all runs. The solid lines correspond to the mass-weighted mean of the distribution and the shaded regions show the standard deviation. The dashed vertical lines correspond to the times at the pericenter passage and the time range is the same as in Fig.~\ref{fig:radio_power_zoom}.}
    \label{fig:tracers_vs_time}
\end{figure}

We start by visualizing the merger at pericenter passage. In Fig.~\ref{fig:proj_maps}, we show slice plots of the density and magnetic field strength of the run initialized with $v_i=2400$ km~s$^{-1}$, $b=250$ kpc, $\beta_p=200$. Other runs look visually similar. Runs with higher relative initial velocities result in the generation of stronger shocks, which deposit more energy into heat and CRe, and more strongly amplify the seed turbulent magnetic field. In Fig.~\ref{fig:distributions}, we show some examples of the dynamical differences between our runs. In the top panel of Fig.~\ref{fig:distributions}, we show the time evolution of the sonic Mach number distribution from the two shocks generated by the binary cluster merger in two runs with different initial relative velocities: $v_i=2400$ km~s$^{-1}$ (solid lines) and $v_i=3000$ km~s$^{-1}$ (dashed lines). We show the time evolution well past the pericenter passage. For reference, we show the time at the pericenter passage for both runs with black lines. Naturally, the run with $v_i=3000$ km~s$^{-1}$ drives stronger shocks propagating through the ICM. 
Finally, we note that well past pericenter passage ($\sim 600$--$650$ Myr), the shocks can reach relatively high Mach numbers as they propagate to larger cluster-centric radii where the ambient gas temperature and sound speed are lower. Both the shape of the Mach number distribution and the maximum values achieved are sensitive to the initial conditions of the binary merger \citep[see e.g.][for a recent study]{2025arXiv251207214K}. Furthermore, due to the idealized nature of our simulations, the resulting Mach number distribution can develop a relatively flatter high-Mach tail at late times, in contrast to the distributions typically found in cosmological simulations \citep[][]{2001ICRC....6.2062M,2003ApJ...593..599R,2009MNRAS.395.1333V,2011MNRAS.418..960V,Wittor_2021}. As shown by \citet{2011MNRAS.418..960V}, the detailed shape of the Mach number distribution is also influenced by numerical methods and resolution. 

\begin{deluxetable*}{ccccc}
\tablecaption{Data from LOFAR observations at 144\,MHz \label{tab:obs}}
\tablehead{
\colhead{uv-taper} &
\colhead{Flux density [mJy]} &
\colhead{Radio power [$10^{25}$ W Hz$^{-1}$]} &
\colhead{$\theta_{\rm maj}\times\theta_{\rm min}$} &
\colhead{$\sigma_{\rm rms}$ [$\mu$Jy beam$^{-1}$]}
}
\startdata
None & 37.814 / $27.601 \pm 1.016$ & 3.25 / 3.71 & 11.02\arcsec\ $\times$ 5.69\arcsec  & 107 \\
8    & 37.476 / $33.590 \pm 1.188$ & 3.96 / 4.51 & 16.59\arcsec\ $\times$ 10.26\arcsec & 165 \\
10   & 37.542 / $34.202 \pm 1.258$ & 4.03 / 4.60 & 17.56\arcsec\ $\times$ 12.17\arcsec & 182 \\
15   & 37.006 / $34.809 \pm 1.371$ & 4.10 / 4.67 & 20.33\arcsec\ $\times$ 18.21\arcsec & 194 \\
30   & 37.026 / $34.922 \pm 1.226$ & 4.11 / 4.69 & 39.07\arcsec\ $\times$ 34.45\arcsec & 341 \\
\enddata
\tablecomments{
We measure the radio flux density in an elliptical region (see Fig.~\ref{fig:lofar}) with a LLS (major axis) of 162$^{\prime\prime}$ or 1032 kpc. At the redshift of the cluster ($z=0.5456$), the angular-to-linear scale conversion is 6.37 kpc/$^{\prime\prime}$ and, the luminosity distance is $D_L=3137.8$ Mpc. The latter is computed assuming the same cosmology as in \citet{2020Giovannini}: $H_0 = 71$ km~s$^{-1}$~Mpc$^{-1}$, $\Omega_m = 0.27$, and $\Omega_{\Lambda} = 0.73$. We report the flux density within the ellipse region (first value) and also the corresponding flux density above a $2\sigma_{\rm rms}$ (second value). We report the radio power with a K-correction considering $\alpha=-1.0$ (typical for radio relics; left value) and  $\alpha=-1.3$ (typical for radio halos; right value).}
\end{deluxetable*}

 In the bottom panel of Fig.~\ref{fig:distributions}, we show the time evolution of the magnetic field strength within a sphere of 4.5 Mpc radius. We considered the center of the sphere to be the minimum of the gravitational potential throughout the merger time evolution. We show two runs with different initial plasma beta parameters: $\beta_p=200$, and $\beta_p=100$. As the merger progresses, the magnetic field strength distribution develops a high-end tail due to the shock-induced magnetic amplification. The run with $\beta_p=100$, i.e., an average stronger magnetic field, reaches higher values than the $\beta_p=200$ run.

\begin{figure*}[t]
  \centering

  % Row 1
  \hspace{-4em}
  \begin{minipage}[t]{0.35\textwidth}\centering
    \includegraphics[width=\linewidth]{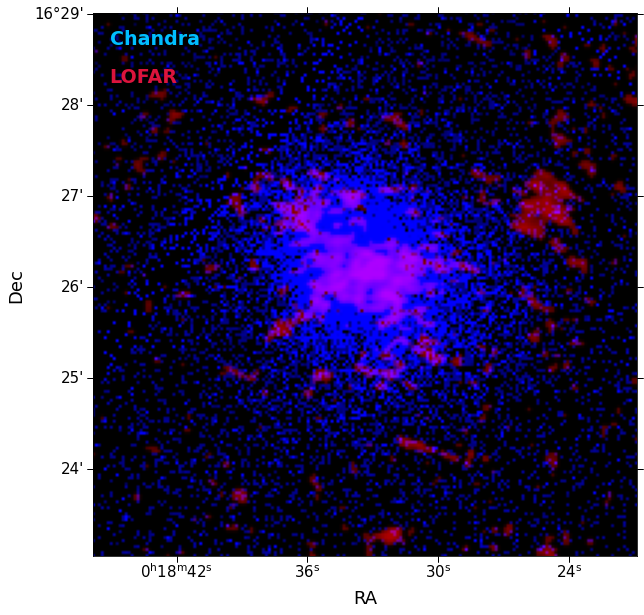}
  \end{minipage}
  \hspace{3em}
  \begin{minipage}[t]{0.35\textwidth}\centering
    \includegraphics[width=\linewidth]{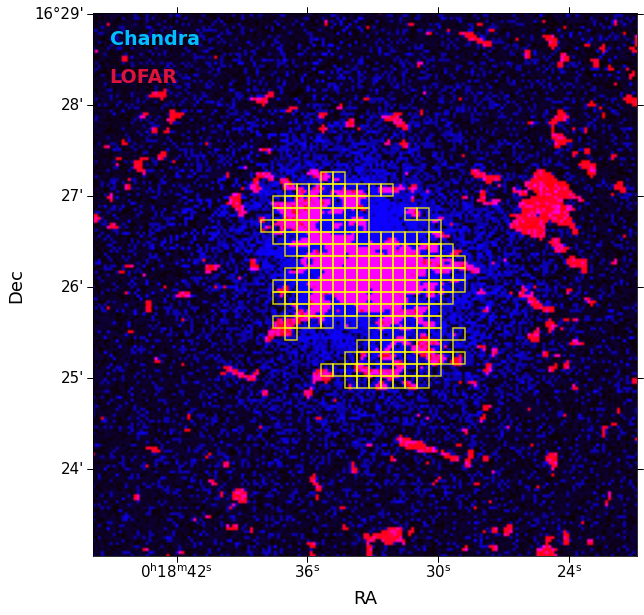}
  \end{minipage}

  \vspace{0.6em}

  % Row 2
  \begin{minipage}[t]{0.41\textwidth}\centering
    \includegraphics[width=\linewidth]{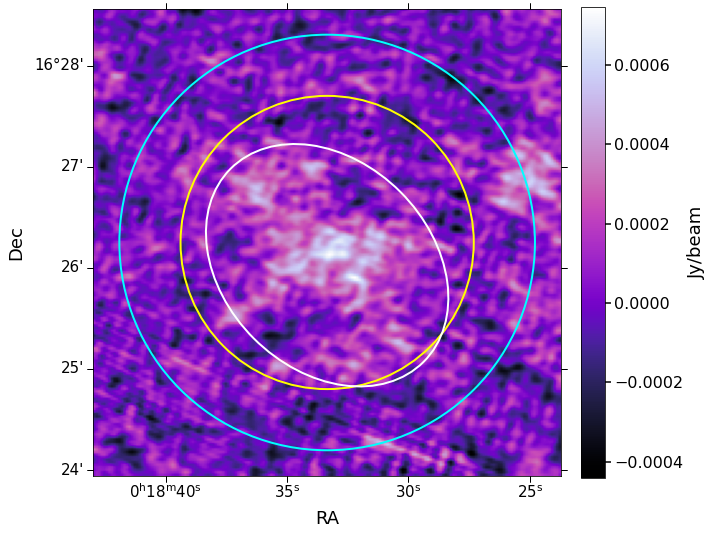}
  \end{minipage}
  \hspace{-0.5em}
  \begin{minipage}[t]{0.41\textwidth}\centering
    \includegraphics[width=\linewidth]{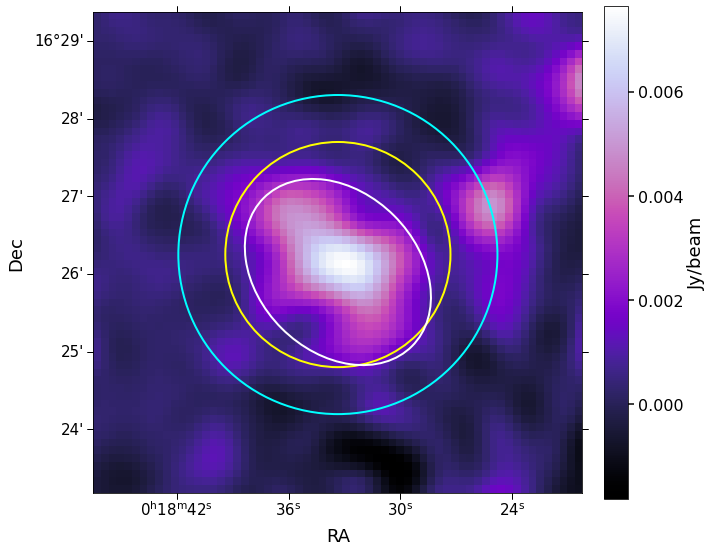}
  \end{minipage}

  \vspace{0.6em}

  % Row 3
  \begin{minipage}[t]{0.41\textwidth}\centering
    \includegraphics[width=\linewidth]{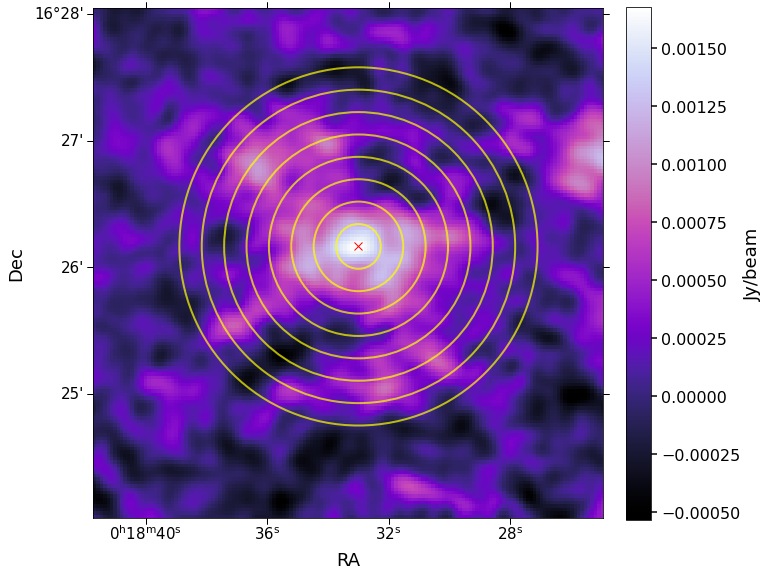}
  \end{minipage}
  \hspace{-0.5em}
  \begin{minipage}[t]{0.41\textwidth}\centering
    \includegraphics[width=\linewidth]{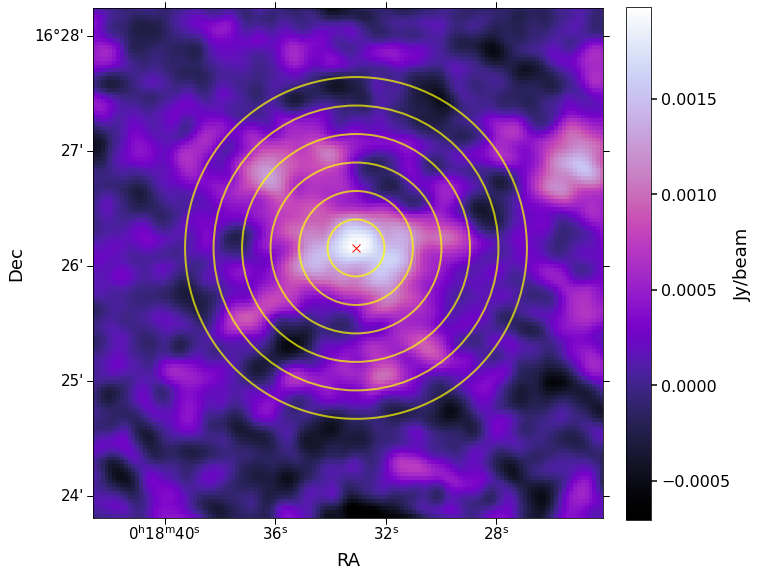}
  \end{minipage}

  \caption{\textit{Upper row}: Composite images of Chandra at 0.5-7 keV band (blue) and LOFAR at 144 MHz with a beam size of 11.02" $\times$ 5.69" (red). In the left hand panel the radio emission is shown with reduced color saturation. In the right hand panel we overplot square regions with the width of the beam. We select these regions within the ellipse region (middle row) and where the emission is above $2\sigma_{\rm rms}$. \textit{Middle row}: Radio images with beam sizes 11.02" $\times$ 5.69" (left) and 39.07" $\times$ 34.45" (right). The ellipse region (white) that we used to select the radio flux density. Circular regions with radii 554.2 kpc (yellow) and 785.4 kpc (cyan) to match the LLS reported in \citet{2000NewA....5..335G} and \citet{2020Giovannini}, respectively. \textit{Lower row}: Radio images with beam sizes 16.59" $\times$ 10.26" (left) and 17.56" $\times$ 12.17" (right). Annuli with width equal to the corresponding beam size are shown with yellow solid lines.
  }
  \label{fig:lofar}
\end{figure*}

%%%%%%%%%%%%%%%%%%%%%%%%%%%%%%%
\subsubsection{Tracers}
%%%%%%%%%%%%%%%%%%%%%%%%%%%%%%%%

As mentioned in Sec.~\ref{sec:num_setup_tracers}, we have passive tracers that follow the MHD fluid and we interpolate the relevant fluid variables onto them. For our analysis, we select only tracer particles that have encountered shocked cells as tagged by the on-the-fly AREPO shock finder \citep[see][for the details on the implementation and App.~\ref{app:shock_energy}]{2015MNRAS.446.3992S}. Specifically, we consider tracers located in cells where the Mach number is non-zero and exceeds the threshold, $\mathcal{M} \geq \mathcal{M}_{\rm th} > 0$ (see Sec.~\ref{sec:num_set-up_MHD}). Therefore, some of the dynamical differences between runs discussed in the previous section are also seen in the variables stored in the tracers. In Fig.~\ref{fig:tracers_vs_time}, we show the time evolution of the sonic Mach number for all our runs. In agreement with the upper panel of Fig.~\ref{fig:distributions}, stronger shocks are generated for those runs with $v_i=3000$ km~s$^{-1}$ than for runs with $v_i=2400$ km~s$^{-1}$. Variations in $\beta_p$ do not affect the mean value of the sonic Mach number. However, variations in the impact parameter induce slight changes in this time evolution. For example, the run with $v_i=2400$ km~s$^{-1}$ and $b=100$ kpc generates stronger shocks than the run with $b=250$ kpc.

%%%%%%%%%%%%%%%%%%%%%%%%%%%%%%%%
\subsection{Observations}\label{sec:observations}
%%%%%%%%%%%%%%%%%%%%%%%%%%%%%%%%

We used LOFAR observations of MACS J0018.5+1626 as our observational reference for comparison with our simulations. This galaxy cluster was covered by the LOFAR Two-metre Sky Survey \citep[LoTSS;][Shimwell et al. submitted]{2019A&A...622A...1S, 2022A&A...659A...1S} in the 120-168 MHz frequency range with the LOFAR high-band antenna (HBA). For studying the extended radio diffuse emission, the target field was extracted and calibrated by adopting the same strategy of \citet{2021A&A...651A.115V,2022A&A...660A..78B}. We obtained images with various weighting schemes and Gaussian uv-taper\footnote{This implies applying a Gaussian weighting in the Fourier (uv) plane, suppressing long baselines to trade angular resolution for improved sensitivity to diffuse emission.}, resulting in beam sizes of 11.02" $\times$ 5.69", 16.59" $\times$ 10.26", 17.56" $\times$ 12.17", 20.33" $\times$ 18.21", and 39.07" $\times$ 34.45", where discrete sources were removed in the uv-plane. The radio images are shown in Fig.~\ref{fig:lofar}. In the first row, we show a composite image of the Chandra observation in the 0.5-7 keV band, and the LOFAR observation with beam size of 11.02" $\times$ 5.69". We only show emission above $2\sigma_{rms}$. The Chandra observations correspond to ObsId 520. We refer the reader to Section 2.2 of \citet{2024ApJ...968...74S} for a complete description of the X-ray data reduction. We see that the diffuse radio emission is co-located with the X-ray emission and would typically be classified as a radio halo. In the second row of Fig.~\ref{fig:lofar}, we show the radio images at the highest and lowest resolution. We also overplot the contours of different regions. The gray ellipse shows the region selected to extract the radio flux density. We report the radio flux densities derived with LOFAR at 144 MHz in Tab.~\ref{tab:obs}. We assumed an uncertainty on the absolute flux scale of 10\% \citep{2022A&A...659A...1S}.

The radio flux density values for the different Gaussian uv-taper give consistent results of $\sim37$ mJy for the diffuse emission. The total radio power in the ellipse region is computed with the following relation, 
\begin{equation}
P_{\nu} = 4\pi D_{L}^{2} S_{\nu} (1 + z)^{-(\alpha+1)}
\end{equation}
in W Hz$^{-1}$ where $D_L$ is the luminosity distance and $\alpha$ is the integrated radiospectral index. In Tab.~\ref{tab:obs}, we report the K-correction with two different values of $\alpha$ associated to typical values for radio relics and radio halos. 

Finally, in the third row of Fig.~\ref{fig:lofar}, we show radio images at intermediate resolutions corresponding to beam sizes of 16.59" $\times$ 10.26" and 17.56" $\times$ 12.17". We overplot annuli to show the regions that we selected for computing radial profiles.

We note that \citet{2020Giovannini} report a VLA flux density of 9.95 mJy and a corresponding radio power of $1.175 \times 10^{25}$ W Hz$^{-1}$ at 1.5 GHz, while \citet{2000NewA....5..335G} report a flux density of 24.95 mJy and a radio power of $0.89 \times 10^{25}$ W Hz$^{-1}$ at 1.4 GHz. However, we do not use these values in our analysis for several reasons. In both cases, key information is missing, including details on the applied K-correction, the method used to extract the integrated flux density or radio power, and the procedure adopted for source subtraction. In addition, the VLA analysis presented in \citet{2020Giovannini} may include emission from the north-western (NW) region of the cluster, given the reported largest linear scale (LLS) of 1570 kpc, whereas \citet{2000NewA....5..335G} report a smaller LLS of 1100 kpc. For reference, the yellow and cyan circles shown in Fig.~\ref{fig:lofar} correspond to diameters of 1570 kpc and 1100 kpc, respectively, illustrating these reported LLS values. The NW emission is not connected to the diffuse halo-like emission shown in the LOFAR images, and would also fall out of the region of interest that we are simulating. Finally, we note that combining the LOFAR flux density with the reported VLA measurements yields unusually flat radio spectral indices. Using the 1.5 GHz flux density, we obtain a spectral index of $\alpha \simeq -0.56$, while combining LOFAR with the VLA measurement at 1.4 GHz results in $\alpha \simeq -0.18$. These values are significantly flatter than the typical spectral indices of diffuse cluster emission, which are generally $\alpha \lesssim -1$. This discrepancy further highlights the uncertainties associated with the existing high-frequency measurements.
Therefore, deeper and higher-quality radio observations at higher frequencies, for example with MeerKAT (L-band), would provide better constraints in the spectral index of MACS J0018.5+1626. For these reasons, throughout this work we restrict ourselves to using the LOFAR measurements as our primary observational benchmark.

%%%%%%%%%%%%%%%%%%%%%%%%%%%%%%%
\subsection{Simulated radio emission}
%%%%%%%%%%%%%%%%%%%%%%%%%%%%%%%%

\subsubsection{Radio and X-ray maps}

\begin{figure*}
    \centering
    \includegraphics[width=\textwidth]{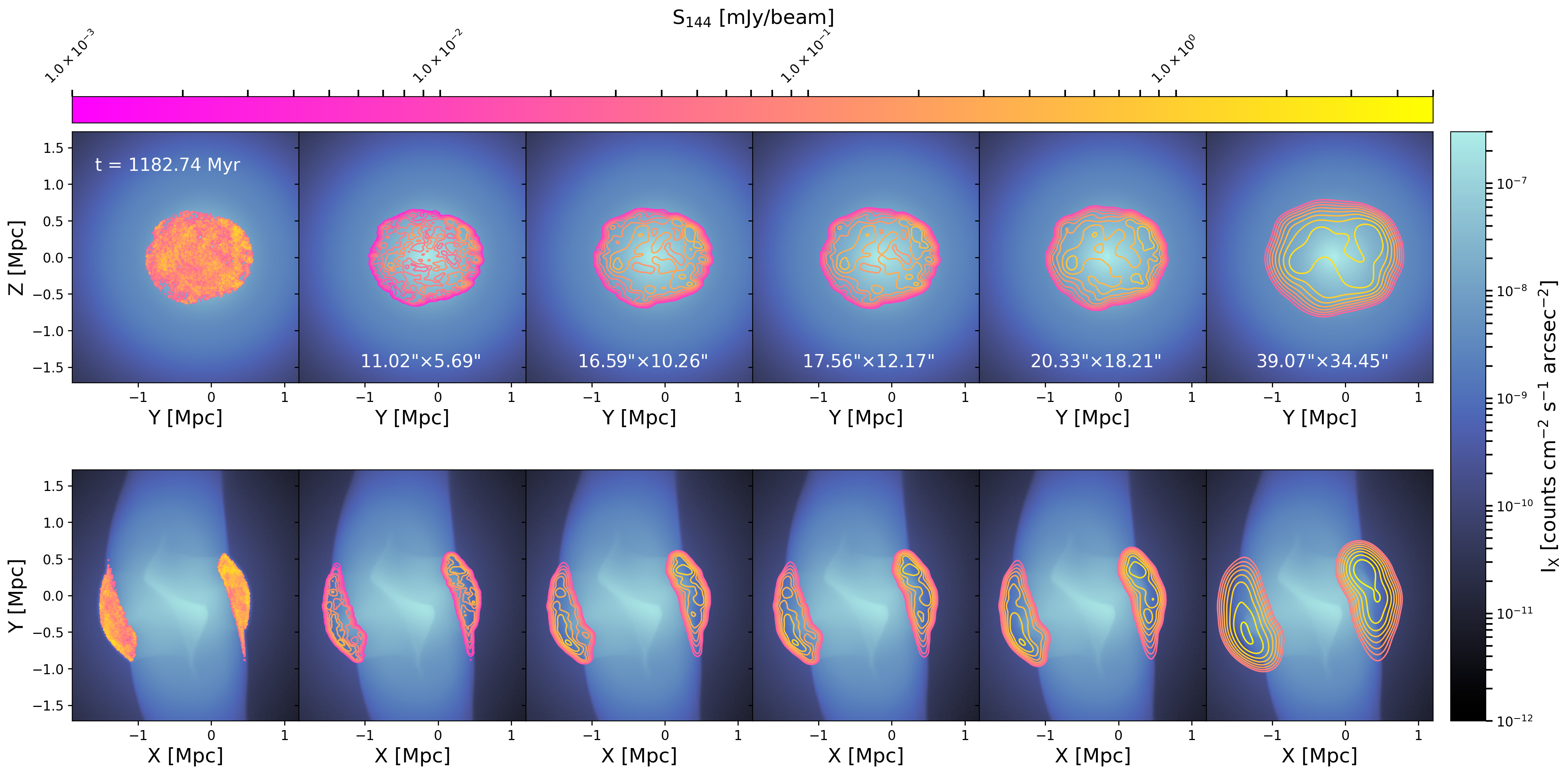}
    \caption{Projected X-ray and radio emission maps for the run initialized with $v_i=2400$ km~s$^{-1}$, $b=250$ kpc, $\beta_p=200$ and the shock acceleration efficiency model from \citet{Kang_2007}{}. \textit{Upper panels}: View with the merger along the LOS (or face-on view). \textit{Lower panels}: View with the merger in the POS. The contours of the radio maps change in each column according to the assumed beam (see white text in each top panel). We use linearly spaced contour levels extending from the maximum intensity down to 1\% of that value.}
    \label{fig:emiss_maps}
\end{figure*}

In Fig.~\ref{fig:emiss_maps}, we show an example of the simulated radio and X-ray emission. The simulated X-ray emission is computed with the pyXSIM\footnote{\url{https://hea-www.cfa.harvard.edu/~jzuhone/pyxsim.html}} package \citep[][]{ZuHone2016pyxsim}. We assume that the ICM plasma is in collisional ionization and use the Astrophysical Plasma Emission Code \citep[APEC, v3.1.2][]{Smith2001,Foster2012} model to produce the X-ray emission from the gas cell quantities. We assume a constant metallicity $Z = 0.3~Z_{\odot}$ as input to the APEC model, since metallicity is not explicitly tracked in our simulations. Fig.~\ref{fig:emiss_maps} shows the projection maps of the X-ray photon emissivity in the 0.5-7 keV band in units of [photons s$^{-1}$ cm$^{-2}$ arcsec$^{-2}$] at the redshift of MACS J0018.5+1626. We also overplot the radio specific intensity contours for the run initialized with $v_i=2400$ km~s$^{-1}$, $b=250$ kpc, $\beta_p=200$. We show the simulated radio and X-ray emission projected along the merger axis (hereafter ``face-on'') in the upper panels and projected along a LOS perpendicular to the merger axis (POS) in the lower panels. In the first column, we show the radio maps without smoothing. For the rest of the radio maps, we apply a Gaussian filter to take into account the effect of the LOFAR beam. For this, we considered the beams reported in Tab.~\ref{tab:obs}. The plane-of-sky (POS) view shows the classical view of radio relic-like emission from a binary cluster merger. The face-on view shows a circular morphology with a diameter of $\sim$1 Mpc. The epoch shown in Fig.~\ref{fig:emiss_maps} corresponds to the time at which the radio power reaches its approximate maximum (see Fig.~\ref{fig:radio_power}). The matching criteria of the ICM-SHOX pipeline disfavours that MACS J0018.5+1626 is a galaxy cluster merger observed along a LOS perpendicular to the merger axis. The likely viewing angle is $\sim$27$^{\circ}$–
40$^{\circ}$ from the merger axis \citep[see Section 5.3 in][]{2024ApJ...968...74S}. The LOFAR observations (see Fig.~\ref{fig:lofar}) also show extended diffuse emission at the center of the cluster and there is no clear signature of two radio relics, as would be the case in the typical POS view. Therefore, in the remainder of the paper we shall mainly focus on the face-on view of the radio emission. The effects of the viewing angle will be discussed in Sec.~\ref{sec:view_angle} and ~\ref{sec:ptp}.

\begin{figure*}
    \centering
    \includegraphics[width=\textwidth]{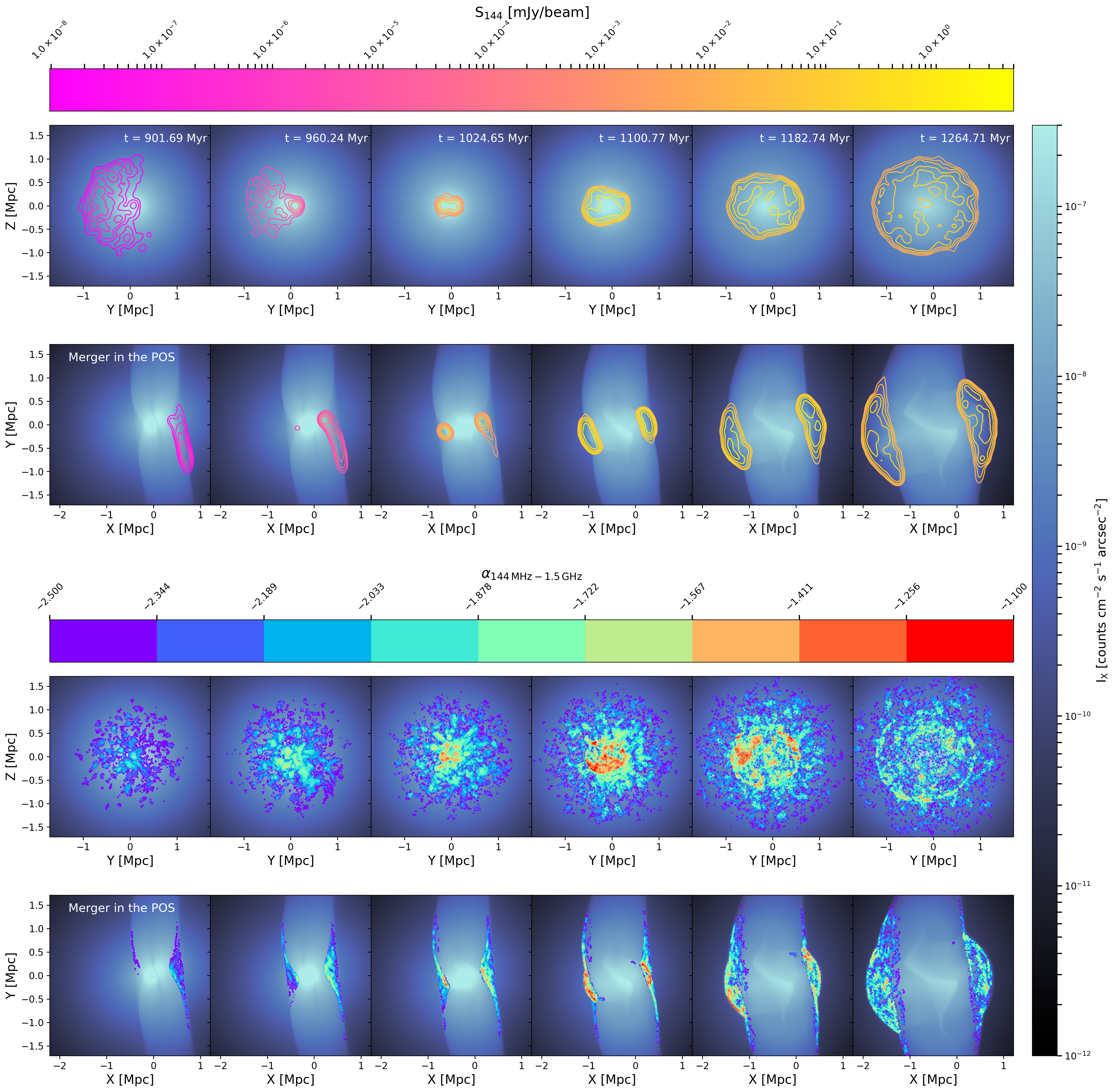}
    \caption{\textit{Upper panels}: Same as Fig.~\ref{fig:emiss_maps}, but here each panel shows a different merger epoch. The radio contours are computed considering the largest beam, 39.07"$\times$ 34.5" and, again, the contour levels extend from the maximum intensity down to 1\% of that value. \textit{Lower panels}: Same X-ray projection maps as the upper panels, but here we overplot the contours of the radio spectral index map between 144 MHz and 1.5 GHz.}
    \label{fig:emiss_maps2}
\end{figure*}

As soon as there are shocks in the binary merger system, there is injection of shock primaries via the DSA mechanism (see Sec.~\ref{sec:fkp}), and therefore there is radio emission. However, in an early epoch, the mean Mach number of the shock is low (see Fig.~\ref{fig:tracers_vs_time}) and therefore, the radio emissivity, intensity, and power are low as well. As the merger progresses, the radio power increases. We illustrate this evolution of the radio emission in Figs.~\ref{fig:emiss_maps2} and \ref{fig:radio_power}. In the upper panels of Fig.~\ref{fig:emiss_maps2} we show the radio surface brightness as a function of time for the run initialized with $v_i=2400$ km~s$^{-1}$, $b=250$ kpc, and $\beta_p=200$, which show radio emission with a prevalent circular morphology that grows in size. We note that at early times, the peak of the radio emission is not co-located with the peak of the X-ray emission. This occurs due the fact that the binary merger is not a perfect head-on collision ($b=0$), but with a non-negligible impact parameter of $b=250$ kpc. In the lower panels of Fig.~\ref{fig:emiss_maps2}, we also show the radio spectral index maps computed with Eq.~\ref{eq:spectral_index} between the frequencies 144 MHz (LOFAR) and 1.5 (VLA). There are some variations in the spectral index maps. The maximum values are between $\sim -1.88$ and $\sim -1.13$. 
We note that the highest values of the spectral index are not necessarily at the center. For example, this is visible at later times when the shock is more extended.
 
%

%%%%%%%%%%%%%%%%%%%%%%%%%%%%%
\subsubsection{Radio power}\label{sec:radio_power}
%%%%%%%%%%%%%%%%%%%%%%%%%%%%%

\begin{figure}
    \centering
    \includegraphics[width=\columnwidth]{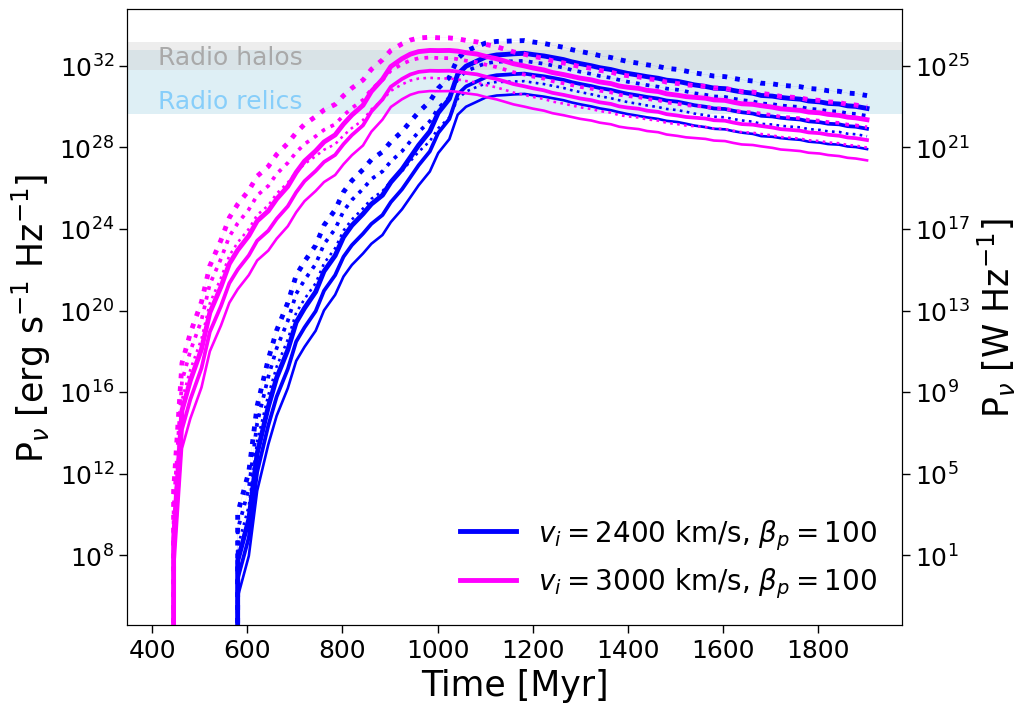}
\caption{144 MHz radio power as a function of time. We show the runs initialized with $v_i=2400$ km~s$^{-1}$, $b=250$ kpc, $\beta_p=100$ (blue) and  $v_i=3000$ km~s$^{-1}$, $b=250$ kpc, $\beta_p=100$ (magenta). The solid and dashed lines correspond to the shock acceleration models with the Mach number-dependent efficiency and constant efficiency, respectively. The increasing line widths correspond to $\eta_0=10^{-1}, 10^{-2}, 10^{-3}$. The shaded blue and grey areas denote the typical observed values of radio relics and radio halos, respectively.}
    \label{fig:radio_power}
\end{figure}

We show the evolution of the radio power at 144 MHz in Fig.~\ref{fig:radio_power} for two runs with two different initial relative velocities. We show all the shock acceleration models that were analyzed (see Tab.~\ref{tab:runs}) with different line styles. The overall evolution is similar for all the models. The first phase is associated with a combination of continually injected shock primary CRes and the effect of adiabatic energy gain of CRe through compression. The peak radio power corresponds to the epoch at which the shock energy dissipation reaches its maximum. After this, the radio power decays slowly. This is related to a combination of continuous injection of the shock primaries and the effect of adiabatic energy losses through expansion. Note that the mean Mach number at later times decreases slightly (see Fig.~\ref{fig:tracers_vs_time}), which leads to the continued injection of shock CRe primaries, but a smaller overall normalization in the particle momentum spectra. We will discuss the details of the particle momentum spectra in Sec.~\ref{sec:spectra}. The runs initialized with $v_i=3000$ km~s$^{-1}$ yield a slightly larger radio power, which is a result of the slightly stronger shocks generated in this case (see Fig.~\ref{fig:tracers_vs_time}). We show as a reference the typical observed radio power from radio relics and radio halos \citep[see e.g.,][]{2017MNRAS.470..240N,Wittor_2021,2023A&A...672A..43C}{}. From Fig.~\ref{fig:tracers_vs_time} it becomes evident that even though shocks are generated early during a galaxy cluster merger, these are very weak in the ICM, which would remain largely undetected in the radio band. We also show the differences between models with constant and Mach number-dependent shock acceleration efficiencies with different line styles. Overall, the free parameter $\eta_0$ controls the normalization of the particle spectra and therefore, larger $\eta_0$ lead to larger radio power in both shock acceleration models.

\begin{figure*}
    \centering
    \includegraphics[width=\columnwidth]{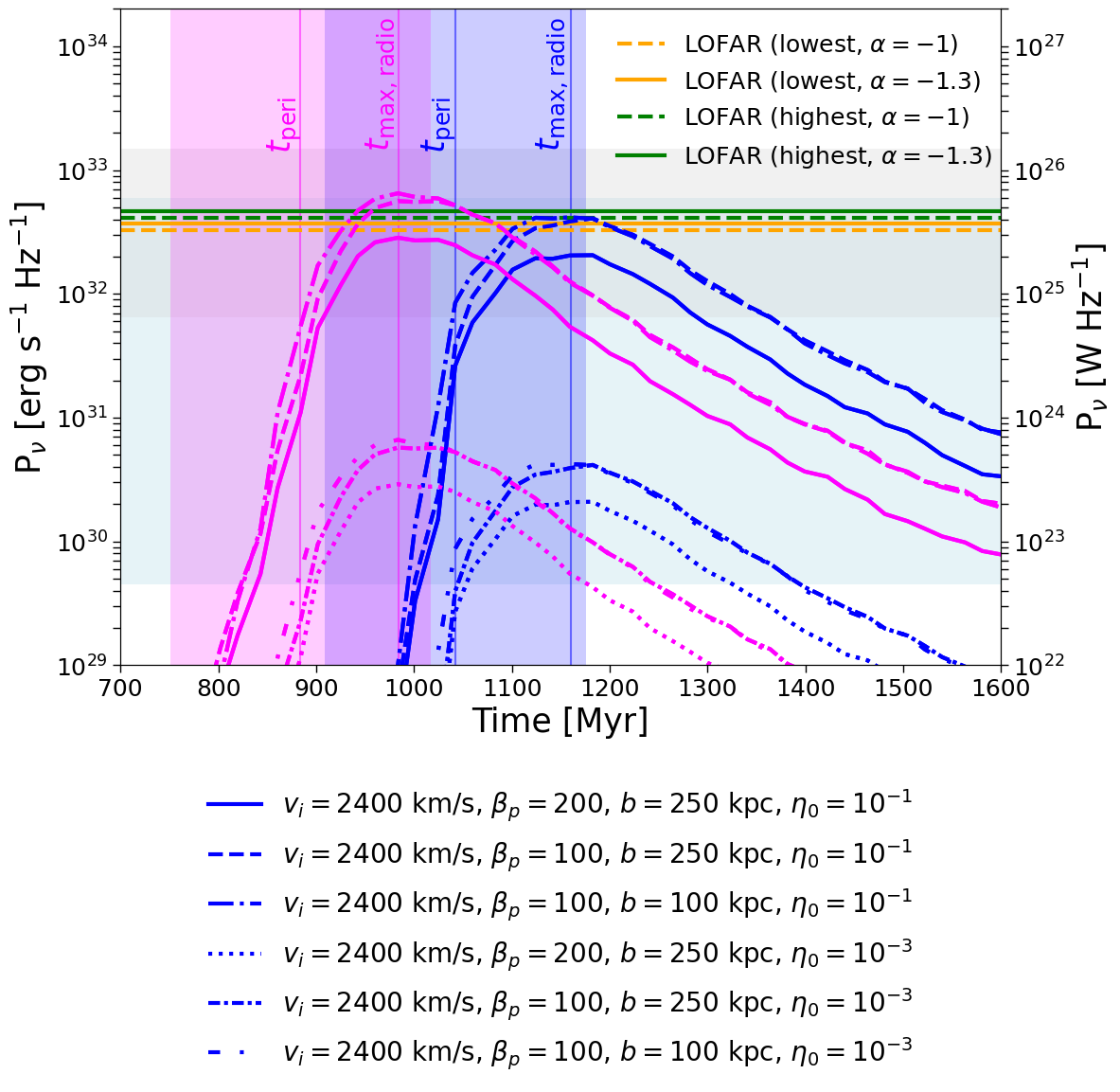}
    \includegraphics[width=\columnwidth]{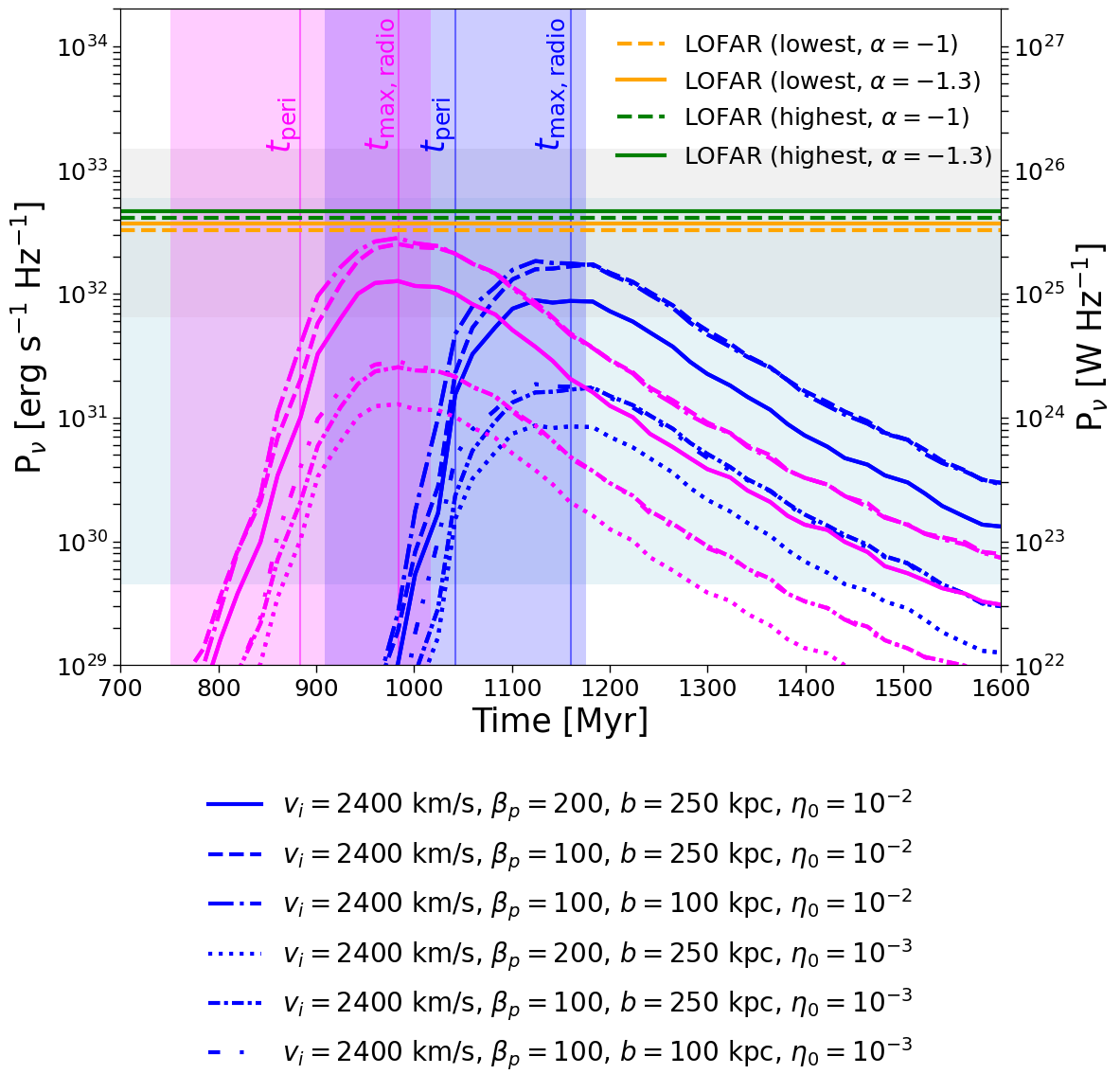}
    \caption{144 MHz radio power as a function of time showing a comparison between all simulation runs ($v_i=2400$ km~s$^{-1}$ in blue and $v_i=3000$ km~s$^{-1}$ in magenta) and various shock efficiency models. We zoom into the epoch of interest from Fig.~\ref{fig:radio_power}. \textit{Left panel}: Runs with a shock acceleration efficiency depending on the Mach number (\citealt{Kang_2007}). \textit{Right panel}: Runs with a constant shock efficiency. The horizontal shaded areas are the same as in Fig.~\ref{fig:radio_power}. The horizontal yellow and green lines show the observed radio power of MACS J0018.5+1626 with LOFAR (see Tab.~\ref{tab:obs}). Highest (lowest) refers to the values from the radio image with no uv-taper (uv-taper 30). The vertical lines show the time of the pericenter passage and the time at which the radio power reaches its maximum. The shaded vertical regions show the likely $t-t_{\rm peri}$ constrained by the ICM-SHOX pipeline. Specifically, we considered $\pm 3 \sigma$ from the median of the weighted distribution of epochs for simulation snapshots that pass the matching criteria (see Fig. 6 of \citealt{2024ApJ...968...74S}).}
    \label{fig:radio_power_zoom}
\end{figure*}

Fig.~\ref{fig:radio_power_zoom} shows a zoom into the period at which the radio power reaches its maximum. The maximum is always reached after pericenter passage, $\sim150$ Myr and $\sim 100$ Myr for runs initialized with $v_i=2400$ km~s$^{-1}$ and $v_i=3000$ km~s$^{-1}$, respectively. We also note that the time of the maximum radio power is highly dependent on the initial conditions of the merger as shown in Fig.~\ref{fig:radio_power_zoom} for runs with different $v_i$, however other factors such as $R$, $b$, and the initial radial profiles of the clusters are expected to play a role here. This general evolutionary behavior, a rise and a subsequent decay in radio power following core passage, is qualitatively consistent with radio relic light curves from cosmological hydrodynamical simulations. For example, \citealt{2024A&A...690A.146N}, find similar trends in the 300 simulation suite, where merger shock emission was modeled using a magnetic field scaling relation and a DSA prescription, without tracking the spectral evolution of the CRe. Quantitatively, however, \citealt{2024A&A...690A.146N} find that the radio power peaks $\sim500$ Myr after core passage for clusters with $\mathrm{M}_{200,1}\sim 10^{15} \, \mathrm{M}_{\odot}$ and $R \lesssim 3.3$. This difference could be attributed to the fact that models without electron spectral aging allow downstream electrons to continue contributing to the emission over longer timescales, effectively delaying the peak. Although a one-to-one comparison is not straightforward, in future work it would be interesting to revisit these differences in a controlled setup matching the initial conditions of MACS J0018.5+1626, and in particular to disentangle how the initial merger parameters and the treatment of electron spectral evolution each contribute to the period at which the radio power maximum is reached after pericenter passage.

In Fig.~\ref{fig:radio_power_zoom}, we also show with vertical shaded regions the epochs of interest coming from the matching criteria from the ICM-SHOX pipeline. In particular, \citet{2024ApJ...968...74S} showed that varying the initial gas profiles and parameters $R$, $b$, and $v_i$, there is a
strong selection for a particular range in time of simulation snapshots that pass the matching criteria to reproduce all the observables. The distribution has a median of $t-t_{\rm peri}=30$ Myr \citep[see Fig.~6 in][]{2024ApJ...968...74S}. In both panels of Fig.~\ref{fig:radio_power_zoom} we show a range lying within $\pm 3 \sigma$ of that median, where $\sigma$ is estimated by approximating the distribution as Gaussian. The peak radio power lies within these likely epochs to explain other observables of MACS J0018.5+1626 for all the runs. We also show the radio power derived from the LOFAR observations at 144 MHz as reference with horizontal lines. To compare to the observational values, we show the results of the two shock acceleration models. In the left panel of Fig.~\ref{fig:radio_power_zoom}, we show the runs with the shock efficiency that have a dependency on the Mach number with two values of $\eta_0$. In the right panel of Fig.~\ref{fig:radio_power_zoom}, we show the runs with the constant shock efficiency with two values of $\eta_0$ as well. We discuss in the following the detailed results of these models:

\begin{itemize}
    \item[i)] \textit{Constant shock acceleration efficiency}: None of these models explain the observed radio power. For the model with $\eta_0 = 10^{-2}$, the electron shock acceleration efficiency required to reach the observed radio power is larger by a factor of $\sim 2$--4.6. This computed range accounts for variations among our models as well as the lowest and highest values quoted in Tab.~\ref{tab:obs}. 
    \item[ii)] \textit{Mach number-dependent shock acceleration efficiency}: In this case, the most optimistic shock acceleration model with parameter $\eta_0=10^{-1}$ leads to epochs at which the simulated radio power is comparable to the observed radio power. We note that $\eta_0=10^{-1}$ corresponds to $\eta_e \sim 10^{-3}$--$10^{-2}$ for the average Mach number found in the simulations in that time frame: $\mathcal{M}_s \sim 2$--3  (see Fig.~\ref{fig:tracers_vs_time}). Beyond the average value, the shock can reach Mach numbers of $\mathcal{M}_s \sim 3.8$--5 after the time of pericenter passage. This corresponds to electron acceleration efficiencies of $\eta_e \sim 2$--3$\times 10^{-2}$ (see Fig.~\ref{app:CRe_vs_Mach}). The radio power from both runs with $v_i=2400$ km~s$^{-1}$ and $v_i=3000$ km~s$^{-1}$ can reach the observational radio power However, this only occurs for those runs with stronger magnetic fields, $\beta_p=100$. In terms of the impact parameter, we see that both of the $b=250$ kpc and $b=100$ kpc runs can reach the observed radio power with the runs with $b=100$ kpc reaching slightly higher values. 
\end{itemize}

The results from both models indicate that reproducing higher radio powers requires relatively large electron acceleration efficiencies and strong magnetic fields. In the case of Mach-number-dependent shock acceleration efficiency models, we find typical values of $\eta_e \sim 10^{-3}$--$10^{-2}$, corresponding to the mean of the Mach number distribution, $\mathcal{M}_s \sim 2$--3. These efficiencies lie within a reasonable, albeit optimistic, range. For example, recent PIC simulations suggest electron acceleration efficiencies of approximately $0.1$--$2\%$ for weak-Mach-number shocks in the ICM \citep[see Fig.~8 in][]{2024ApJ...976...10G}.  However, regions of the shock reaching higher Mach numbers, $\mathcal{M}_s \sim 3.8$--5, with corresponding efficiencies of $\eta_e \sim (2$--$3)\times 10^{-2}$, can dominate the synchrotron emissivity. Such efficiencies are comparable to the most optimistic values inferred from gamma-ray upper limits in galaxy clusters, namely $\eta_{e,\rm max} \sim 0.05$ \citep[see, e.g.,][]{2003ApJ...585..128K,2010MNRAS.409..449P}. We therefore regard these runs as optimistic scenarios. In Sec.~\ref{sec:discussion}, we will discuss additional possibilities that can predict a larger radio power in our models such as shock re-acceleration and/or turbulent re-acceleration.

The four runs with the shock acceleration models with a Mach number dependent efficiency varying between $\eta_e \sim 10^{-3}$--$10^{-2}$ (corresponding to the parameter $\eta_0=10^{-1}$) define our set of most optimistic models that could potentially explain the observed radio emission from MACS J0018.5+1626. We show a summary of those models in Tab.~\ref{tab:matched} including the merger epochs at which these runs match the observed radio power. In the remainder of the paper, we will mainly discuss these runs. 

\begin{deluxetable}{ccccc}
\tablecaption{Best models for MACS J0018.5+1626\label{tab:matched}}
\tablehead{
\colhead{$v_i$ [km s$^{-1}$]} &
\colhead{$\beta_p$} &
\colhead{$b$ [kpc]} &
\colhead{$t_1$, $t_2$ [Myr]} &
\colhead{$t_{\rm rel,1}$, $t_{\rm rel,2}$ [Myr]}
}
\startdata
2400 & 100 & 250 & $\sim 1120$--$1150$ & $\sim 78$--$108$ \\
     &     &     & $\sim 1210$          & $\sim 168$       \\
\hline
2400 & 100 & 100 & $\sim 1111$--$1125$ & $\sim 69$--$83$ \\
     &     &     & $\sim 1210$          & $\sim 168$      \\
\hline
3000 & 100 & 250 & $\sim 925$--$942$   & $\sim 40$--$58$ \\
     &     &     & $\sim 1053$--$1093$ & $\sim 169$--$209$ \\
\hline
3000 & 100 & 100 & $\sim 925$--$942$   & $\sim 40$--$58$ \\
     &     &     & $\sim 1053$--$1093$ & $\sim 169$--$209$ \\
\enddata
\tablecomments{
Best-matching models that can explain the observed radio emission of MACS J0018.5+1626 under the assumption of standard DSA. The shock acceleration efficiency depends on the Mach number; here we adopt $\eta_0 = 10^{-2}$ (corresponding to $\eta_e = 10^{-3}$--$10^{-2}$). The fourth column lists the approximate epochs, $t_1$ and $t_2$, at which the radio power matches the LOFAR observations before and after the peak emission (see Fig.~\ref{fig:radio_power_zoom}). The last column lists the same epochs measured relative to the time of pericentre passage, $t_{\rm rel,i} = t_i - t_{\rm peri}$.
}
\end{deluxetable}

\begin{figure}
    \centering
   \includegraphics[width=0.9\columnwidth]{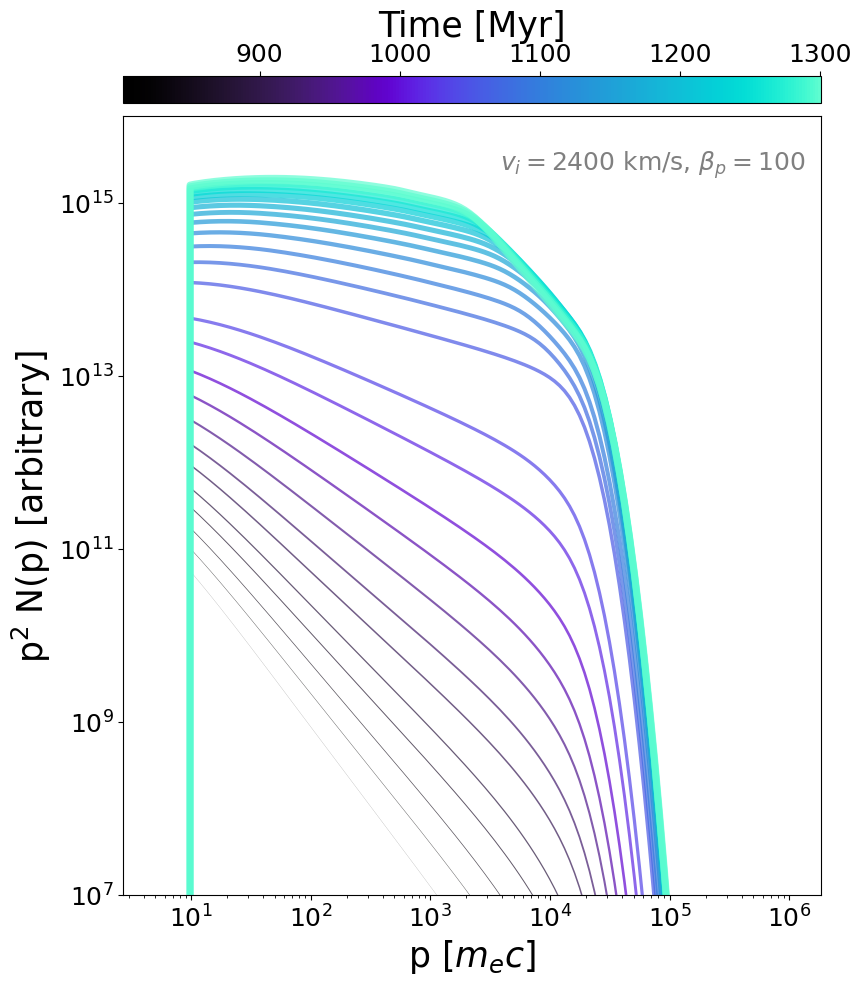}
    \includegraphics[width=0.9\columnwidth]{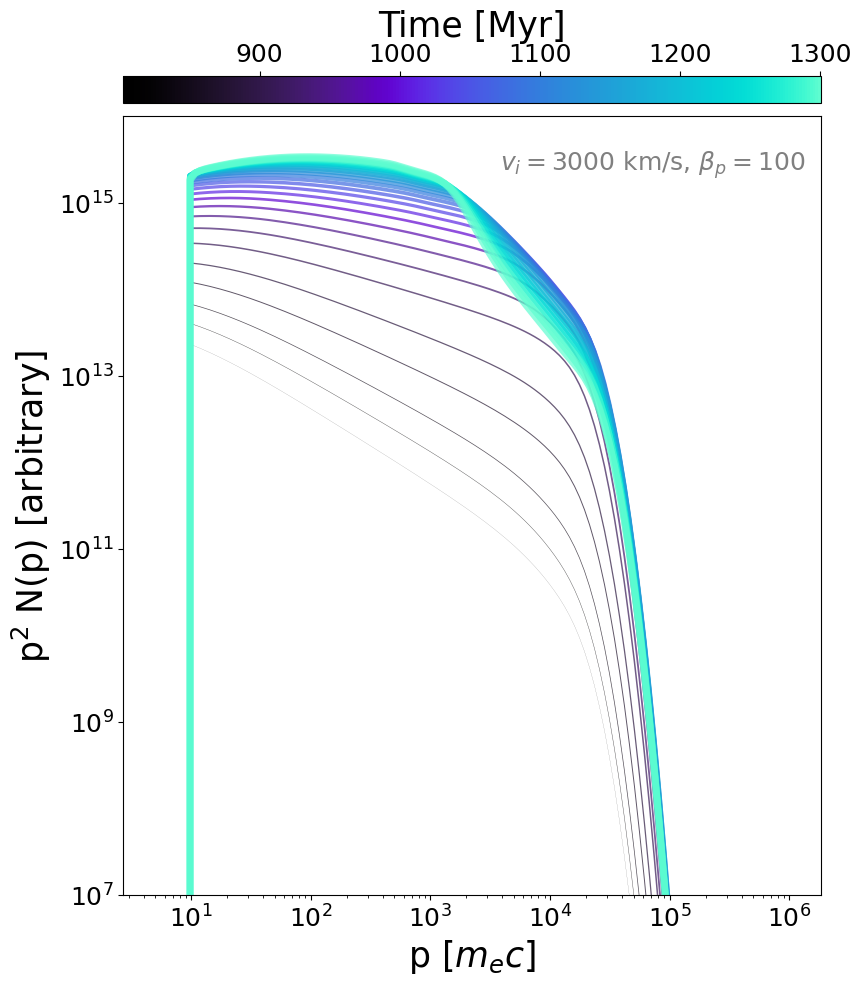}
    \caption{Evolution of total spectra (integrated over all particles) as a function of momentum, showing the comparison between runs initialized with different initial relative velocities: $v_i=2400$ km~s$^{-1}$ (\textit{upper panel}) and $v_i=3000$ km~s$^{-1}$ (\textit{lower panel}). }
    \label{fig:spectra}
\end{figure}

%%%%%%%%%%%%%%%%%%%%%%%%%%%%%%%%%%%%%%%%%%%%
\subsubsection{Spectral properties}\label{sec:spectra}
%%%%%%%%%%%%%%%%%%%%%%%%%%%%%%%%%%%%%%%%%%%%

In Fig.~\ref{fig:spectra}, we show the time evolution of the CRe momentum spectra for two examples: runs initialized with $v_i=2400$ km~s$^{-1}$, $b=250$ kpc, $\beta_p=100$ and with $v_i=3000$ km~s$^{-1}$, $b=250$ kpc, $\beta_p=100$. Runs with $b=100$ kpc show similar trends. The spectra are obtained by summing the individual CRe momentum spectra of all tracer particles over each momentum bin. We focus on the time interval of interest, defined as the epoch during which the radio power reaches its peak (see Figs.~\ref{fig:radio_power} and \ref{fig:radio_power_zoom}). We can clearly see how the evolution of the CRe momentum spectra defines the evolution of the radio power as discussed in Sec.~\ref{sec:radio_power}. In both runs, the initial increase in the normalization is mainly driven by adiabatic energy gain and continuous injection of shock primaries. At high momenta, the curvature of the CRe momentum spectra reflects the impact of energy losses. At later times, as the shocks expand, energy losses at $p/m_e c \sim 10^{3}$–$10^{4}$ become increasingly important. This behavior explains the subsequent decrease in the radio power observed at later epochs in Figs.~\ref{fig:radio_power} and \ref{fig:radio_power_zoom}. We note that energy losses are stronger in runs with $v_i = 3000$ km~s$^{-1}$, as expected.

\begin{figure}
    \centering
    \includegraphics[width=0.9\columnwidth]{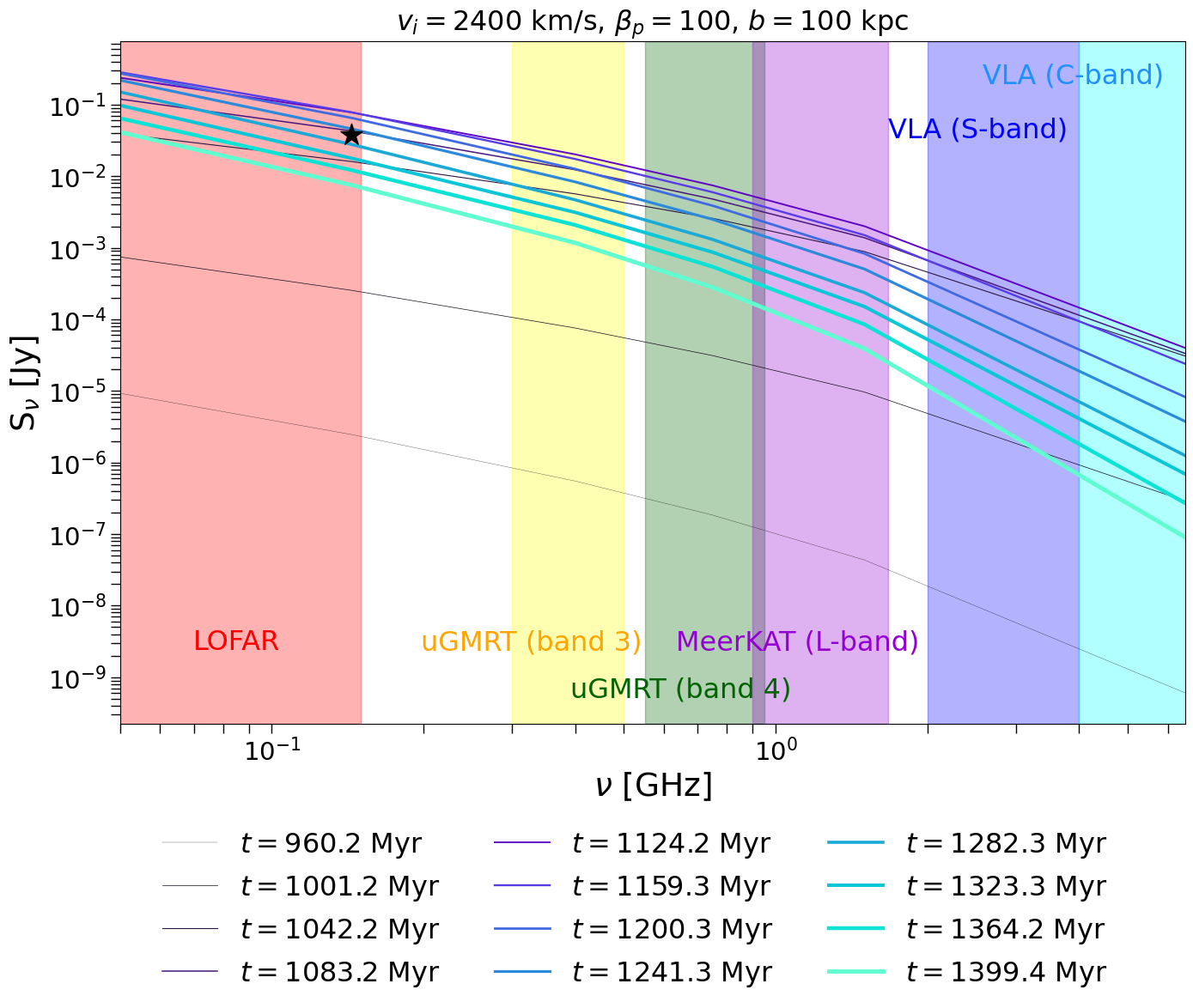}
    \includegraphics[width=0.9\columnwidth]{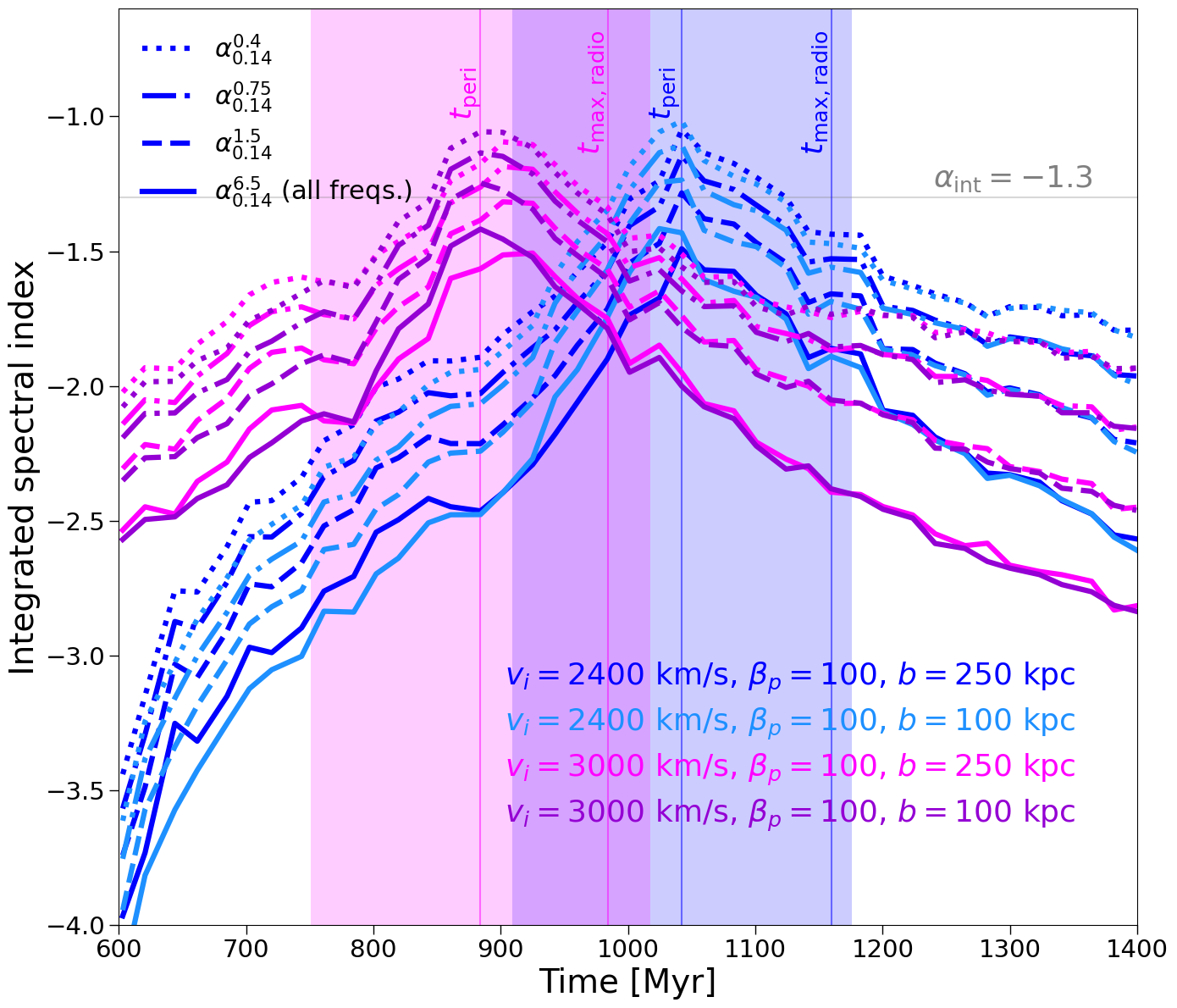}
    \caption{\textit{Upper panel}: Flux density as a function of frequency for the run initialized with $v_i=2400$ km~s$^{-1}$, $b=100$ kpc, $\beta_p=100$ in Tab.~\ref{tab:matched}, with shock acceleration efficiency parameter of $\eta_0=10^{-2}$ ($\eta_e=10^{-3}-10^{-2}$). We show the time evolution with different colors. The black star refers to the measured LOFAR flux density, $\sim 37.8$ mJy, at 144 MHz. The shaded vertical regions indicate the frequency ranges covered by other telescopes, such as uGMRT (band 3 and 4), MeerKAT (L-band), and the VLA (L- and C-band). \textit{Lower panel}: Integrated radio spectral index, $-\alpha_{\rm int}$, as a function of time. The results from the linear fit in log–log space using all frequencies (0.05, 0.144, 0.4, 0.75, 1.5, 6.5 GHz) is shown in solid lines. The other line-styles show the results from using only two frequencies. We show the evolution for all runs in Tab.~\ref{tab:matched} with different line colors. The shaded vertical regions and vertical lines are the same as in Fig.~\ref{fig:radio_power_zoom}. The horizontal gray line shows the typical spectral index for radio halos, $\alpha_{\rm int}=-1.3$.}
    \label{fig:spectral_index}
\end{figure}

In the upper panel of Fig.~\ref{fig:spectral_index}, we show the flux density as a function of frequency and time for the run initialized with $v_i=2400$ km~s$^{-1}$, $b=100$ kpc, $\beta_p=100$ and the shock acceleration efficiency depending on the Mach number with $\eta_0=10^{-2}$ ($\eta_e=10^{-3}-10^{-2}$). We computed the synchrotron emissivity for six frequencies: 0.05, 0.144, 0.4, 0.75, 1.5, 6.5 GHz. The flux density values are obtained from the surface brightness maps projected along the merger axis (face-on view) and convolved with a beam size of 11.02" $\times$ 5.69". We show the flux density measured in the highest-resolution LOFAR image, which has the same beam size. The time evolution of the flux density reflects the trends already discussed for the radio power and the CRe momentum spectra. As a result, we note that at $t\gtrsim1200$ Myr, the spectrum shows some steepening, which becomes more pronounced with time. However, the degree of curvature at earlier epochs is modest and, depending on the frequency coverage and measurement uncertainties of a given observation, may be consistent with an apparently power-law spectrum as typically reported for radio relics.

An important diagnostic in radio observations is the integrated radio spectral index, $\alpha_{\rm int}$. This is usually computed by performing a linear fit in log–log space to the flux density as a function of frequency. In the lower panel of Fig.~\ref{fig:spectral_index}, we show the integrated radio spectral index as a function of time for all our runs with $\beta_p=100$ (see Tab.~\ref{tab:matched}). The evolution is similar across all runs. During the initial phase, $\alpha_{\rm int}$ gradually flattens, reaching its flattest value at pericenter passage, after which it progressively steepens at later times. We note that variations in the impact parameter $b$ do not lead to significant differences between the runs.

The integrated radio spectral index generally depends on the frequency range and data points included in the linear fit. This behavior is expected for diffuse cluster emission, which often exhibits power-law spectra with curvature at high frequencies (see, e.g., Fig.~\ref{fig:spectra}). Including high-frequency data points, where the CRe momentum spectrum is more strongly affected by radiative losses and spectral curvature, typically results in a more negative value (steeper) of $\alpha_{\rm int}$.
As a reference, we show the results obtained from the linear fits using all six frequencies with solid lines in the lower panel of Fig.~\ref{fig:spectral_index}. However, such wide frequency coverage is available only for a limited number of well-studied clusters. In most observational cases, the integrated spectral index is derived from only two frequency measurements. For this reason, we also present $\alpha_{\rm int}$ computed using two data points, one of which is always the LOFAR observing frequency at 144 MHz. This approach is particularly relevant for future radio observations of MACS J0018.5+1626.
From the lower panel of Fig.~\ref{fig:spectral_index}, we can see that the integrated radio spectral index ranges from $\sim-1.0$ to $\sim -1.8$ from pericenter passage to the epoch at which the radio power peaks depending on the frequencies used for the fit\footnote{See App.~\ref{app:resolution}, for a discussion on resolution effects.}. 
For example, when combining LOFAR observations with future uGMRT measurement predictions at 400 or 750~MHz, evaluated near the epoch of peak radio power, we obtain integrated spectral indices of $\alpha_{\rm int} \simeq -1.4$-$1.5$ for runs initialized with $v_i = 2400~\mathrm{km\,s^{-1}}$, and $\alpha_{\rm int} \simeq -1.3$-$1.4$ for runs with $v_i = 3000~\mathrm{km\,s^{-1}}$.
All the values discussed above fall within the typical range reported in radio observations. For example, integrated radio spectral indices associated with radio relics usually span from approximately $-1.0$ to $-1.5$ \citep[see][and references therein]{2019SSRv..215...16V}. In the case of radio halos, \citet{2009A&A...507.1257G} find integrated spectral indices ranging from about $-1.1$ to $-1.4$ between 325 MHz and 1.4 GHz, with a mean value of $\alpha_{\rm int} \simeq -1.3$. Similarly, \citet{2021A&A...651A.115V} report a mean integrated spectral index of $\alpha_{\rm int} \simeq -1.2$ between 150 MHz and 1.4 GHz. New observational catalogs are becoming increasingly available \citep[see][e.g.,\ a recent MeerKAT catalog]{2025MNRAS.543.1638K}, which will provide improved constraints on radio halo spectral indices. 

In summary, we show that face-on radio shocks or relics lead to integrated spectral indices of $\alpha_{\rm int} \lesssim -1$. For MACS~J0018.5+1626, we find that during the most likely epoch, coincident with the peak of the radio power, the integrated spectral index ranges from $\sim -1.3$ to $-1.8$, depending on the specific model and the frequency range (or data points) used for the fit. New observations at higher radio frequencies will be essential to further constrain our models.

\begin{figure}
    \centering
    \includegraphics[width=0.9\columnwidth]{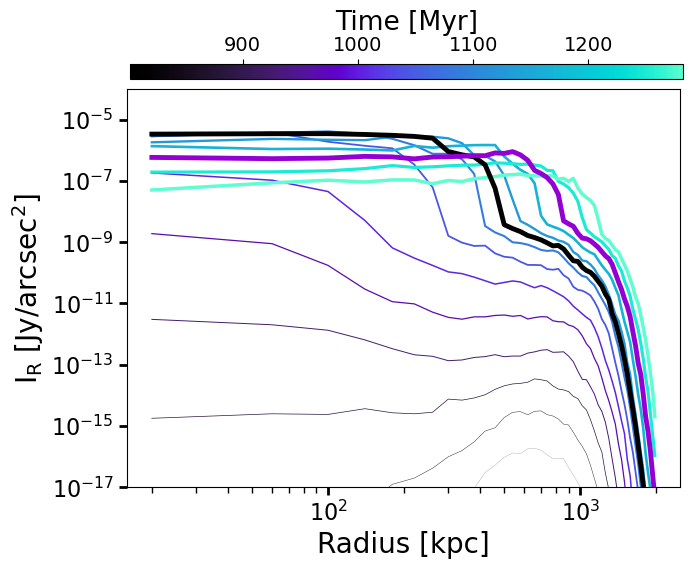}
    \includegraphics[width=0.9\columnwidth]{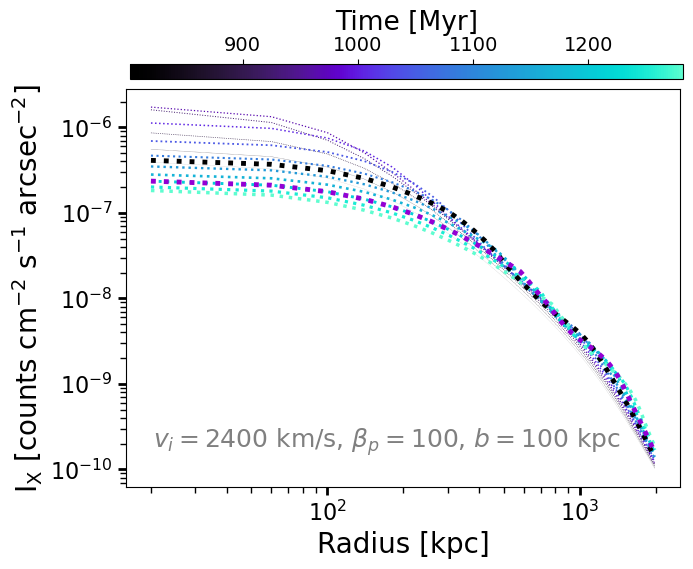}
    \caption{Evolution of radio (144 MHz; \textit{upper panel}) and X-ray (0.5-7 keV band; \textit{lower panel}) radial profiles for the run initialized with $v_i=2400$ km~s$^{-1}$, $b=100$ kpc, $\beta_p=100$ in Tab.~\ref{tab:matched}. Thick black and dark-violet lines show times $t_1$ and $t_2$ from Tab.~\ref{tab:matched}. The radial profiles are computed on surface brightness maps obtained by projecting along the merger axis (face-on view).}
    \label{fig:profiles_evolution}
\end{figure}

%%%%%%%%%%%%%%%%%%%%%%%%%%%%%%%%%%%%%%%%%%%%
\subsubsection{Radial profiles}\label{sec:view_angle}
%%%%%%%%%%%%%%%%%%%%%%%%%%%%%%%%%%%%%%%%%%%%

\begin{figure*}
    \centering
    \includegraphics[width=0.95\columnwidth]{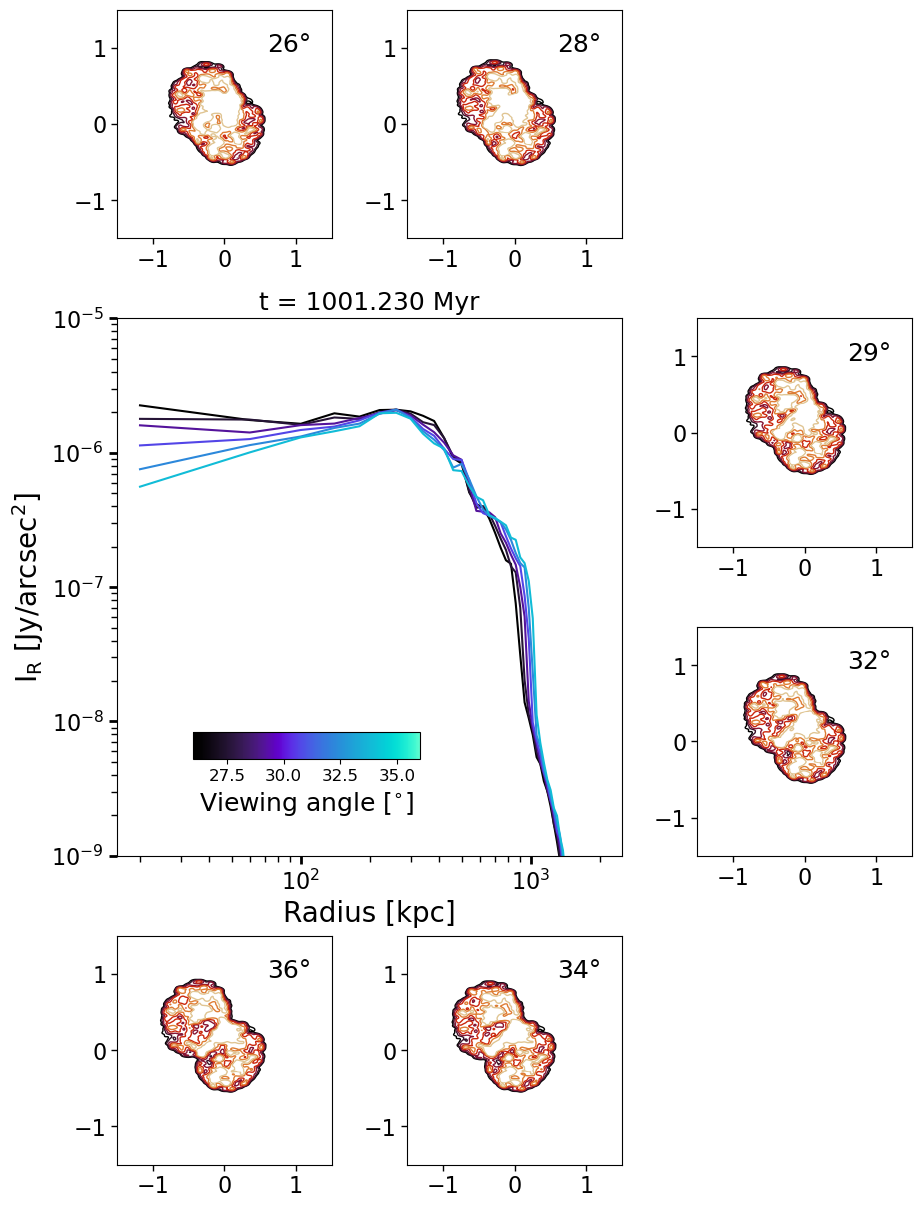}
    \hspace{15mm}
    \includegraphics[width=0.95\columnwidth]{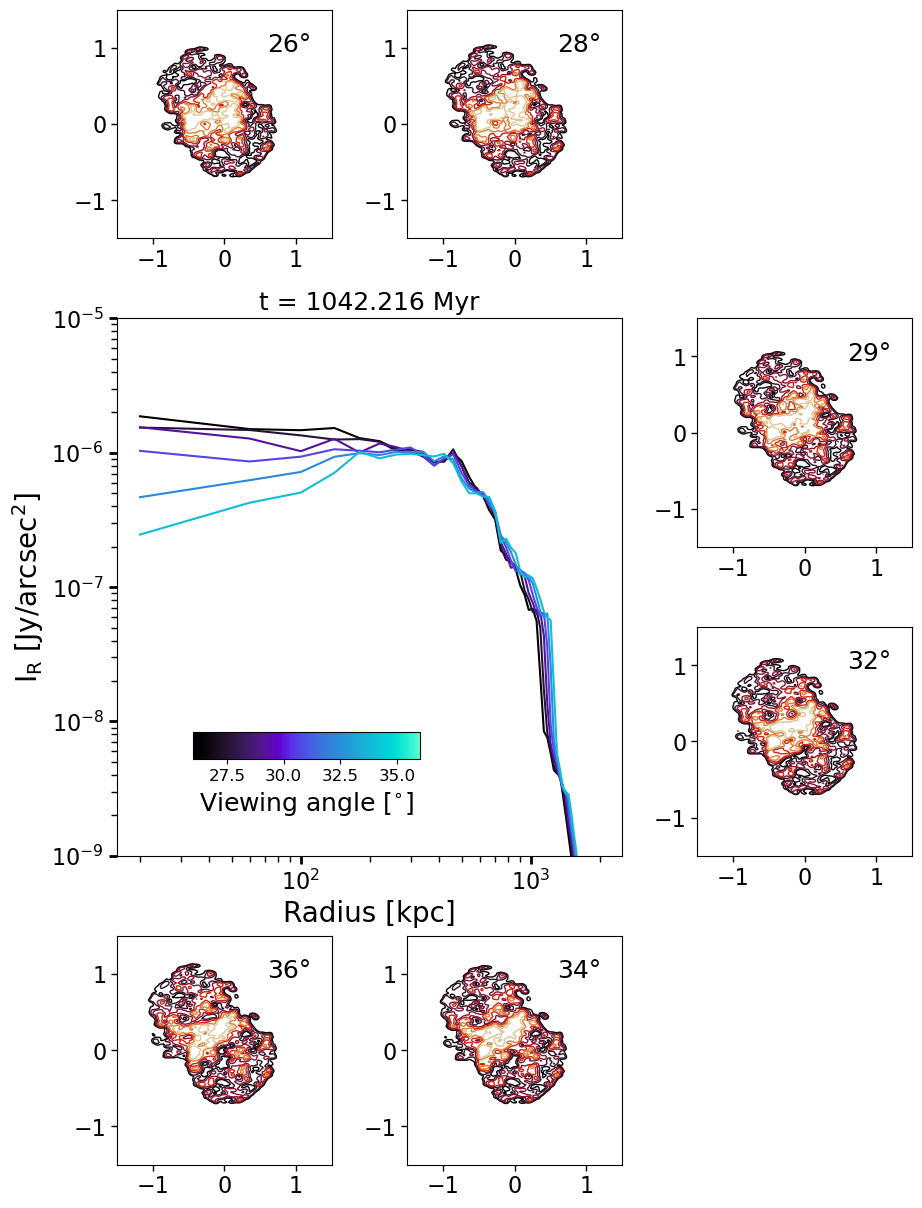}
    \caption{Variations of the radio profiles (\textit{central panels}) and surface brightness maps (\textit{upper}, \textit{right} and \textit{lower sub-panels}) with the viewing angle. We show the run initialized with $v_i=3000$ km~s$^{-1}$, $b=100$ kpc, $\beta_p=100$ in Tab.~\ref{tab:matched}. The radio surface brightness sub-panels are in units of Jy/beam. Contours are shown at $3\sigma\sqrt{N_L}$ levels, with $N_L = 1, 2, 4, 8, 16, 32, 64$, corresponding to progressively higher reference levels for extended emission, where $\sigma = 1.8 \, \mu$Jy/beam for the purposes of better visualization. We show two different times (\textit{left} and \textit{right panels}) of the same run. 
}
    \label{fig:multipanel_profiles}
\end{figure*}

In Fig.~\ref{fig:profiles_evolution}, we show an example of the time evolution of the radio and X-ray profiles for one of our optimistic runs, namely, the run initialized with $v_i=2400$ km~s$^{-1}$, $b=100$ kpc, $\beta_p=100$. We compute the profiles from surface-brightness maps obtained by projecting along the merger axis (face-on view). In this case, we do not apply any Gaussian smoothing to the radio images. The time evolution of the radio profiles reflects the time evolution of the radio power in Fig.~\ref{fig:radio_power} and the total particle spectrum in Fig.~\ref{fig:spectra}. As the merger progresses and the radio emission builds up, the radial profiles show a progressive enhancement toward the central region of the cluster. At later times, the profiles begin to flatten in the central region and exhibit a decline at large radii. The radius where this transition between the two behaviors occurs gradually shifts to increasingly larger radii with time, reflecting the outward propagation and expansion of the shock waves. The X-ray profiles, on the other hand, show a less complex time evolution. In general, the X-ray surface brightness exhibits an initial enhancement followed by a decline. These variations occur primarily in the central region of the cluster ($\lesssim 300$ kpc). Overall, the radio and X-ray profiles show a clear sensitivity to the merger dynamics across the different simulation runs. We show these differences in Fig.~\ref{fig:radial_profiles_all} of App.~\ref{app:variations}.

As mentioned before, the likely viewing angle of MACS J0018.5+1626 is $\sim 27^{\circ}$–$40^{\circ}$ from the merger axis \citep[see Section 5.3 in][]{2024ApJ...968...74S}. Hence, we analyzed the effect of the viewing angle in the surface brightness maps and the corresponding radial profiles. In the LOFAR observations we see that the emission is more extended in the NE direction (see Fig.~\ref{fig:lofar}). To match the orientation of the observed system, we explore viewing-angle variations in the POS. We consider lines of sight inclined $45^\circ$ from the $+X$ axis in the image coordinates, corresponding to the NE direction in the projected maps. We show an example of these variations in Fig.~\ref{fig:multipanel_profiles} for the run initialized with $v_i=3000$ km~s$^{-1}$, $b=100$ kpc, $\beta_p=100$ at two different epochs to illustrate the general effects of the viewing angle. For both epochs, a larger viewing angle (e.g., $36^{\circ}$) leads to a decrease in the radio surface brightness toward the cluster center. The radio surface-brightness maps (sub-panels of Fig.~\ref{fig:multipanel_profiles}) show that this behavior arises due to the fact that at larger viewing angles the emission from the two shocks no longer significantly overlap. As a result, the radio emission appears more extended in a single direction. At large radii ($\gtrsim 200$ kpc) the variations in the radial profiles due to the viewing angle are minimal.

\begin{figure}
    \centering
    \includegraphics[width=\columnwidth]{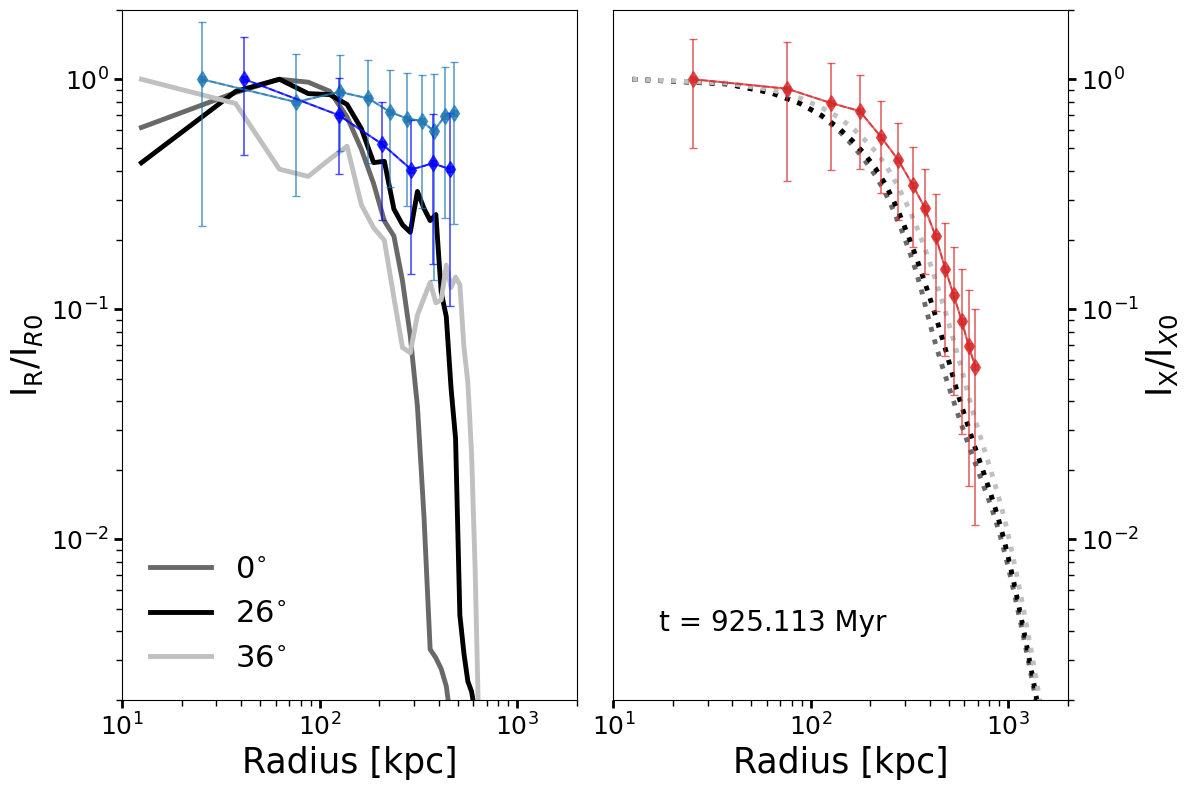}
    \includegraphics[width=\columnwidth]{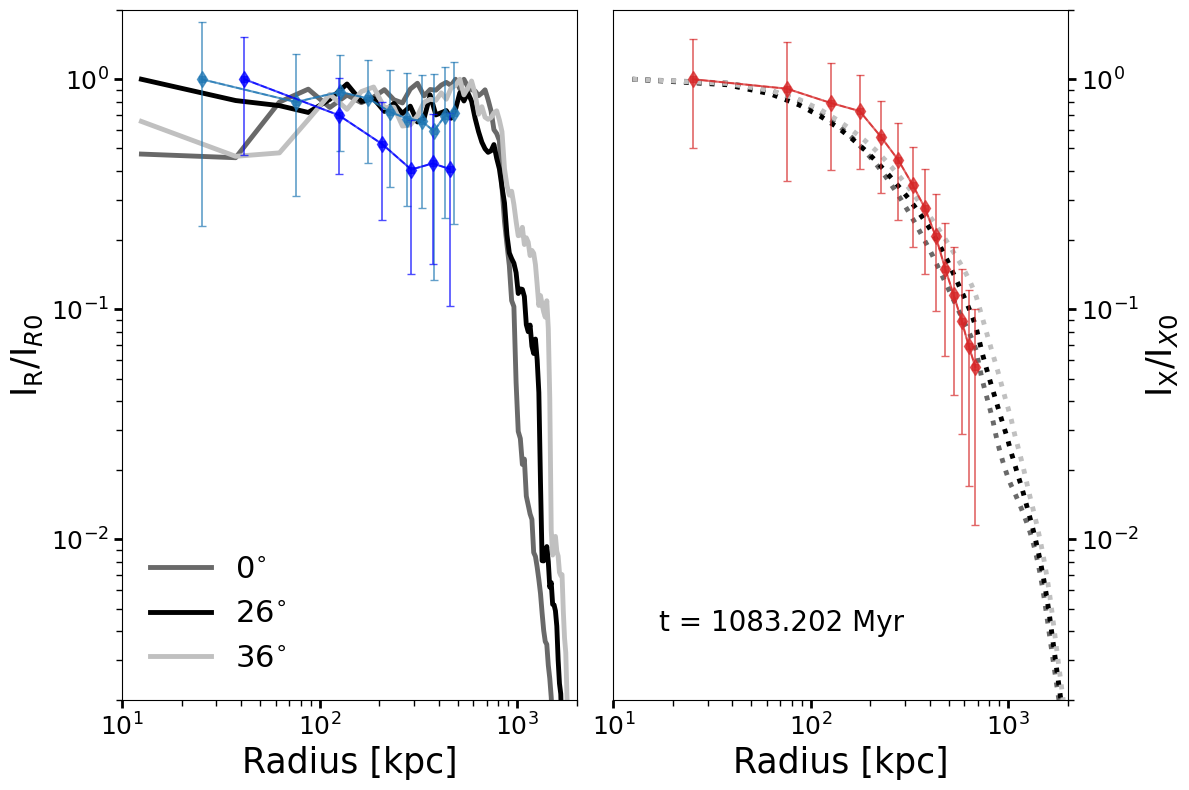}
    \caption{Normalized radio (left-axis and solid lines) and X-ray (right-axis and dotted lines) radial profiles for two epochs of the run initialized with $v_i=3000$ km~s$^{-1}$, $b=100$ kpc, $\beta_p=100$ in Tab.~\ref{tab:matched}. We overplot the data from Chandra in red and from LOFAR in blue. We use the radio images with a beam size of 11.02" $\times$ 5.69" (light blue) and  16.59" $\times$ 10.26" (dark blue). We show the profiles for the face-on view, $0^{\circ}$, and for $26^{\circ}$ and $36^{\circ}$ viewing angles.}
    \label{fig:profiles_with_data}
\end{figure}

Next, we computed radial profiles from the LOFAR radio images and the Chandra X-ray image to compare with the results of our simulations. An example of the annuli used to compute the profiles is shown in the lower panels of Fig.~\ref{fig:lofar}. The width of each annulus matches the radio beam size. We exclude emission below $2\sigma_{\mathrm{rms}}$ (see Tab.~\ref{tab:obs}). For the Chandra images, we apply the same binning but compute the radial profile out to larger radii. We truncate the profile at the radius where 70\% of the pixels have positive background-subtracted flux, which corresponds to $\sim$640 kpc. We show an example of this comparison in Fig.~\ref{fig:profiles_with_data}. The error bars in the LOFAR surface brightness profiles (in Jy/beam) are $\sigma_{\rm err} = \sigma_{\rm rms}/\sqrt{N_{\rm beam}}$, where $N_{\rm beam}=N_{\rm pix}/A_{\rm beam,pix}$ is the number of beams, $N_{\rm pix}$, is the number of pixels in the selected region (in this case an annulus), and $A_{\rm beam,pix}$ is the beam area in pixels. 
The error bars of the Chandra profile are Poissonian, computed as the square-root of the summed counts in each annulus and scaled by the annulus area.

In Fig.~\ref{fig:profiles_with_data}, we show the normalized radial profiles from both the observations and the simulations. Even by visual inspection, certain simulation epochs and viewing angles appear to better reproduce the shape of the observed radio and X-ray profiles. For the specific case shown here, corresponding to the run initialized with $v_i = 3000~\mathrm{km \,s^{-1}}$, $b = 100~\mathrm{kpc}$, and $\beta_p = 100$, the later epoch (lower panel of Fig.~\ref{fig:profiles_with_data}) and a viewing angle of $26^{\circ}$ appear to provide a closer match to the flatter radio profile observed at $r \gtrsim 100~\mathrm{kpc}$ in the high-resolution LOFAR image (beam size $11.02'' \times 5.69''$). The X-ray radial profile also shows improved agreement with the observations at this later epoch. Motivated by these trends, we systematically explore the dependence on epoch and viewing angle across different simulation runs using a simple $\chi^2$–based goodness-of-fit statistic comparing the normalized simulated and observed radial profiles. For each simulation snapshot and viewing angle, the simulated profiles are linearly interpolated onto the radial bins of the observations and compared within the radial range where the observational profiles are reliably detected. The $\chi^2$ is defined as
\begin{equation}
\chi^2 = \sum_i \frac{\left[I_{\rm sim}(r_i) - I_{\rm obs}(r_i)\right]^2}{\sigma_{\rm obs}^2(r_i)} ,
\end{equation}
where $I_{\rm sim}$ and $I_{\rm obs}$ are the normalized simulated and observed surface brightness profiles, respectively, and $\sigma_{\rm obs}$ denotes the observational uncertainties. For the radio profiles, this corresponds to $\sigma_{\rm err}$. We use the reduced $\chi^2_\nu = \chi^2/N_{\rm dof}$ as a relative goodness-of-fit metric to rank different epochs and viewing angles. In the following, we focus on the radio radial profiles. The $\chi^2$ statistics are computed by comparing the simulations to the profiles derived from the highest-resolution (no uv-taper) and an intermediate-resolution (uv-taper 8) LOFAR images (see Tab.~\ref{tab:obs}). We show the results of the statistics for all the optimistic runs in Tab.~\ref{tab:matched} in App.~\ref{app:variations}.

In Fig.~\ref{fig:chi_obs00} of App.~\ref{app:variations}, we show a two-dimensional grid of reduced $\chi^2_\nu$ values as a function of viewing angle and simulation epoch. The viewing angles lie within the likely viewing angle of MACS J0018.5+1626, $\sim 27^{\circ}$–$40^{\circ}$, and the epochs lie within approximately $t_1$ and $t_2$ in Tab.~\ref{tab:matched}. As expected, the reduced $\chi^2_\nu$ values vary across different epochs and viewing angles changes due to variations in the morphology of the shock surfaces during the merger. Overall, lower $\chi^2_\nu$ values, corresponding to a closer match in the shape of the radial profiles, are obtained when comparing the simulations to the radial profiles derived from the highest-resolution LOFAR images. This is mainly due to enhanced radio emission at larger radii, $\gtrsim 100$ kpc, as can be seen in Fig.~\ref{fig:profiles_with_data}.
From these comparisons, several trends emerge. For runs initialized with $v_i = 2400~\mathrm{km\,s^{-1}}$, viewing angles $\lesssim 30^{\circ}$ and epochs closer to $t_2$ provide better agreement with the observed radio profiles. For runs initialized with $v_i = 3000~\mathrm{km\,s^{-1}}$, viewing angles $\lesssim 30^{\circ}$ and epochs closer to $t_2$ are favored when comparing with the highest-resolution LOFAR images, while viewing angles $\lesssim 36^{\circ}$ and epochs closer to $t_1$ yield better agreement when using the intermediate-resolution LOFAR images.

Each cell corresponds to the comparison between a simulated and an observed normalized radio radial profile for a given snapshot and projection. Lower values of $\chi^2_\nu$ indicate a closer match in profile shape. For visual guidance, we highlight the subset of models belonging to the lowest 10\% of $\chi^2_\nu$ values with white star symbols, identifying regions of parameter space that provide the best agreement with the observations.

We emphasize that $\chi^2_\nu$ is used here as a relative goodness-of-fit metric, and not as an absolute test of the statistical acceptability of a given model for MACS~J0018.5+1626.

%%%%%%%%%%%%%%%%%%%%%%%%%%%%%%%%%%%%%%%%%%%%
\subsubsection{Point to point correlation functions}\label{sec:ptp}
%%%%%%%%%%%%%%%%%%%%%%%%%%%%%%%%%%%%%%%%%%%%

\begin{deluxetable}{ccccc}
\tablecaption{Radio-X-ray correlation\label{tab:ptp}}
\tablehead{
\colhead{uv-taper} &
\colhead{Slope} &
\colhead{$r_P$} &
\colhead{$\rho_S$} &
\colhead{$\theta_{\rm FWHM}$ [$^{\prime\prime}$]}
}
\startdata
None & $0.12 \pm 0.02$ & 0.41 & 0.44 & 7.92  \\
8    & $0.31 \pm 0.04$ & 0.60 & 0.66 & 13.04 \\
10   & $0.38 \pm 0.05$ & 0.62 & 0.68 & 14.62 \\
15   & $0.51 \pm 0.07$ & 0.75 & 0.76 & 19.24 \\
30   & $0.61 \pm 0.11$ & 0.89 & 0.90 & 36.69 \\
\enddata
\tablecomments{Best linear fit to the observed point-to-point correlation between
radio (LOFAR; 144 MHz) and X-ray (Chandra; 0.5-7 keV) surface brightness. We report the fitted slopes and Pearson ($r_P$) and Spearman ($\rho_S$) correlation coefficients of the observations described in Sec.~\ref{sec:observations}. See also Fig.~\ref{fig:lofar}. The uncertainty in the slope represents the $1\sigma$ error from the least-squares regression. The last column shows the beam FWHM which defined the width of the squared regions.}
\end{deluxetable}

\begin{figure}
    \centering
   \includegraphics[width=0.9\columnwidth]{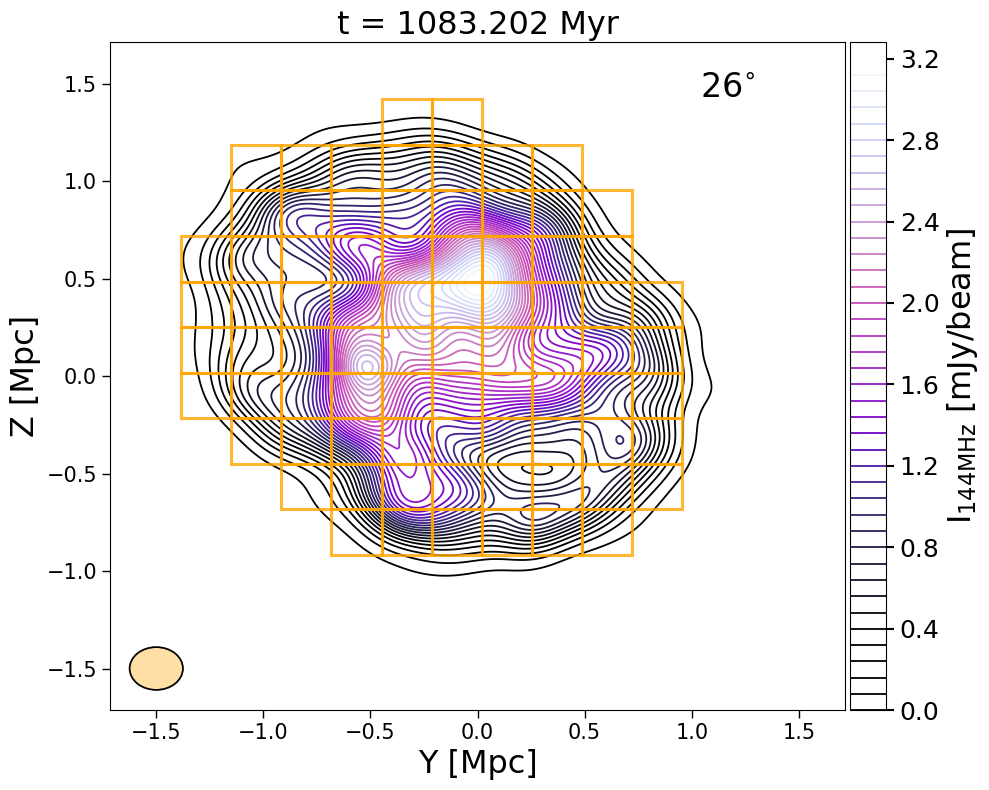}
  \includegraphics[width=0.9\columnwidth]{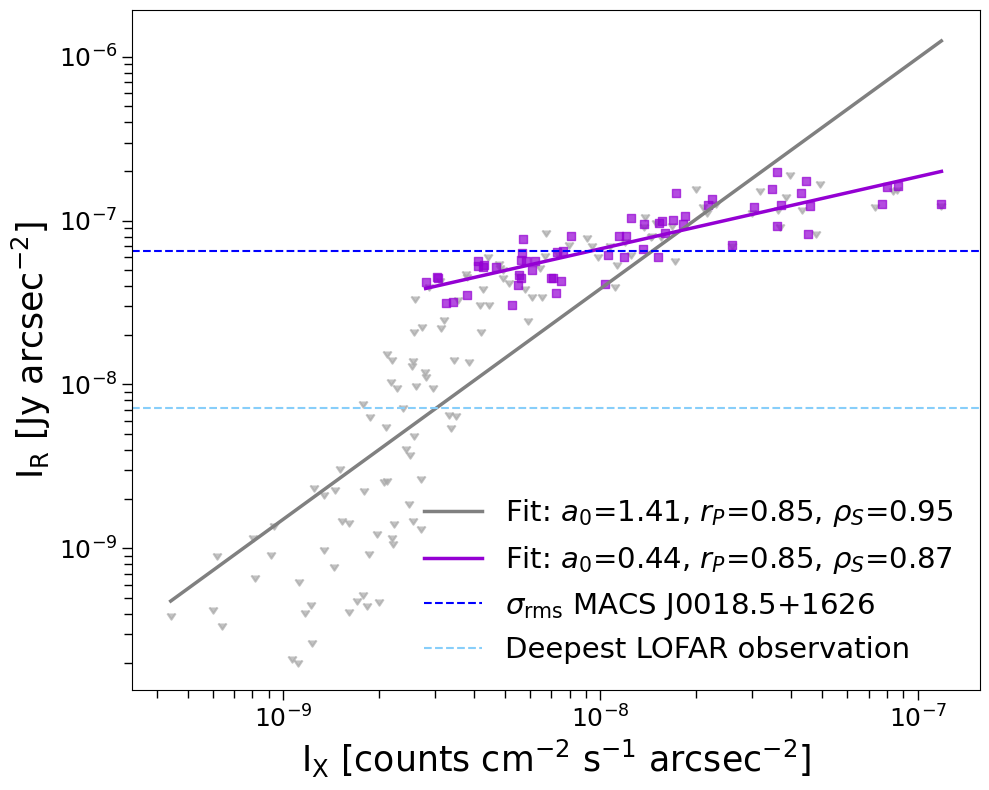}
    \caption{Example of the point-to-point correlation function for the run initialized with $v_i=3000$ km~s$^{-1}$, $b=100$ kpc, $\beta_p=100$ and the shock acceleration efficiency model from \citealt{Kang_2007} with $\eta_0=10^{-1}$. \textit{Upper panel}: Contour radio surface brightness contours as seen with a viewing angle of 26$^{\circ}$. Here $X$ and $Y$ are the image plane coordinates. On the left lower corner we show the size of the beam, 39.07" $\times$ 34.45", which corresponds to that of the lowest resolution LOFAR image. We overplot the squared regions used to compute the correlation function above the threshold (see text). \textit{Lower panel}: Radio vs X-ray surface brightness scatter plot. We show the linear fits for all the data points (gray) and those above the threshold (violet). The horizontal light-blue line indicates the lowest value achieved in the deepest LOFAR image published by \citet{2025A&A...695A..80S}.}
    \label{fig:ptp_squares}
\end{figure}

In this section we discuss the point-to-point correlation between radio and X-ray surface brightness from the observations and simulations. This is a typical analysis reported for diffuse radio emission in galaxy clusters in the form of radio halos. We fitted the relation between the radio and X-ray surface brightness using a linear model in logarithmic space,
\begin{equation}
\log I_{\rm R} = a_0 + b_0\,\log I_{\rm X}.
\end{equation}
To compute the correlation we identified squared regions within the ellipse region defined in Fig.~\ref{fig:lofar} and computed the mean surface brightness in each square. The width of each square is selected according to the beam FWHM associated to each radio image. We only considered radio surface brightness values larger than $2\sigma_{\rm rms}$ (see Tab.~\ref{tab:obs}). We define the same squared regions in the Chandra image to compute the point-to-point correlation. In Tab.~\ref{tab:ptp} we show the results from the least-squares regression. The strength of the correlation is quantified using both the Pearson and Spearman coefficients. 

The radio and X-ray surface brightnesses are correlated independently of the resolution of the radio images. The strongest correlation appears for the radio image with the lowest resolution (uv-taper 30), with a Pearson coefficient and a Spearman coefficient $r_P=0.89$ and $\rho_S=0.9$, respectively. On the other hand, at the highest resolution radio image (i.e., no uv-taper) shows just a moderate correlation with the X-ray surface brightness. These correlations are sublinear. Overall, the linear slope of the correlation varies between 0.12 and 0.61. 

We use a similar methodology to compute the point-to-point correlation between radio and X-ray surface brightness in the simulations. First, the simulated radio images are smoothed with a Gaussian beam. We generate different radio images with the same beam sizes as those of the LOFAR observations (see Tab.~\ref{tab:obs}). Then, we grid the simulated radio images with squared regions with widths corresponding to the different beam sizes from the LOFAR images (see Tab.~\ref{tab:obs} and \ref{tab:ptp}). For the X-ray simulated images we define the same squared regions. We reject the information from squared regions with no radio emission from this analysis. We compute the least-squares regression twice: one without putting any threshold on the radio surface brightness values and a second one with a conservative threshold. For the latter, and given the variety of runs and models that we have, we considered as a conservative threshold a $10^{\rm th}$ of the maximum value of the radio surface brightness value in each of the radio images\footnote{A threshold of one tenth of 
the peak value is broadly consistent with the typical dynamic 
range used in point-to-point correlation analyses of observed 
diffuse radio emission \citep[see e.g.][for applications to samples of 
clusters and radial profiles, respectively]{2024A&A...686A...5B, 
2025ApJ...992...88R}}. 
An example of this procedure can be seen in Fig. \ref{fig:ptp_squares}. We show an example of the run initialized with $v_i=3000$ km~s$^{-1}$, $b=100$ kpc, $\beta_p=100$ at $t=1083.2$ Myr. In the upper panel, we show the grid with squared regions. In the lower panel, we show the point-to-point correlation. In this case, the linear fit indicates a sublinear correlation for those data points above the selected threshold and a superlinear correlation for all data points. 

\begin{figure}
    \centering
    \includegraphics[width=0.92\columnwidth]{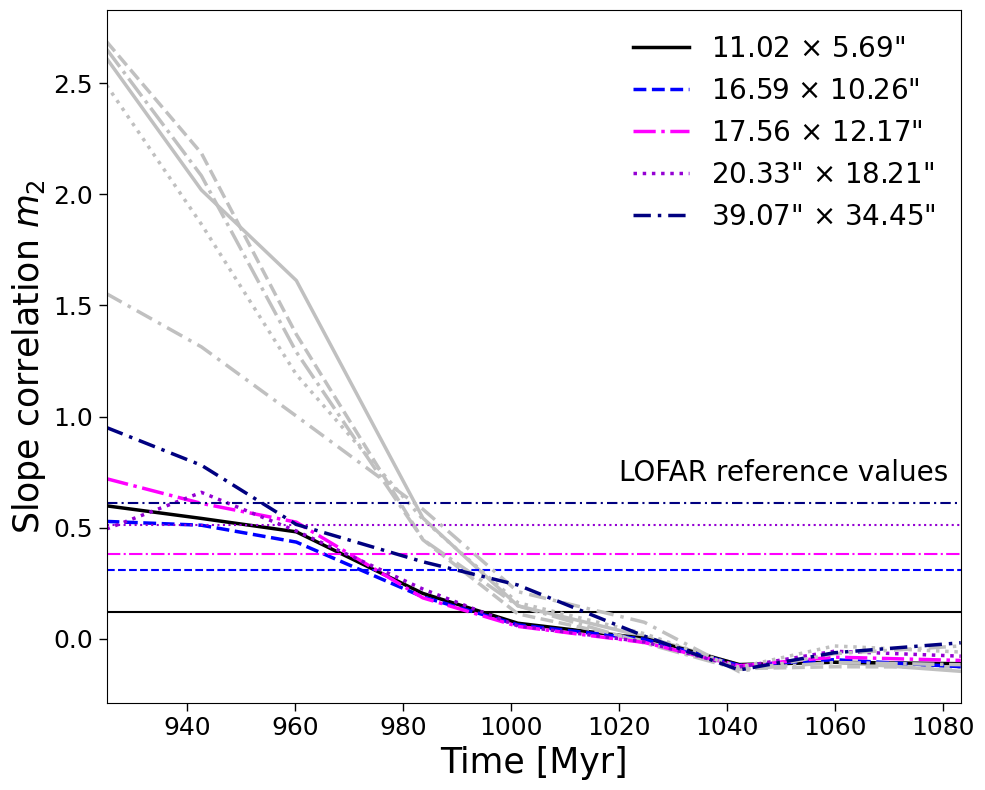}
    \includegraphics[width=0.92\columnwidth]{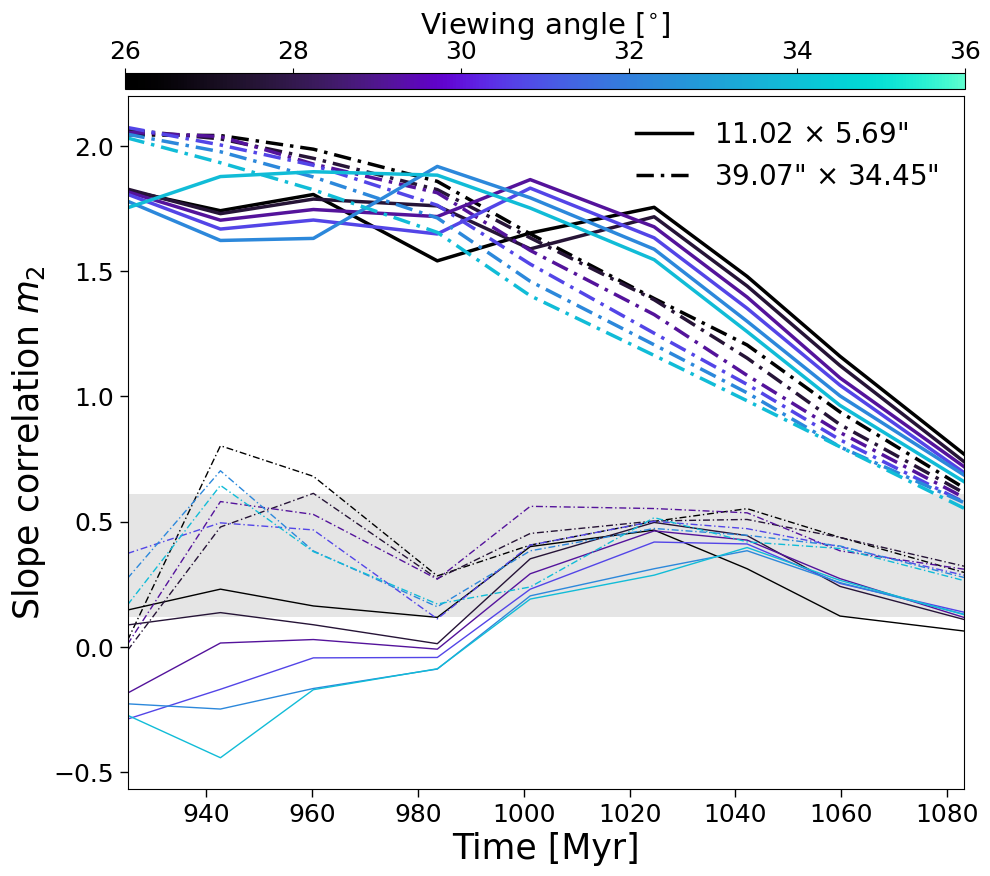}
    \caption{Evolution of the slope ($m_2$) of the radio vs X-ray point-to-point correlation function for the run initialized with $v_i=3000$ km~s$^{-1}$, $b=100$ kpc, $\beta_p=100$ and the shock acceleration efficiency model from \citealt{Kang_2007} with $\eta_0=10^{-2}$. \textit{Upper panel (face-on view)}: We show the slope for the simulated data points without (gray) and with (colors) a threshold. The different beam sizes are shown with different line-styles (see Tab.~\ref{tab:obs}). The horizontal lines correspond to the observational values (see Tab.~\ref{tab:ptp}). \textit{Lower panel (off-axis view)}: Same evolution as upper panel but showing the effect of different view angles. We only show the slope values corresponding to the analysis of the lowest and highest resolution simulated images. The gray shaded region shows the highest of lowest LOFAR reference values, similar to the upper panel. The slopes for the simulated data points without and with a threshold are shown with thicker and thinner line widths, respectively.}
    \label{fig:slopes_2PL}
\end{figure}

We note that, however, following this observational method may be misleading. As can be seen in the upper panel of Fig.~\ref{fig:slopes_2PL}, the data points without a threshold would be better fitted by two power laws instead of one. The separation in trends clearly distinguishes a fresh population of accelerated electrons through DSA and an already old population of electrons. For this reason, we fit the data with a continuous broken power-law model characterized by two asymptotic slopes $m_1$ and $m_2$, a break scale $x_b$, and a smoothness parameter $\Delta$:
\begin{equation}
y(x)
 = y_0
\left(\frac{x}{x_b}\right)^{m_1}
\left[
1 + \left(\frac{x}{x_b}\right)^{1/\Delta}
\right]^{(m_2 - m_1)\,\Delta}
\end{equation}
In this case, $m_1$	corresponds to the slope of the fainter data points, which trace an older population of electrons that has experienced more cooling\footnote{We additionally checked that the data where ${\rm I}_{\rm R} \lesssim 10^{-8}$ in Fig.~\ref{fig:ptp_squares} corresponds to an older population of electrons by inspecting the corresponding spectral index map. We find that $\alpha_{\rm 1.5 \, GHz}^{144 \, MHz}\lesssim-2.5$.}. In contrast, $m_2$ corresponds to the slope of the brighter data points, which represent the emission that is more readily detected in observations. Recent studies of radio halos have started to explore this approach of fitting a broken power-law to the point-to-point correlation function \citep[see][]{2024A&A...686A...5B}.

In the upper panel of Fig.~\ref{fig:slopes_2PL}, we show how the slope ($m_2$) of the point-to-point correlation function changes as a function of time for the same run. We start by showing the evolution for the radio images obtained by projecting along the merger axis (face-on). 
The correlation is superlinear and then it transitions to a sublinear function. This trend is closely related to the evolution of the shocks and their morphology. While the morphology of each of the two shocks during this period can be fairly described by two spherical shells, the curvature changes. In an earlier phase, the shocks are smaller and with higher curvature. More importantly, the highest Mach numbers are located at the shock caps (or apex regions), leading to enhanced radio emissivity in those regions (see Sec.~\ref{sec:DSA}). In the face-on projection, this leads to peaked radio emission. We see a hint of this already in the emission maps shown in Fig.~\ref{fig:emiss_maps2} and the radial profiles in Fig.~\ref{fig:profiles_evolution}.  
Later on, as the shocks propagate outward, their size increases and their curvature decreases. The regions of high Mach number progressively extend over a larger
fraction of the shock surface, reducing the Mach number contrast along the shock surface. Therefore, this also reduces the contrast between the brightest
radio emitting regions and the rest of the shock leading to radio emission that becomes more
spatially uniform. Overall, in the face-on projection view, this evolution starting from an initially peaked surface brightness distribution that gradually
broadens and flattens, naturally explains the transition from a superlinear to a
sublinear radio-X-ray correlation. We note that the transition between superlinear and sublinear is seen in the results of both fits using all data points (shown with gray lines) and only those above the threshold.

In the upper panel of Fig.~\ref{fig:slopes_2PL}, we can also see the effect of the beam size. The largest beam size erases signatures of substructure and smooths out the contrast between the brightest
radio emitting regions and the rest of the shock. This generally will flatten the slope of radio-X-ray correlation. We show as a reference the slope values obtained with the LOFAR and Chandra images and reported in Tab.~\ref{tab:ptp}. Between $\sim980$ and $\sim1020$ Myr, the slope values derived from the simulated images are consistent with the observationally inferred values for MACS J0018.5+1626. 

Next, we show that the viewing angle also affects the point-to-point correlation between the radio and X-ray surface brightness\footnote{See App.~\ref{app:another_angle} for an example viewing angle where the two shocks no longer overlap.}. This is illustrated in the lower panel of Fig.~\ref{fig:slopes_2PL}. When no surface-brightness threshold is applied, the radio–X-ray correlation remains superlinear for all data points for most of the time interval, independent of the beam size. In contrast, when only data points above the threshold are considered, the correlation becomes predominantly sublinear. At the highest resolution, the fitted slope can become negative, indicating an apparent anti-correlation. 

\begin{figure}
    \centering
    \includegraphics[width=\columnwidth]{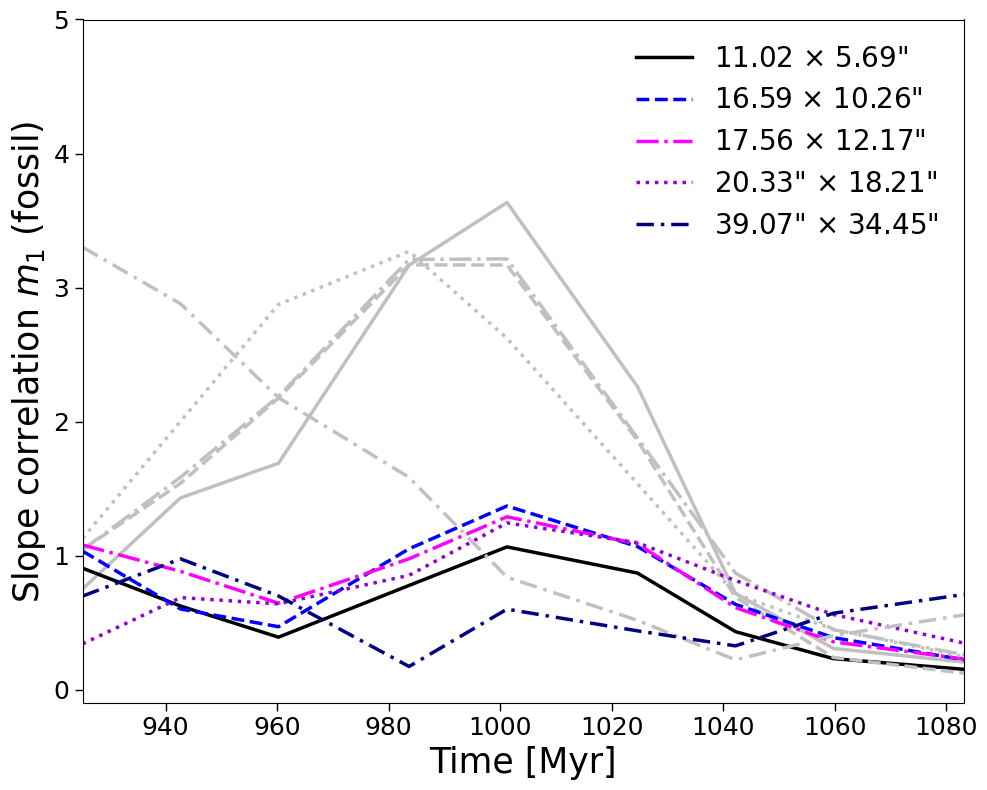}
    \caption{Evolution of the slope ($m_1$) of the radio vs X-ray point-to-point correlation function of the older population of electrons. We show the face-on view and same run as in Fig.~\ref{fig:slopes_2PL}.}
    \label{fig:slope_older}
\end{figure}

Finally, in Fig.~\ref{fig:slope_older}, we show the same time evolution as in the upper panel of Fig.~\ref{fig:slopes_2PL} but for the slope $m_1$, corresponding to the older population of electrons. In this case, we also observe that the correlation can transition from superlinear to sublinear when no threshold is applied, whereas it remains either linear or sublinear when a threshold is imposed. We suggest that the broken power-law fitting approach should only be regarded as a good first-order approximation for identifying the presence of two populations of electrons. However, this method alone is not accurate enough to unambiguously define the boundary between older and freshly injected electron populations. Furthermore, at a later period, well beyond the time of pericenter passage, we expect the older population of electrons to show little or no correlation\footnote{Note that we expect this to hold more generally for fossil electron populations of different origins, such as aged AGN plasma or electrons left over from previous shock or turbulent re-acceleration episodes. Therefore, in the case of DSA re-acceleration, a correlation is not necessarily expected.}. This is already hinted in Fig.~\ref{fig:slope_older} at $t\gtrsim 1060$ Myr.

Overall, this highlights the difficulty of disentangling older from fresher electron populations based solely on the interpretation of the radio–X-ray point-to-point correlation. In general, we find that $m_1 \neq m_2$ for radio shocks viewed face-on; however, in practice, observational thresholds may obscure this distinction.

In summary, we show that face-on radio shocks or relics exhibit a clear correlation between radio and X-ray surface brightness. The slope of this point-to-point correlation varies with time and depends on both the beam size and the viewing angle, leading to trends that are not uniquely sublinear or superlinear. Interestingly, radio halos are also observed to exhibit radio–X-ray point-to-point correlations, with slopes most often found to be sublinear
 \citep[see e.g.,][]{2001A&A...376..803G, 2019A&A...622A..20H,2020A&A...636A...3X, 2021A&A...646A.135R,2021A&A...650A..44B, 2022ApJ...933..218B,2022MNRAS.515.1871R}, with some cases showing linear \citep[see e.g.,][]{2020ApJ...897...93B} and superlinear correlations \citep[see e.g.,][]{2018ApJ...852...65R}. Therefore, our results indicate that such radio–X-ray correlations are not exclusive to radio halos, but can also arise in merger-driven shock and relic systems, depending on viewing geometry, resolution, and merger stage. This highlights the importance of accounting for these effects when interpreting observed correlation slopes. We will further discuss these results in Sec.~\ref{sec:discussion}.

%%%%%%%%%%%%%%%%%%%%%%%%%%%%%%%%%%%%%%%%%%%%
\section{Discussion}\label{sec:discussion}
%%%%%%%%%%%%%%%%%%%%%%%%%%%%%%%%%%%%%%%%%%%%

MACS J0018.5+1626 is a massive galaxy cluster, ${\rm M}_{500c}\sim10^{15} \, {\rm M}_{\odot}$, at a relatively high redshift of $z_{\rm obs}=0.5456$. Clusters of this mass typically host diffuse radio emission classified as radio halos \citep[see][for recent LOFAR studies]{2022A&A...660A..78B,2023A&A...672A..43C}. As shown in Sec.~\ref{sec:observations}, the observed diffuse radio emission exhibits a roughly round morphology that closely follows the X-ray emission and, based on morphology alone, would indeed be classified as a radio halo. In App.~\ref{app:radio_mass_relation}, we show the radio power–mass relation derived from LOFAR observations for both radio halos and radio relics along with our measured radio power (see Tab.~\ref{tab:obs}). The radio power of MACS J0018.5+1626 lies below the radio power–mass relation of radio halos. However, it is known that this relation has a scatter which can be attributed to various factors such as merger activity and/or particle acceleration mechanisms \citep[see discussion in][and references therein]{2023A&A...680A..30C}. On the other hand, the radio power of MACS J0018.5+1626 lies closely on the radio power–mass relation of radio relics.

Based on radio data alone, we cannot exclude a turbulent origin for the diffuse radio emission in MACS J0018.5+1626, and hence a radio halo interpretation. However, other observables point to the fact that this cluster is observed along a LOS almost aligned with the merger axis and very close to pericenter passage as found in the multi-wavelength analysis performed by \citet{2024ApJ...968...74S}. In this configuration, the presence of two merger-driven shocks is unavoidable, as demonstrated in this work. Furthermore, under these conditions, standard DSA is expected to operate. In addition to the two merger shocks, turbulence is likely to contribute to the observed emission. Radio halos are powered by turbulence injected in the ICM during mergers, and the turbulent cascade is expected to develop on timescales comparable to the dynamical time of galaxy clusters. Two scenarios are possible: either the ICM of MACS J0018.5+1626 was already turbulent prior to the current merger, or the turbulence is being driven by this merger. In the first scenario, a prior merger must have occurred 1--2 Gyr ago, as this is the typical timescale for turbulence to decay in the ICM, though such a prior event would be difficult to constrain observationally. In the second scenario, \citet{2013MNRAS.429.3564D,Donnert_2013} showed that turbulent re-acceleration reaches a maximum approximately 1.3--2 Gyr after core passage. Since the ICM-SHOX pipeline constrains the current merger to be at an earlier stage than this, the ongoing merger alone would not yet have had sufficient time to drive the level of turbulence required. Additionally, it is worth noting that only some clusters with double relics host radio halo emission, which may be related to the epoch at which these merging systems are observed. \citet{2017MNRAS.470.3465B} analyzed various double relic systems with and without radio halos and estimated the time since the merger, $t_{\rm merger}$, using the projected separation between the relics as a proxy for the actual time after core passage. From their Fig.~7, systems with confirmed radio halos all have $t_{\rm merger} \gtrsim 0.2$ Gyr, while for MACS J0018.5+1626 the radio power maximum is reached at $\sim$100--150 Myr post-pericenter. Although statistics are limited, this would suggest that a radio halo may not yet have had time to develop in MACS J0018.5+1626, further supporting the interpretation that the observed emission is shock-dominated at this epoch. On the other hand, systems such as El Gordo and MACS J1149.5+2223, which have comparable masses and mass ratios to MACS J0018.5+1626 and reported $t_{\rm merger} \sim 0.2$ Gyr, do host radio halos. Given the uncertainties in $t_{\rm merger}$, their true time since core passage could be comparable to the $\sim100$--$150$ Myr found here, which would suggest that a halo contribution in MACS J0018.5+1626 cannot be entirely ruled out. Directly constraining the level of turbulence in the ICM of this system through future analysis of X-ray surface-brightness fluctuations or XRISM observations  would be needed to assess this more carefully.
We will explore the role of turbulent re-acceleration in a follow-up paper.

Another point to discuss is that only a limited number of radio relics can be fully explained by standard diffusive shock acceleration (DSA). For example, \citet{2020A&A...634A..64B} showed that, even under the assumption of optimistic magnetic field strengths, only a subset of radio relics can be reproduced with electron shock efficiencies $\eta_e < 0.1$. As discussed above, the shock acceleration efficiency required in our scenario --- in which two merger shocks are viewed nearly face-on --- needs to be of the order of $\eta_e \sim 10^{-3}$--$10^{-2}$ to explain the observed emission. 
In radio relics, the synchrotron emissivity is dominated by the high-Mach-number tail of the distribution \citep[][]{dominguezfernandez2020morphology,Wittor_2021}. 
In our models, the shock Mach numbers can reach values of $\mathcal{M}_s \sim 3.8$--5, corresponding to electron acceleration efficiencies of $\eta_e \sim (2$--$3)\times 10^{-2}$. Such high values of $\eta_e$ are generally regarded as overly optimistic.
Recently, \citet{2024ApJ...976...10G} carried out a systematic study of electron acceleration efficiencies in quasi-parallel, non-relativistic shocks using PIC simulations. They found that the electron acceleration efficiency is typically $\lesssim 2\%$. In particular, for shocks with $\mathcal{M}_s \sim 2.7$ and $\mathcal{M}_s \sim 5.5$\footnote{We note that an additional factor accounting for the adiabatic index, $\sqrt{2\gamma_0}$, should be included in their Mach number definition to ensure consistency with ours.}, they report efficiencies of $\eta_e \sim 2 \times 10^{-3}$ and $\eta_e \sim 4 \times 10^{-3}$, respectively (see Fig.~\ref{app:CRe_vs_Mach}). It is worth noting that, in PIC simulations, the inferred acceleration efficiency depends on the extent of the non-thermal tail, i.e., how far the downstream cutoff lies above $p_{\rm inj}$, which is often limited by the finite simulation run time. Because PIC simulations typically capture only about one decade of the non-thermal tail, the resulting efficiencies should be regarded as lower limits. Given the expected logarithmic scaling with this ratio, an increase by a factor of $\sim 1$--4 over the quoted values would be plausible. Overall, these results suggest that adopting more conservative shock acceleration efficiencies in our simulations, for example $\eta_e \lesssim 10^{-3}$, would require additional mechanisms to reproduce the observed radio power of MACS J0018.5+1626. As discussed previously, turbulence may play an important role by further accelerating electrons \citep[see e.g.,][]{2014IJMPD..2330007B}. Given the merger configuration, this is likely to take place in the downstream of the shocks. Alternatively and as mentioned before, the presence of fossil electrons prior to shock passage could also contribute via shock re-acceleration or cluster-wide turbulent re-acceleration. We plan to explore these possibilities in a future work. In this context, new radio observations at higher frequencies will be crucial to further constrain these scenarios.

Another important factor in determining the radio power is the magnetic field strength. Higher magnetic field values can increase the synchrotron emissivity and therefore, increase the radio power. In our simulations, an initial average plasma beta of $\beta_p = 100$, combined with electron shock efficiencies $\eta_e \sim 10^{-3}$--$10^{-2}$, is sufficient to reproduce the observed radio power. While it would in principle be possible to explore models with lower values of $\beta_p$ \citep[see][for cases with $\beta_p \sim 50$]{2022MNRAS.512.2157C}, such configurations would result in unrealistically high magnetic field strengths at the time of pericenter passage and afterwards, reaching values of order tens of $\mu$G. At present, we lack independent observational diagnostics to constrain the magnetic field configuration in MACS J0018.5+1626, such as Faraday rotation measure information. Consequently, we refrain from further tuning the magnetic field strength and adopt $\beta_p = 100$ as a conservative and physically motivated choice. Future Faraday Rotation measure observations will be needed to better constrain the magnetic field properties in this system.

Finally, we discuss two subtle aspects that affect the radio--X-ray point-to-point correlation: geometry and resolution. First, we showed that the correlation found in our analysis is largely driven by projection effects and by the geometry of the shock surfaces. Accounting for more complex shock morphologies, such as those naturally produced in cosmological simulations, would therefore be a natural extension of this work. Alternatively, fossil electrons re-accelerated by the shock or by turbulence may contribute to the observed radio emission. These mechanisms can imprint different signatures on the observed radio--X-ray point-to-point correlation than what we observed in Section \ref{sec:ptp}. These aspects motivate a more systematic investigation beyond the specific case of MACS J0018.5+1626, which we defer to future work. Second, we note that observationally the radio--X-ray correlation appears to be resolution dependent (see Tab.~\ref{tab:ptp}). Part of this behavior may be attributed to different $\sigma_{\rm rms}$ thresholds adopted in radio images. However, changes in the slope due to different resolutions have been reported in few other systems, such as the Coma cluster \citep[see, e.g., Tab.~5 in][]{2022ApJ...933..218B}, and Abell 2744 \citep[][]{2021A&A...654A..41R}. While only a limited number of radio observations currently explore point-to-point correlations across multiple resolutions, we expect that forthcoming radio surveys will increasingly enable such analyses.

%%%%%%%%%%%%%%%%%%%%%%%%%%%%%%%%%%%%%%%%%%%%
\section{Summary and conclusions}\label{sec:conclusions}
%%%%%%%%%%%%%%%%%%%%%%%%%%%%%%%%%%%%%%%%%%%%

In this work, we carry out a detailed analysis of a simulated radio relic seen face-on motivated by previous observations and simulations of the galaxy cluster MACS J0018.5+1626.
We carried out 3D MHD simulations of binary cluster mergers in combination with tracer particles and a Fokker-Planck solver to model the radio emission from this cluster. Our analysis builds on results from the ICM-SHOX project, which, by matching X-ray, tSZ, kSZ, optical and lensing observables to simulations, indicates that MACS J0018.5+1626 is undergoing a binary merger close to pericenter passage and is being observed along a line of sight nearly aligned with the merger axis. We explore variations in the most likely initial conditions within the ICM-SHOX range, such as the relative velocity of the two clusters, $v_i$, and the impact parameter, $b$. 
We find that, under these conditions, two prominent shocks form with an average Mach number of $\mathcal{M}_s \sim 2$--$3$ and a standard deviation of $\sigma_{\mathcal{M}} \sim 0.5$--$1.5$. Within this scenario, we examine the conditions under which the radio emission from MACS J0018.5+1626 may be reproduced by standard DSA.

We use LOFAR observations of MACS J0018.5+1626 at 144 MHz as our observational benchmark and constrain the conditions necessary to reproduce the observed radio emission. In the following, we summarize these results:
\begin{itemize}
    \item[i)] \textit{DSA shock efficiency}: A shock efficiency model with an explicit dependence on the Mach number and effective electron acceleration efficiencies of $\eta_e \sim 10^{-3}$--$10^{-2}$ is required to reproduce the observed radio power at 144 MHz. 

    \item[ii)] \textit{Magnetic field strength}: An initial effective $\beta_p = 100$ is sufficient to reproduce the radio power at 144 MHz, largely independent of the other parameters, $v_i$ and $b$. At the epochs when the radio power peaks, this implies mean magnetic field strengths of order $\sim$0.8--1 $\mu$G with a standard deviation of $\sim$1.2 $\mu$G in the shock regions. Models with $\beta_p = 200$ provide a poorer match to the observations.
    
     \item[iii)] \textit{Epoch}: Overall, the radio power increases with time and peaks around the epoch of maximum shock energy dissipation. The epochs that best reproduce the observed radio power at 144 MHz are found during this phase. For runs with $v_i$ = 2400~km~s$^{-1}$, these epochs span $\sim$68--168~Myr after pericenter passage, whereas for runs with $v_i$ = 3000~km~s$^{-1}$, they span $\sim$40--200~Myr after pericenter passage.

    \item[iv)] \textit{Viewing angle}: Based on the morphology of the surface-brightness maps, and the comparison of radial profiles with observations, we find that relatively small viewing angles with respect to the merger axis provide a better match to the observed properties. For runs with $v_i$ = 2400~km~s$^{-1}$, viewing angles $\lesssim 30^{\circ}$ tend to yield improved agreement with the observed radio profiles. For runs with $v_i$ = 3000~km~s$^{-1}$, viewing angles $\lesssim 36^{\circ}$ are similarly preferred. Larger viewing angles result in a significant non-overlap of the two shocks, which is difficult to reconcile with the roundish central morphology observed in the LOFAR images.
\end{itemize}

Additionally, considering the full set of simulations, we can draw general conclusions about the comparison between radio and X-ray observables, which can be further explored in a future work:

\begin{itemize}
    \item[i)] \textit{Radial profiles}: We show that shocks viewed face-on generally lead to morphologically rounded radio emission. The simulated X-ray and radio radial profiles can be fairly similar at earlier epochs. However, at later epochs, due to the existence of more extended shocks, the radio radial profiles can become flatter than the X-ray radial profiles at larger radii. This excess in the radio at large radii can produce a deviation from a single component profile. We also show that the merger state and the viewing angle can change the shape of the radio profiles.
    \item[ii)] \textit{Correlation functions}: We show that the X-ray-radio point-to-point correlation function is subject to variability depending on factors such as the merger state, viewing angle, and angular resolution. As a result, the correlation is not uniquely sublinear or superlinear. We therefore emphasize that care must be taken when interpreting sublinear correlation slopes as uniquely associated to radio halo emission. We also show that more aged electrons can give rise to different slope values.
    % Maybe the last sentence is too vague
\end{itemize}

Independently of the specific particle acceleration mechanism at work, we conclude that since MACS J0018.5+1626 is undergoing a merger close to the pericenter passage and seen almost along the merger axis, this implies the presence of two shocks viewed almost face-on. Under these conditions, standard DSA is expected to operate. In this work, we showed that, within this scenario, the overlap of the two radio relics produces rounded radio emission at the cluster center which could be morphologically similar to radio halo emission. The electron shock efficiency required in this scenario is, however, on the optimistic side. We expect that additional processes, such as the presence of fossil electrons, could facilitate further acceleration of electrons in the environment of this cluster. In particular, shock re-acceleration and turbulent re-acceleration should also play a role. We will explore these alternatives in a future work. 
Additionally, new radio observations of MACS J0018.5+1626 at higher frequencies are needed to further refine our predictions. In particular, radio spectral index and polarization information would be crucial for placing stronger constraints on this system. 

Finally, this work demonstrates the power of the ICM-SHOX pipeline and, more broadly, the strength of combining multi-wavelength observations with simulations to constrain the physics of galaxy cluster mergers. This study presents the first set of simulations that model the radio emission of a cluster from the ICM-SHOX sample. In a future work, we will extend this approach to model the radio emission of the full cluster sample and aim to incorporate the radio mock observables into the ICM-SHOX pipeline.

%%%%%%%%%%%%%%%%%%%%%%%
\section*{Acknowledgments}
%%%%%%%%%%%%%%%%%%%%%%%
% 
%
% 
We thank J.~Donnert, the original developer of the Fokker-Planck 
solver, and C.~Gheller and F.~Vazza, who together with the first author 
form the core development team of the upgraded code.
We acknowledge very useful discussions with K.~Rajpurohit, D.~Caprioli, L.~B\"oss, I.~Zhuravleva, and A.~Heinrich.
We would also like to thank the anonymous reviewer for useful comments that increased the quality of this manuscript.
The simulations
presented in this work made use of computational resources on the
Cannon cluster at Harvard University. P. Dom\'inguez-Fern\'andez acknowledges the Future Faculty Leaders Fellowship at the Center for Astrophysics, Harvard-Smithsonian. 
This paper employs a Chandra dataset, obtained by the Chandra X-ray Observatory, contained in the Chandra Data Collection (CDC) 610\dataset[doi:10.25574/cdc.610]{https://doi.org/10.25574/cdc.610}. 
Support for JAZ was provided by the {\it Chandra} X-ray Observatory Center, which is operated by the Smithsonian Astrophysical Observatory for and on behalf of NASA under contract NAS8-03060. EMS acknowledges support from a National Science Foundation Graduate Research Fellowship (NSF GRFP) under Grant No. DGE‐1745301. EMS and JS acknowledge the support from NASA/80NSSC25K0597. TM acknowledges support from the Agencia Estatal de Investigaci\'on (AEI) and the Ministerio de Ciencia, Innovaci\'on y Universidades (MICIU) Grant ATRAE2024-154740 funded by MICIU/AEI//10.13039/501100011033. JG is supported by the National Science Foundation under Award No. 2401781. S.G. acknowledges financial support for his postdoctoral research at Princeton University through NSF grant PHY-2206607 and Simons Foundation grant MP-SCMPS-00001470.

LOFAR is the Low Frequency Array designed and constructed by ASTRON. It has observing, data processing, and data storage facilities in several countries, which are owned by various parties (each with their own funding sources), and which are collectively operated by the LOFAR ERIC under a joint scientific policy. The LOFAR resources have benefited from the following recent major funding sources: CNRS-INSU, Observatoire de Paris and Université d'Orléans, France; BMFTR, MKW-NRW, MPG, Germany; Science Foundation Ireland (SFI), Department of Business, Enterprise and Innovation (DBEI), Ireland; NWO, The Netherlands; The Science and Technology Facilities Council, UK; Ministry of Science and Higher Education, Poland; The Istituto Nazionale di Astrofisica (INAF), Italy. This research made use of the Dutch national e-infrastructure with support of the SURF Cooperative (e-infra 180169) and the LOFAR e-infra group. The Jülich LOFAR Long Term Archive and the German LOFAR network are both coordinated and operated by the Jülich Supercomputing Centre (JSC), and computing resources on the supercomputer JUWELS at JSC were provided by the Gauss Centre for Supercomputing e.V. (grant CHTB00) through the John von Neumann Institute for Computing (NIC). This research made use of the University of Hertfordshire high-performance computing facility and the LOFAR-UK computing facility located at the University of Hertfordshire and supported by STFC [ST/P000096/1], and of the Italian LOFAR-IT computing infrastructure supported and operated by INAF, including the resources within the PLEIADI special ``LOFAR'' project by USC-C of INAF, and by the Physics Department of Turin university (under an agreement with Consorzio Interuniversitario per la Fisica Spaziale) at the C3S Supercomputing Centre, Italy. This research is part of the project LOFAR Data Valorization (LDV) [project numbers 2020.031, 2022.033, and 2024.047] of the research program Computing Time on National Computer Facilities using SPIDER that is (co-)funded by the Dutch Research Council (NWO), hosted by SURF through the call for proposals of Computing Time on National Computer Facilities. 

\software{ The source codes used for
the simulations of this study, AREPO \citep{2010MNRAS.401..791S,2011MNRAS.418.1392P} is freely available
on \url{https://arepo-code.org/}. The main tools for our analysis are:
        pyXSIM \citep{ZuHone2016pyxsim},
        python \citep{van1995python}, matplotlib \citep{Hunter:2007}, numpy \citep{harris2020array}, Astropy \citep{2013A&A...558A..33A,2018AJ....156..123A,2022ApJ...935..167A}, and  
          the yt analysis toolkit \citep{2011ApJS..192....9T}. 
          %freely available at
          %\url{https://hea-%www.cfa.harvard.edu/~jzuhone/pyxsim.html},
          %\url{https://matplotlib.org/}, \url{https://www.numpy.org} and \url{https://yt-project.org/}. 
          }
% astropy, scipy, pyxsim




\bibliography{paola.bib,sample631}%.bib{}
\bibliographystyle{aasjournal}



\appendix

%%%%%%%%%%%%%%%%%%%%%%%%%%%
\section{Interpolation onto the tracers}
\label{app:interpolation}
%%%%%%%%%%%%%%%%%%%%%%%%%%%%

\setcounter{figure}{0}
\renewcommand{\thefigure}{A.\arabic{figure}}

\begin{figure*}
    \centering
    \includegraphics[width=0.9\columnwidth]{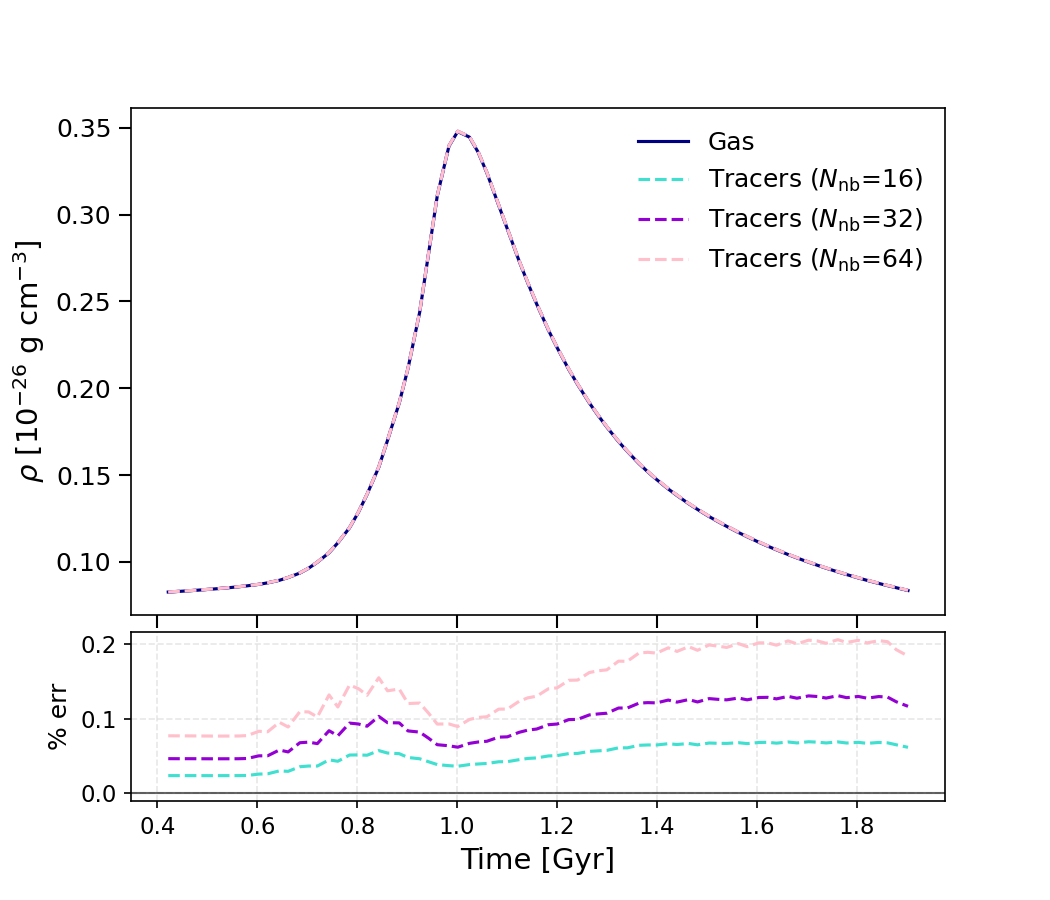}
    \includegraphics[width=0.9\columnwidth]{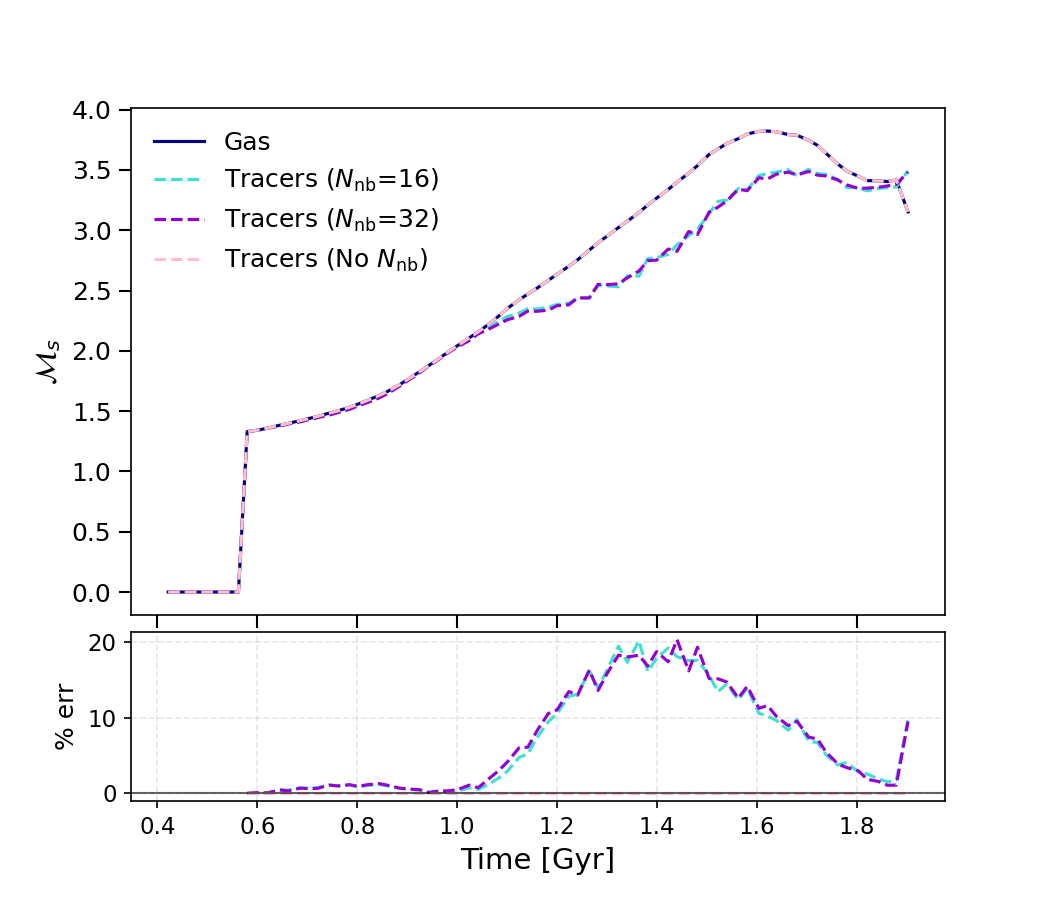}
    \caption{Mean $\rho$ (\textit{left panel}) and $\mathcal{M}_s$ (\textit{right panel}) gas evolution (solid) compared to the interpolated data onto the tracers (dashed) considering $N_{nb}=16,32,64$. In the lower panels of each of the plots we show the percentage relative error. Note that for $\mathcal{M}_s$ we also show the interpolation without accounting for neighbors.} 
    \label{fig:app1}
\end{figure*}

In this section we briefly describe the method to interpolate the information from the gas particles (Arepo \texttt{PartType0}) to the tracer particles (Arepo \texttt{PartType2}). First we identify the IDs and locate the coordinates of the desired gas particles. We build KD-trees for the gas and tracer particle coordinates. Then, using the tracer positions as queries against the gas KD-tree, we find the nearest gas neighbors which give the coincident indices used for tracer–gas matching. Having the coincident indices, we proceed to find the nearest neighbors $N_{\rm nb}$ and finally we compute a kernel-weighted average to the desired variable:
\begin{equation}
    \rho_{\rm tr} = \frac{\Sigma_{j=1}^{ N_{\rm nb}} \rho_{j} \, W(r_{ij},h_i)}{\Sigma_{i=1}^{N_{\rm nb}} \, W(r_{ij},h_i)}
\end{equation}
where $r_{ij}=r_i-r_j$, $h$ is s the distance to the $N_{\rm nb}$-th nearest neighbor and $W(r,h)$ is a smoothing kernel. We use the cubic spline (M4) kernel \citep[][]{1985A&A...149..135M},
\begin{equation}
    W(q) = 
    \begin{cases}
        1- \frac{3 q^2}{2} + \frac{3q^3}{4}, \, 0\leq q \leq 1, \\
        \frac{1}{4} (2-q)^3, \, 1\leq q \leq 2.
    \end{cases}
\end{equation}
In Fig.~\ref{fig:app1}, we show an example of the interpolation considering $N_{\rm nb}=16,32,64$. The relative errors in general remain below $\sim 0.2$\%, with the error increasing with the number of neighbors considered. We note that for sharp discontinuities, such as shocks, this method can lead to a deviation from the true mean gas value and larger relative errors. This is shown in the last panel of Fig.~\ref{fig:app1}. For this reason, we do not compute a kernel-weighted average for the quantities at the shock discontinuity, such as the sonic Mach number, $\mathcal{M}_s$, and instead use the value at the identified gas IDs.

%%%%%%%%%%%%%%%%%%%%%%%%%%%
\section{Energy losses and energy dissipated at the shock}
\label{app:shock_energy}
%%%%%%%%%%%%%%%%%%%%%%%%%%%

\setcounter{figure}{0}
\renewcommand{\thefigure}{B.\arabic{figure}}
\setcounter{equation}{0}
\renewcommand{\theequation}{B\arabic{equation}}

The energy losses considered in the Fokker-Planck solver are:\\
%
%\begin{itemize}
\\    \textit{Synchrotron}: 
    \begin{equation}
    \left\vert \frac{dp}{dt} \right\vert_{\rm syn} = \frac{4}{9} r_0^2 \beta^2 \gamma^2 B^2,
    \end{equation}
    \\
    \textit{Inverse Compton (IC)}:
    \begin{equation}
        \left\vert \frac{dp}{dt} \right\vert_{\rm IC} = \frac{4}{9} r_0^2 \beta^2 \gamma^2 B^2_{\rm CMB} (1+z)^4,
    \end{equation}
    \textit{Coulomb collisions}:
    \begin{equation}
        \left\vert \frac{dp}{dt} \right\vert_{\rm CC} = \frac{4 \pi r_0^2 n_{th} m_e c^2}{\beta_e} \ln \Lambda, 
    \end{equation}
    \textit{Bremsstrahlung (free-free emission)}:
    \begin{equation}
        \left\vert \frac{dp}{dt} \right\vert_{\rm ff} = 8 \alpha_h r_0^2 m_e c^2 n_{\rm th} \gamma \left[ X_{\rm H} + 3X_{\rm He} \right] \left[ \ln(2 \gamma) - \frac{1}{3} \right],
    \end{equation}
    and \textit{Adiabatic}:
    \begin{equation}
        \left\vert \frac{dp}{dt} \right\vert_{\rm adb} = - \frac{1}{3}\left( \mathbf{\nabla} \cdot \mathbf{v} \right) p,
    \end{equation}
where $r_0=q_e^2/(m_ec^2)$ is the classical electron radius, $\beta^2=1-\gamma^{-2}$, $\gamma^2=\left(p/(m_e c)\right)^2+1$, $B_{\rm CMB}$ is the equivalent inverse Compton magnetic field due to the cosmic background
radiation field, $n_{\rm th}$ is the thermal number density, $X_{\rm H}$ is the hydrogen fraction, $X_{\rm He}$ is the helium fraction, and $\alpha_h$ is the fine structure constant. The Coulomb logarithm is defined as
\begin{equation}
    \ln \Lambda = 37.8 + \log \left[ \left(\frac{T}{10^8~\rm K}\right) \left(\frac{n_{\rm th}}{10^3~\rm cm^{-3}} \right)^{-1/2}\right].
\end{equation}

We assume that the energy density of CRs that are injected and accelerated at the shock is given by
\begin{equation} \label{eq:CRe_energy_dens}
   \Delta \varepsilon_{\text{CR},diss} = \eta(\mathcal{M}) \, \Delta \varepsilon_{\mathrm{sh}},
\end{equation}
in [$\mathrm{erg \, cm}^{-3}$], where  $\Delta \varepsilon_{\mathrm{sh}}$ is the shock-dissipated energy density in the downstream in [$\mathrm{erg \, cm}^{-3}$]. We define the latter by computing the difference of the thermal energy densities in the downstream and upstream regimes, corrected for the adiabatic energy increase due to gas compression,
\begin{equation}
    \Delta \varepsilon_{\rm sh} = \epsilon_{\text{th,dw}} - \epsilon_{\text{th,up}} \cdot \sigma^\Gamma, 
\end{equation}
where $\epsilon_{\text{th,dw}}$ and $\epsilon_{\text{th,up}}$, are the downstream and upstream thermal energy densities, respectively, and $\sigma$ is the shock compression factor,
\begin{equation}
    \sigma^{-1} = \frac{1}{\mathcal{M}^2} + \frac{\Gamma-1}{\Gamma+1} \left(1-\frac{1}{\mathcal{M}^2} \right).
\end{equation}
The upstream values needed to compute the upstream thermal energy are computed using the Rankine-Hugoniot jump conditions:
\begin{equation}
    \frac{\rho_{\text{up}}}{\rho_{\text{dw}}} =  \frac{1}{\sigma},
\end{equation}
and
\begin{equation}
    \frac{T_{\text{up}}}{T_{\text{dw}}} =  \frac{(\Gamma + 1)^2 \cdot \mathcal{M}^2}{2\Gamma \mathcal{M}^2 - (\Gamma - 1)} \cdot \frac{1}{(\Gamma - 1) \mathcal{M}^2 + 2}.
\end{equation}

For the identification of the shocks in the MHD simulations we follow the zone-flagging methodology of 
\citet{2015MNRAS.446.3992S}, where 
a cell is flagged as a shock-zone candidate if it simultaneously satisfies 
three criteria: (i)~a converging flow, $\nabla\cdot\mathbf{v} < 0$; 
(ii)~co-alignment of the temperature and density gradients,
$\nabla T \cdot \nabla\rho > 0$, which filters contact 
discontinuities and cold fronts; and (iii)~a temperature ratio consistent with 
$\mathcal{M} \geq \mathcal{M}_\mathrm{min}~=1.3$, i.e.
\begin{equation}
    \log\frac{T_{\rm dw}}{T_{\rm up}} \;\geq\; 
    \left.\log\frac{T_{\rm dw}}{T_{\rm up}}\right|_{\mathcal{M}=\mathcal{M}_\mathrm{min}},
\end{equation}
where the temperature ratio is evaluated from the Rankine--Hugoniot jump condition for an ideal gas. Finally, $\mathcal{M}$ is then obtained by inverting 
temperature jump condition given the pre- and post-shock temperatures 
$T_{\rm up}$ and $T_{\rm dw}$, measured along the direction of the local temperature 
gradient. The shock zone typically spans 2–3 cells in thickness, through which tracers pass in roughly one to a few MHD timesteps. Accordingly, injection into DSA is treated as a one-time event, occurring as the tracer crosses the shock surface cell or the immediately adjacent post-shock cell \citep[see][for a similar method]{2026A&A...706A..39W}.

%%%%%%%%%%%%%%%%%%%%%%%%%%%
\section{Numerical resolution in the momentum grid}
\label{app:resolution}
%%%%%%%%%%%%%%%%%%%%%%%%%%%%

\setcounter{figure}{0}
\renewcommand{\thefigure}{C.\arabic{figure}}
\setcounter{equation}{0}
\renewcommand{\theequation}{C\arabic{equation}}

\begin{figure}
    \centering
    \includegraphics[width=0.9\columnwidth]{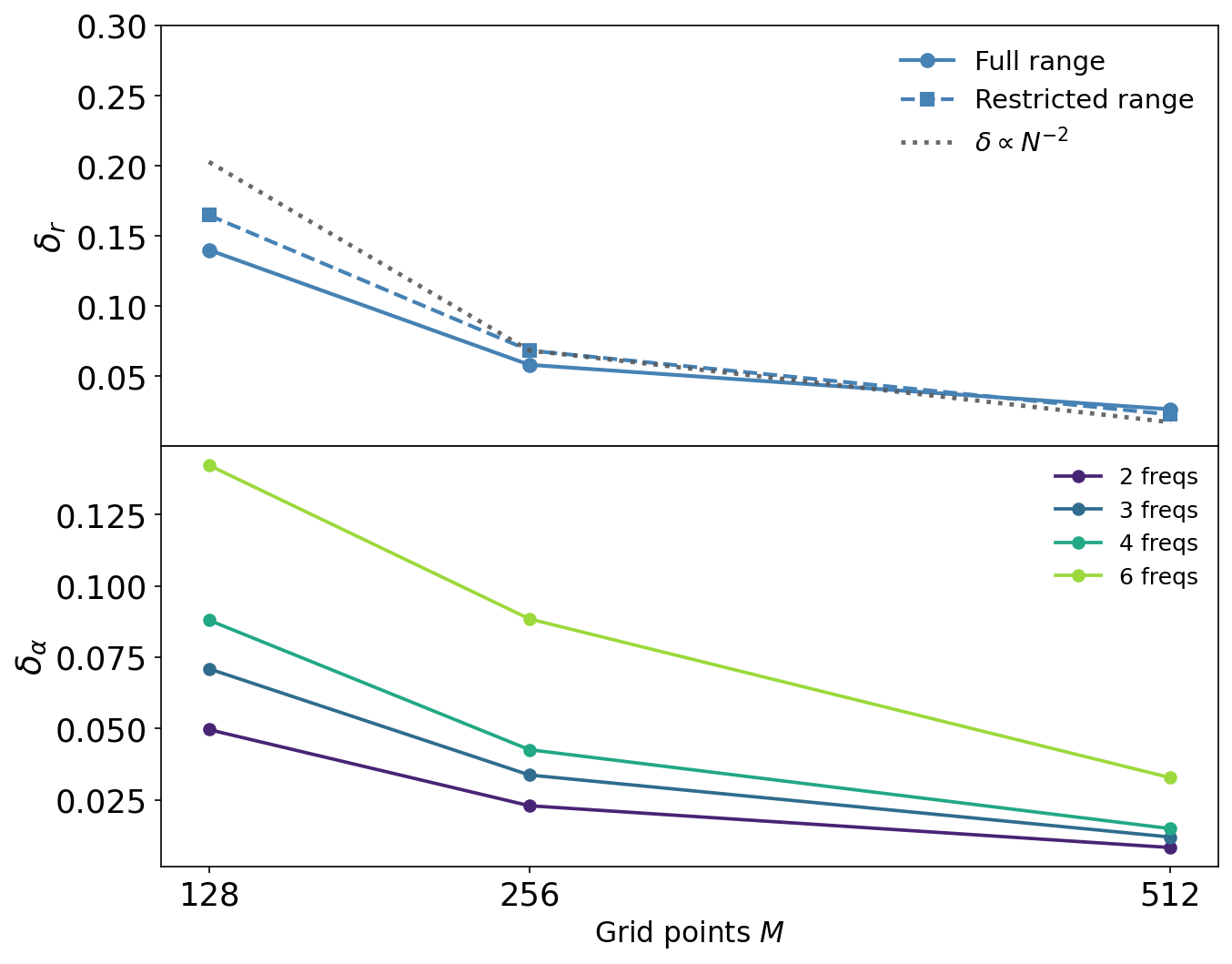}
    \caption{\textit{Upper panel}: Energy-weighted relative error with respect of the highest resolution run with $1024$ grid points. The dotted gray line shows the 
second-order scaling $\delta \propto N^{-2}$. \textit{Lower panel}: Fractional error on the synchrotron spectral index 
$\delta_\alpha$ with respect of the highest resolution run with $1024$ grid points. We fitted $\mathcal{J}_{\rm syn}$ vs $\nu$ over frequency subsets of 2, 3, 4, and 
6 radio frequencies.}
    \label{fig:res}
\end{figure}

In \citet{Donnert_2013} it was shown that the shape of the CRe spectra produced by the numerical method used in this work is relatively insensitive to momentum-space resolution. To assess the impact of momentum-space resolution on our results, we perform  a convergence study using four resolution runs with $M = 128$, 
256, 512, and 1024 momentum grid points per particle for the the run initialized with $v_i=3000$ km~s$^{-1}$, $b=100$ kpc, $\beta_p=100$. We quantify the convergence using two metrics. First, the energy-weighted error on the CR electron spectrum \citep[see][for a similar resolution analysis]{2019MNRAS.488.2235W},
\begin{equation}
    \delta_r = \frac{\int |f_\mathrm{sim}(p) - f_\mathrm{ref}(p)|\, p\, \mathrm{d}p}
    {\int f_\mathrm{ref}(p)\, p\, \mathrm{d}p},
\end{equation}
where $f_\mathrm{ref}$ is the spectrum with $M = 1024$ grid points, evaluated both over the full momentum range and restricted to the radio band 
($p/m_ec \in [10^3, 2\times10^4]$). Second, the fractional error on the synchrotron spectral index, $\alpha$, fitted over different frequency subsets,
%
%\begin{equation}
$ \delta_\alpha = \frac{|\alpha_\mathrm{sim} - \alpha_\mathrm{ref}|}
    {|\alpha_\mathrm{ref}|}.$
%\end{equation}
%
Fig.~\ref{fig:res} shows both metrics for a representative particle (ID=237620) that experienced a shock with $\mathcal{M} \sim 2$ and also energy losses. The spectral error 
$\delta_r$ decreases from $\sim 15$\% at $M = 128$ to $\sim 4$\% at $M = 512$, broadly consistent with second-order 
convergence ($\delta_r \propto N^{-2}$). The spectral index error $\delta_\alpha$ 
is below 15\% at $M = 128$ and below 5\% at $M = 512$ 
for all frequency subsets considered. We note that the CR electron spectrum of this particle exhibits a characteristic 
curvature around the cooling break. Fitting a single power-law to such a curved spectrum inevitably introduces a bias that depends on the frequency range used, as illustrated by the 
different curves in the bottom panel of Fig.~\ref{fig:res}. At higher momentum resolution, the break is better resolved, leading to more accurate spectral index estimates. At lower frequencies, where the spectrum mostly retains its injection power-law shape below the break, the spectral index is less sensitive to resolution 
and $\delta_\alpha$ is smaller. 

In practice, our simulations follow $\sim 10^7$ tracer particles 
simultaneously, which places strong constraints on the affordable 
momentum-space resolution per particle. With $M = 128$ 
momentum bins, the spectral error in the radio band remains below 
$\sim 15$\%. 
Similar to previous works \citep[e.g.][]{Donnert_2013,2023A&A...669A..50V,2026A&A...709A.191N}, we therefore consider $M = 128$, or 16 logarithmic spaced bins per decade, a reasonable compromise between accuracy and computational feasibility.

\setcounter{figure}{0}
\renewcommand{\thefigure}{E.\arabic{figure}}

\begin{figure}
    \centering
    \includegraphics[width=0.85\columnwidth]{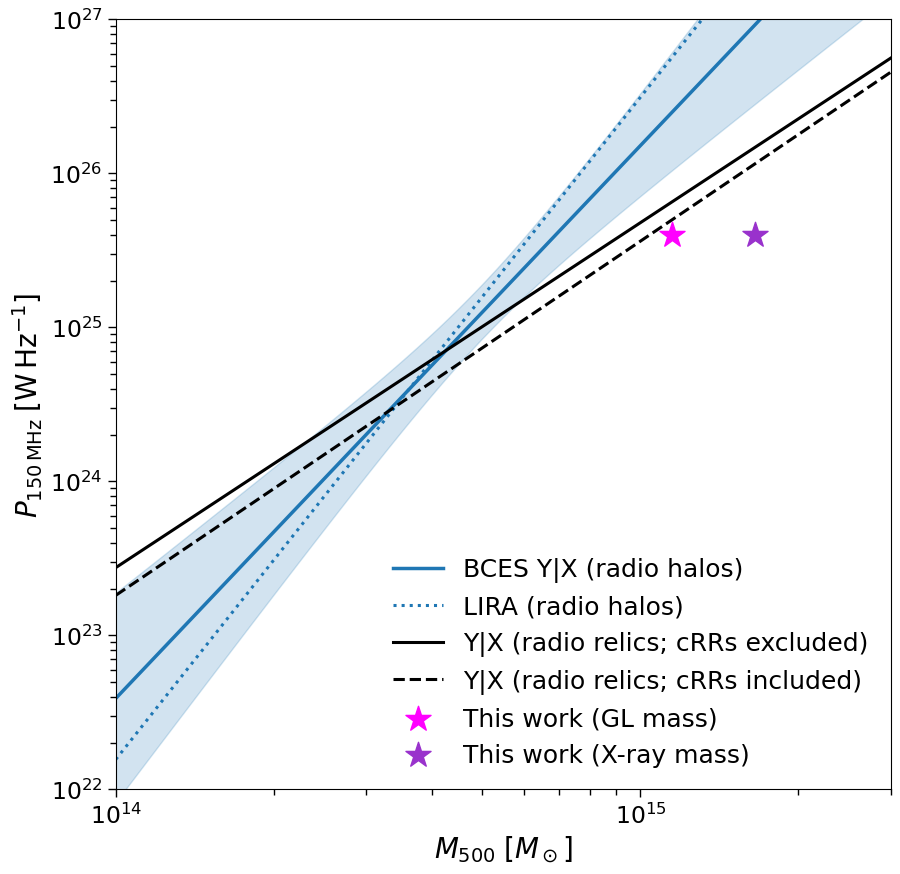}
    \caption{Radio power mass relation of observed radio halos (blue) and radio relics (black). The lines and shaded region are from the relations reported in Tab.~2 of \citet{2023A&A...680A..30C} and Tab.~3 of \citet{2023A&A...680A..31J}.}
    \label{fig:power_relation}
\end{figure}

%%%%%%%%%%%%%%%%%%%%%%%%%%%
\section{Variations in models}
\label{app:variations}
%%%%%%%%%%%%%%%%%%%%%%%%%%%%

\setcounter{figure}{0}
\renewcommand{\thefigure}{D.\arabic{figure}}

\begin{figure*}
    \centering
    \includegraphics[width=0.3\textwidth]{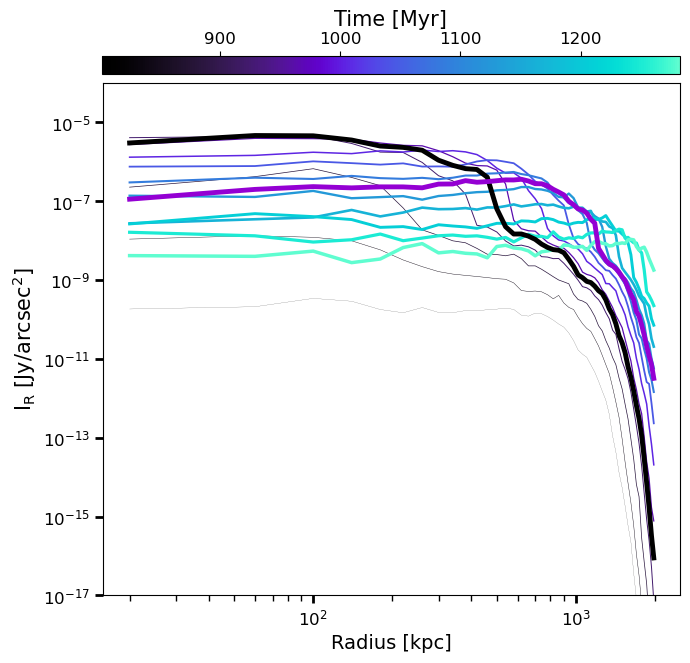}
    \includegraphics[width=0.3\textwidth]{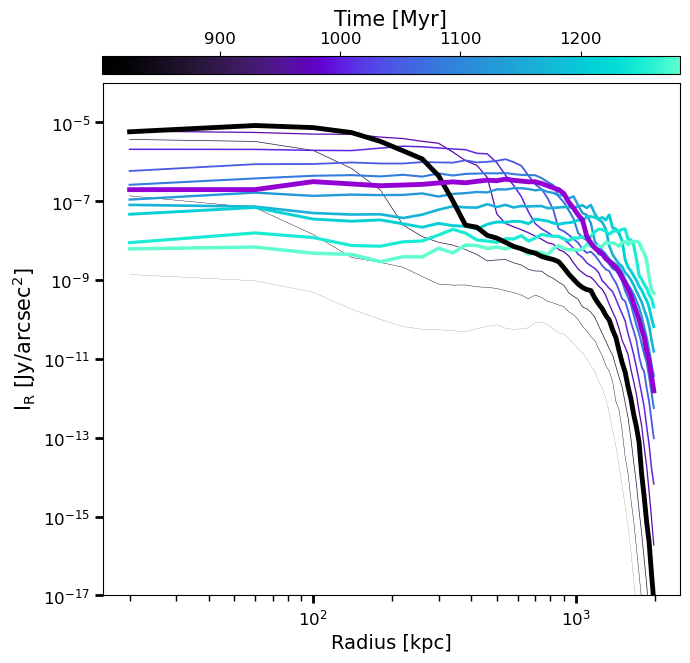} 
    \includegraphics[width=0.3\textwidth]{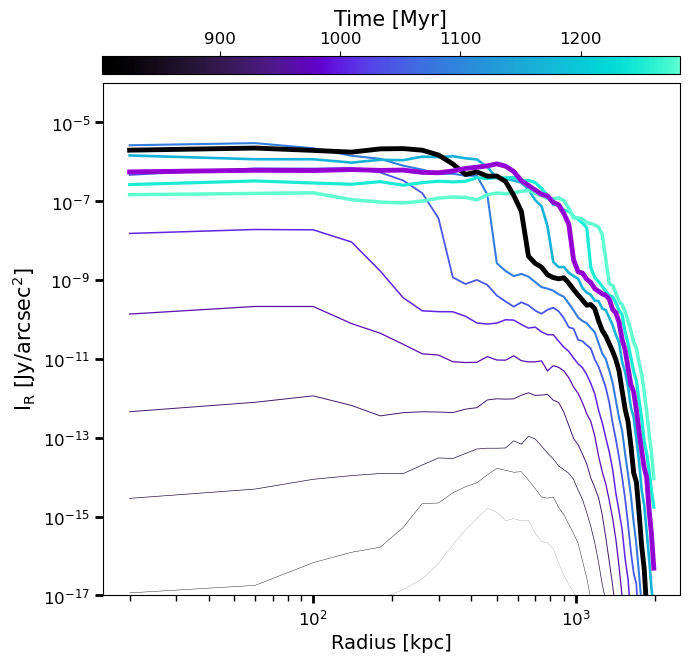} \\
    % X-ray
    \includegraphics[width=0.3\textwidth]{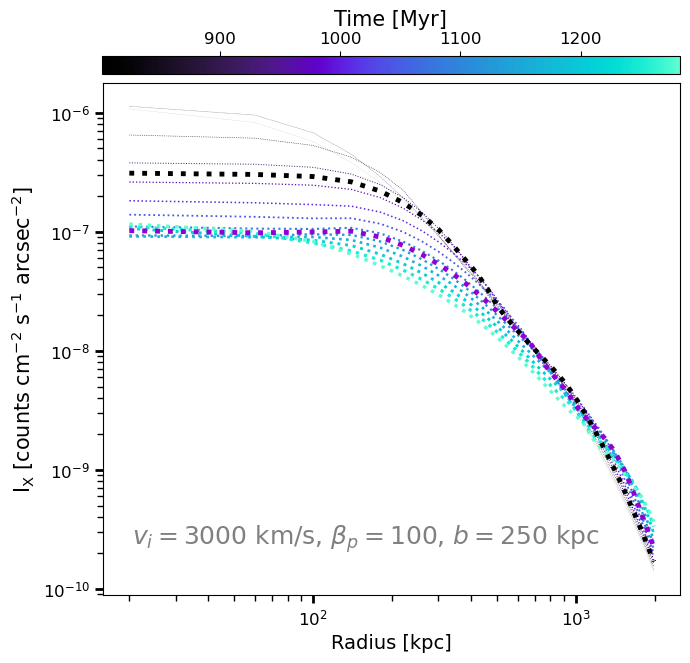}
    \includegraphics[width=0.3\textwidth]{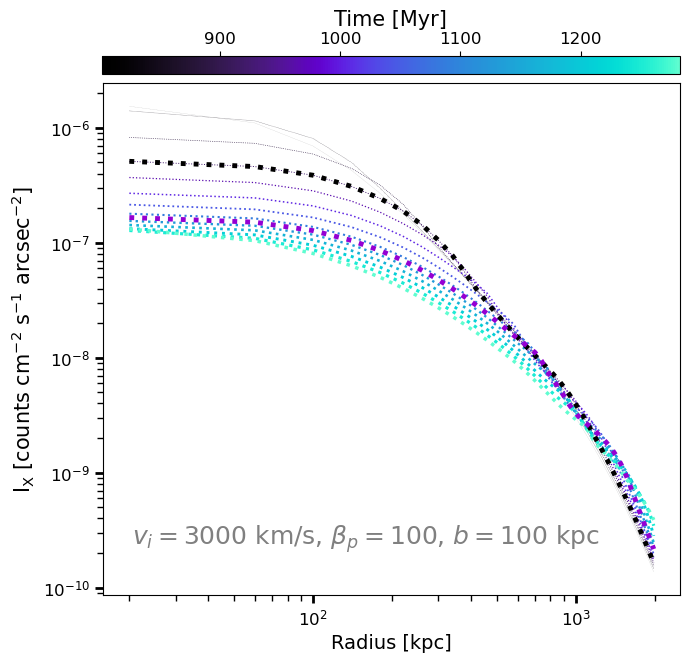}
    \includegraphics[width=0.3\textwidth]{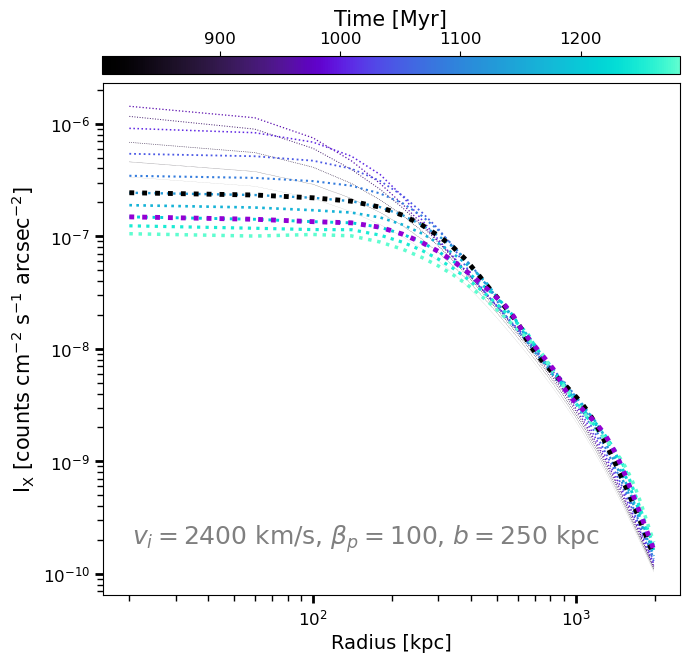}
    \caption{Time evolution of radio (upper panels) and X-ray (lower panels) radial profiles for all optimistic runs. The thick black and darkviolet lines are shown like in Fig.~\ref{fig:profiles_evolution}. For each run we show $t_1$ and $t_2$ from Tab.~\ref{tab:matched}.}
    \label{fig:radial_profiles_all}
\end{figure*}

\setcounter{figure}{0}
\renewcommand{\thefigure}{F.\arabic{figure}}

\begin{figure*}
    \centering
   \includegraphics[width=0.8\columnwidth]{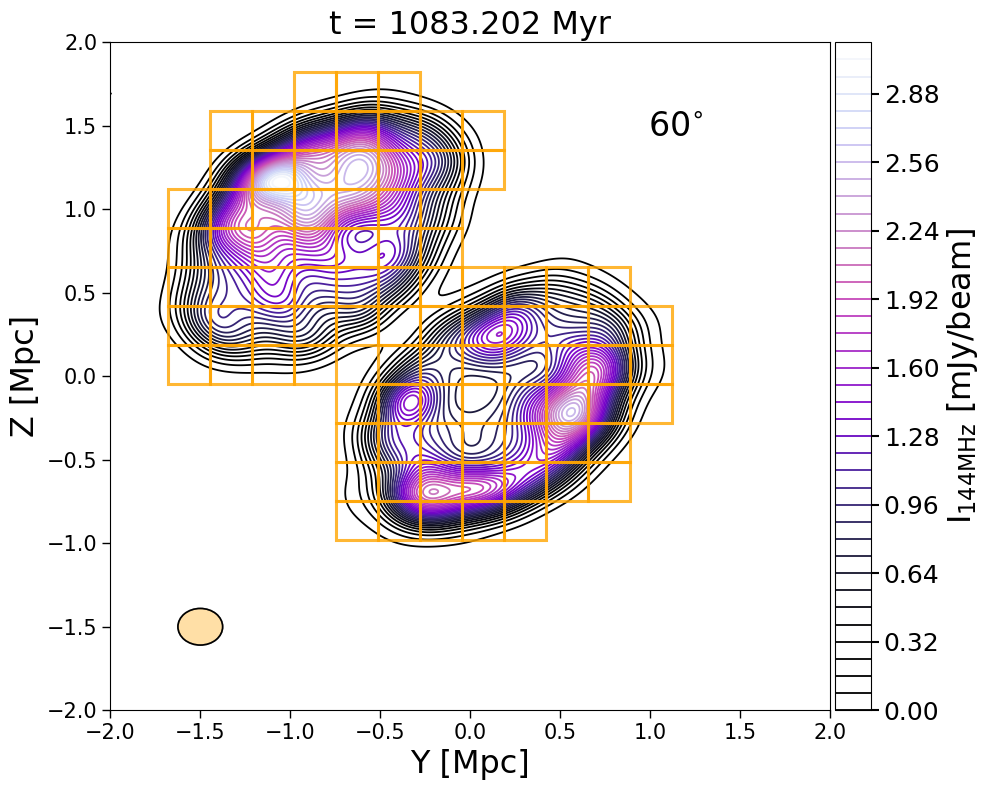}
  \includegraphics[width=0.8\columnwidth]{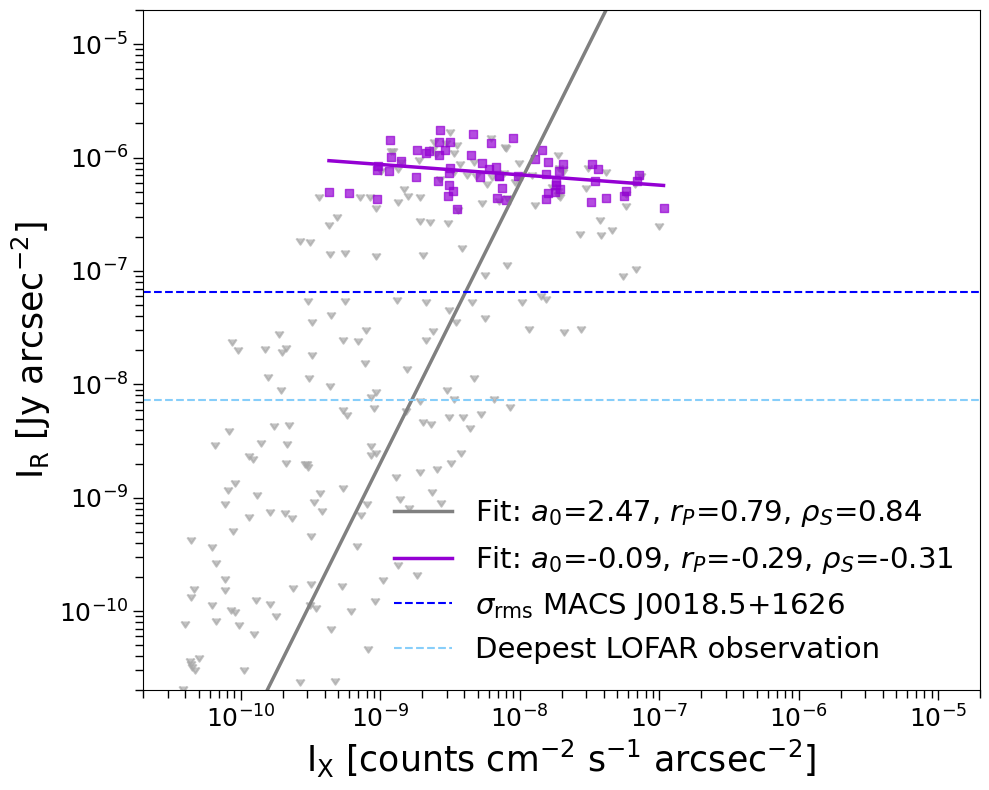}
    \caption{Same as Fig.~\ref{fig:ptp_squares}, but for a viewing angle of $60^{\circ}$.}
    \label{fig:ptp_squares2}
\end{figure*}

In Fig.~\ref{fig:radial_profiles_all}, we show the time evolution of the radio and X-ray radial profiles for all our models. This figure complements the results already shown in Fig.~\ref{fig:profiles_evolution}.

In Fig.~\ref{fig:chi_obs00}, we present the reduced $\chi^2_\nu$ values obtained from fitting the simulated radio surface brightness profiles to the observed one. The figure shows a two-dimensional grid of $\chi^2_\nu$ as a function of viewing angle and simulation epoch. Each cell represents the comparison between a normalized simulated radio radial profile and the observed profile for a given snapshot and line of sight, restricted to the region of parameter space that best describes MACS J0018.5+1626 (see Tab.~\ref{tab:matched}). Lower values of $\chi^2_\nu$ indicate a closer match to the LOFAR observations in profile shape. For visual guidance, we highlight the subset of models belonging to the lowest 10\% of $\chi^2_\nu$ values with white star symbols, identifying regions of parameter space that provide the best agreement with the observations.

\setcounter{figure}{1}
\renewcommand{\thefigure}{D.\arabic{figure}}

\begin{figure*}
    \centering
    \includegraphics[width=0.41\textwidth]{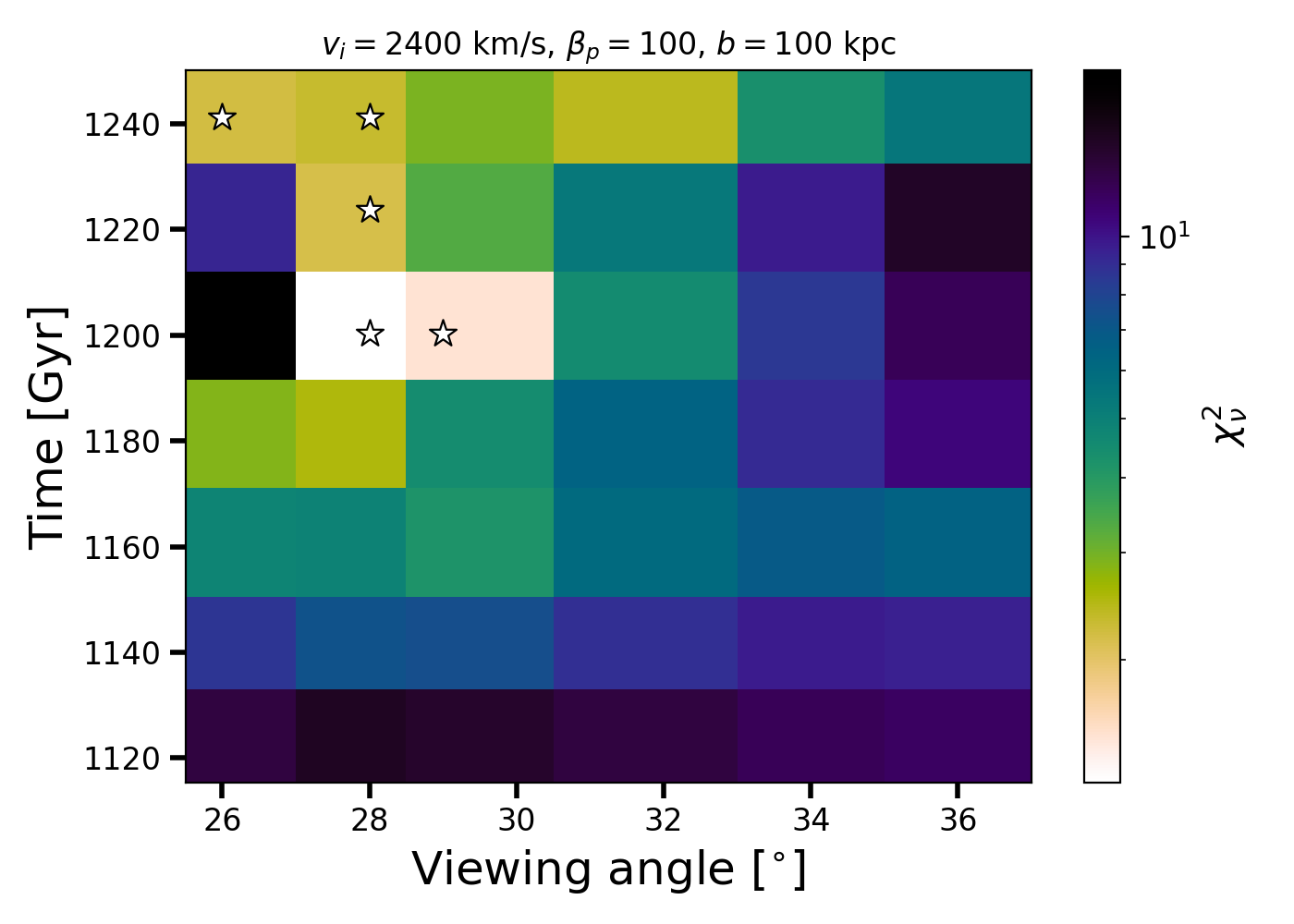}
    \includegraphics[width=0.41\textwidth]{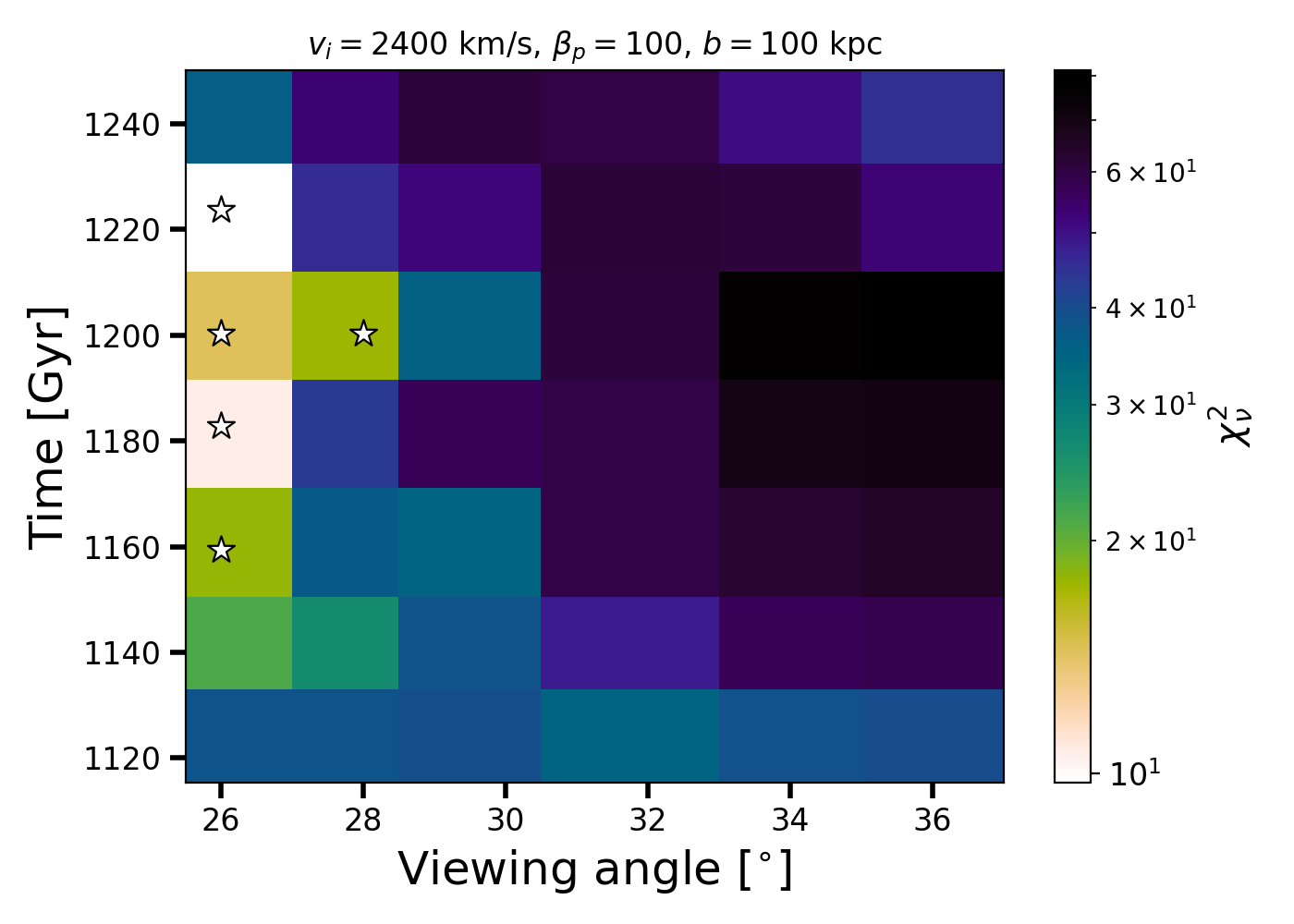} \\
    \includegraphics[width=0.41\textwidth]{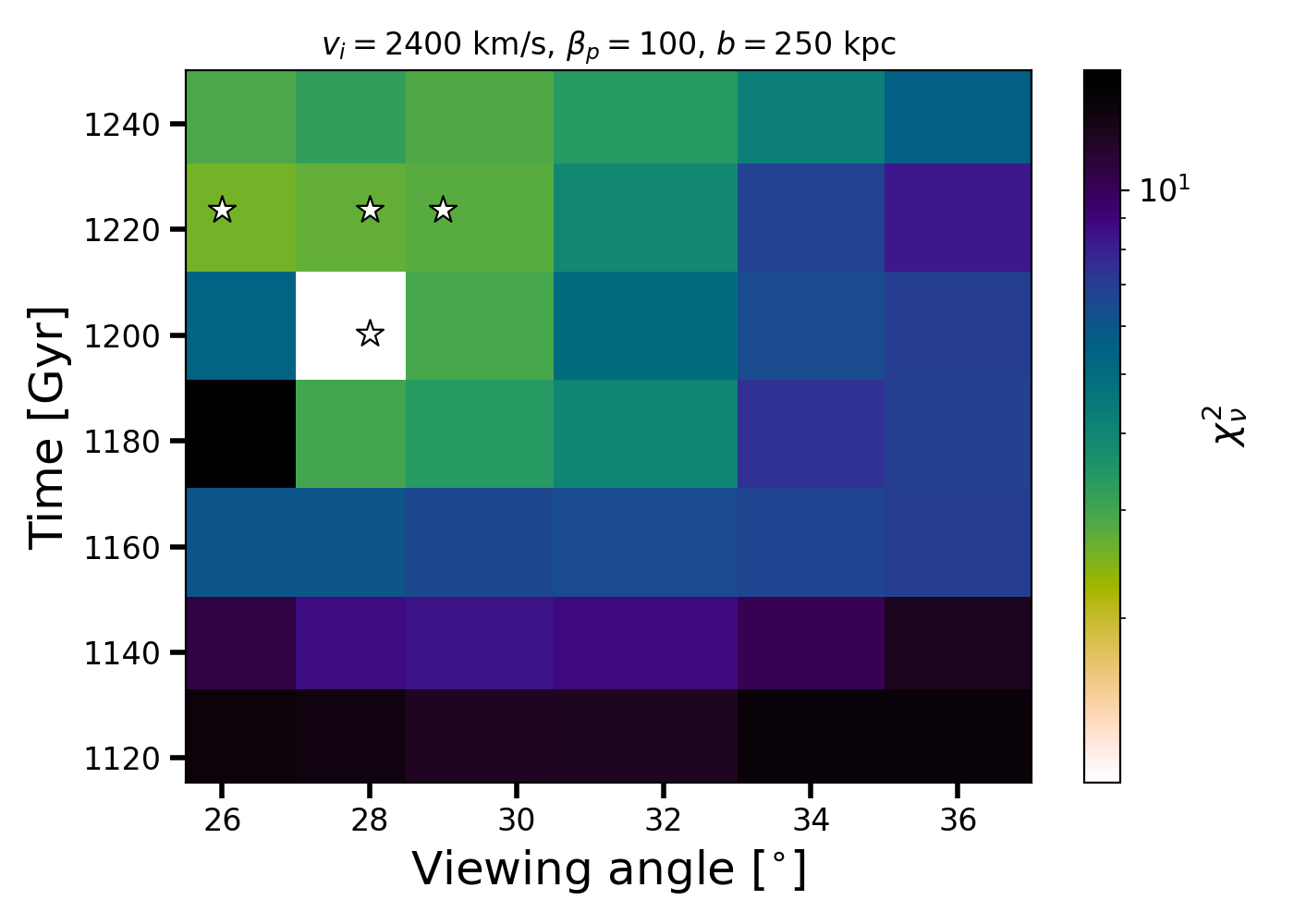}
    \includegraphics[width=0.41\textwidth]{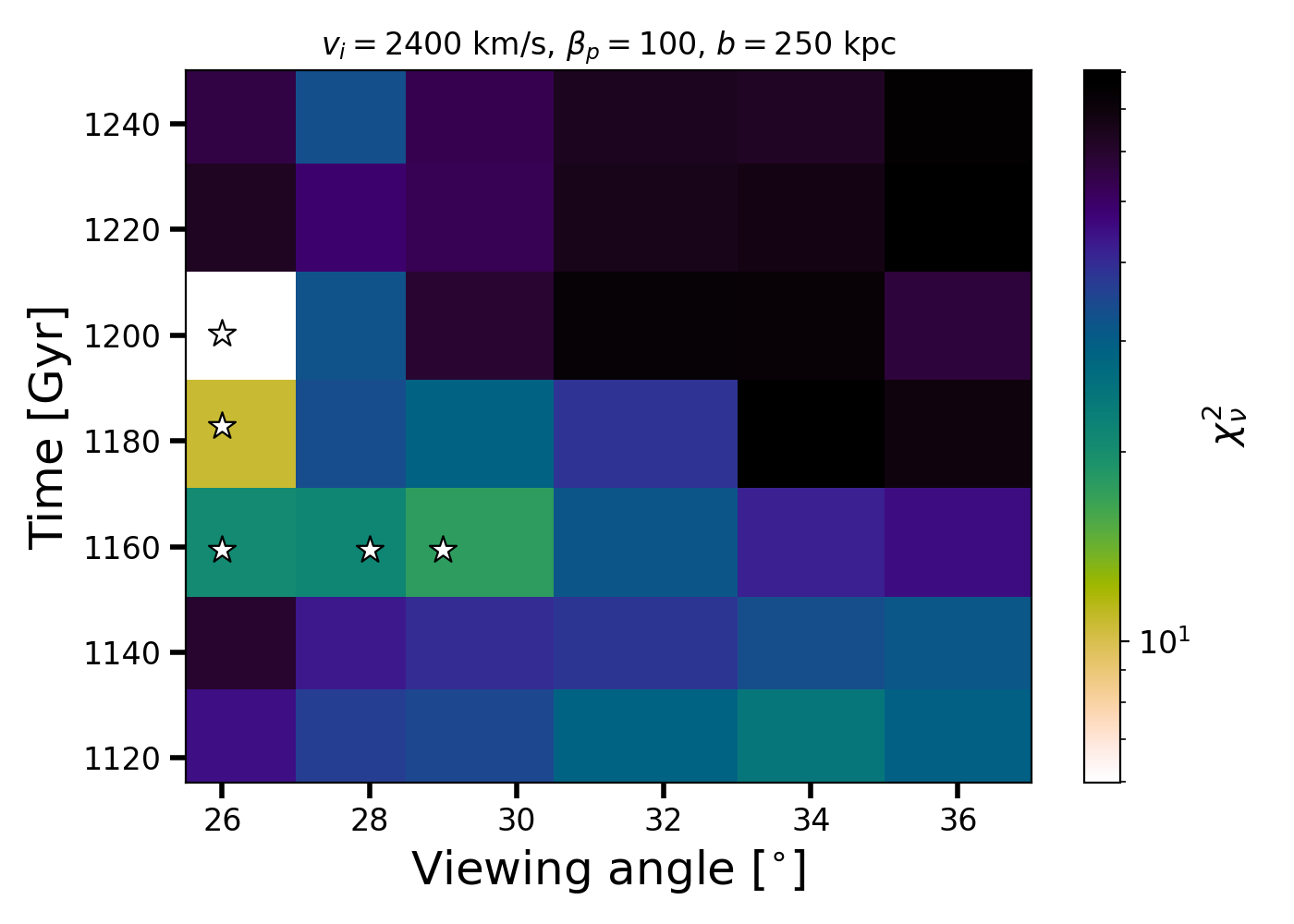} \\
    \includegraphics[width=0.41\textwidth]{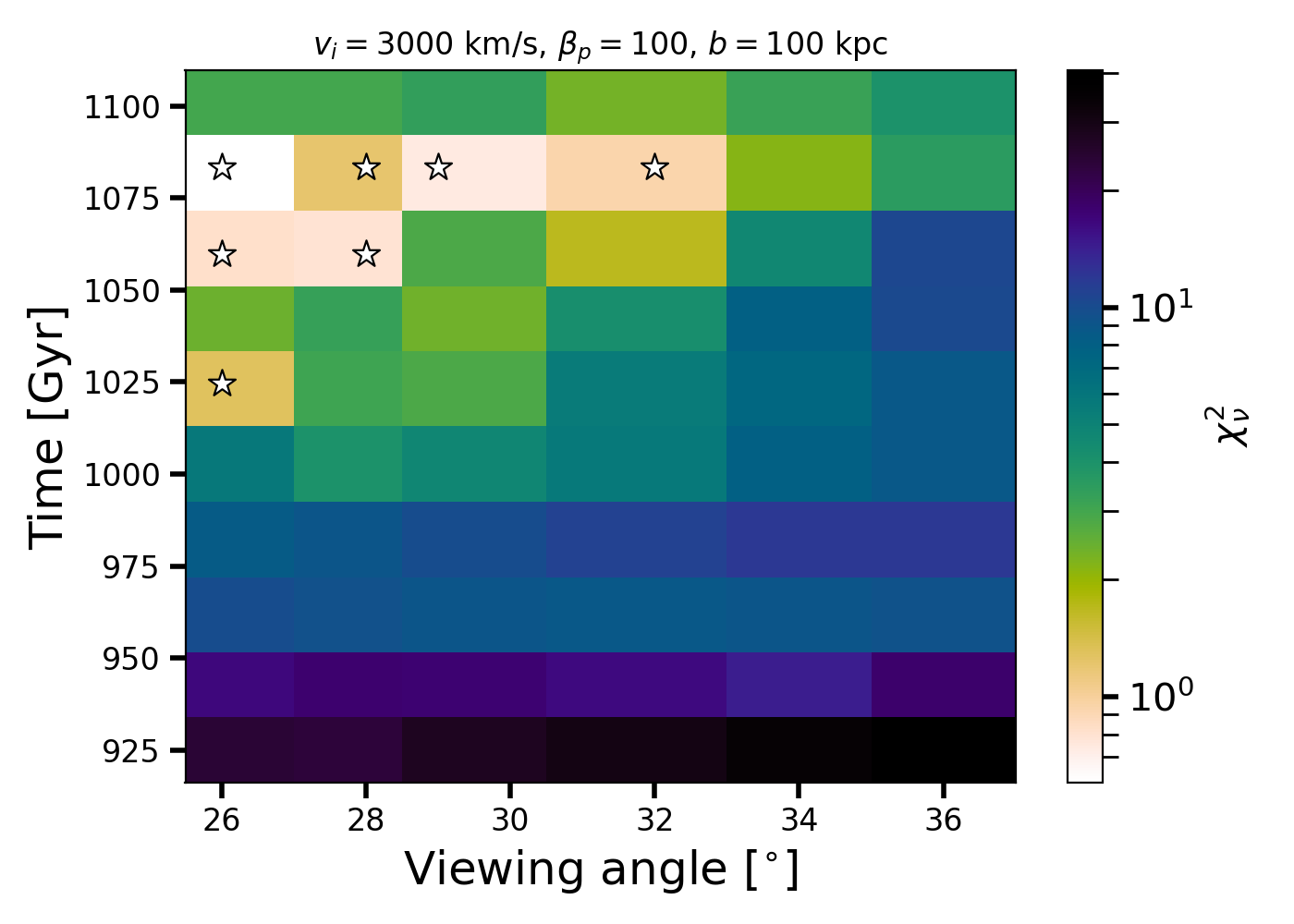} 
    \includegraphics[width=0.41\textwidth]{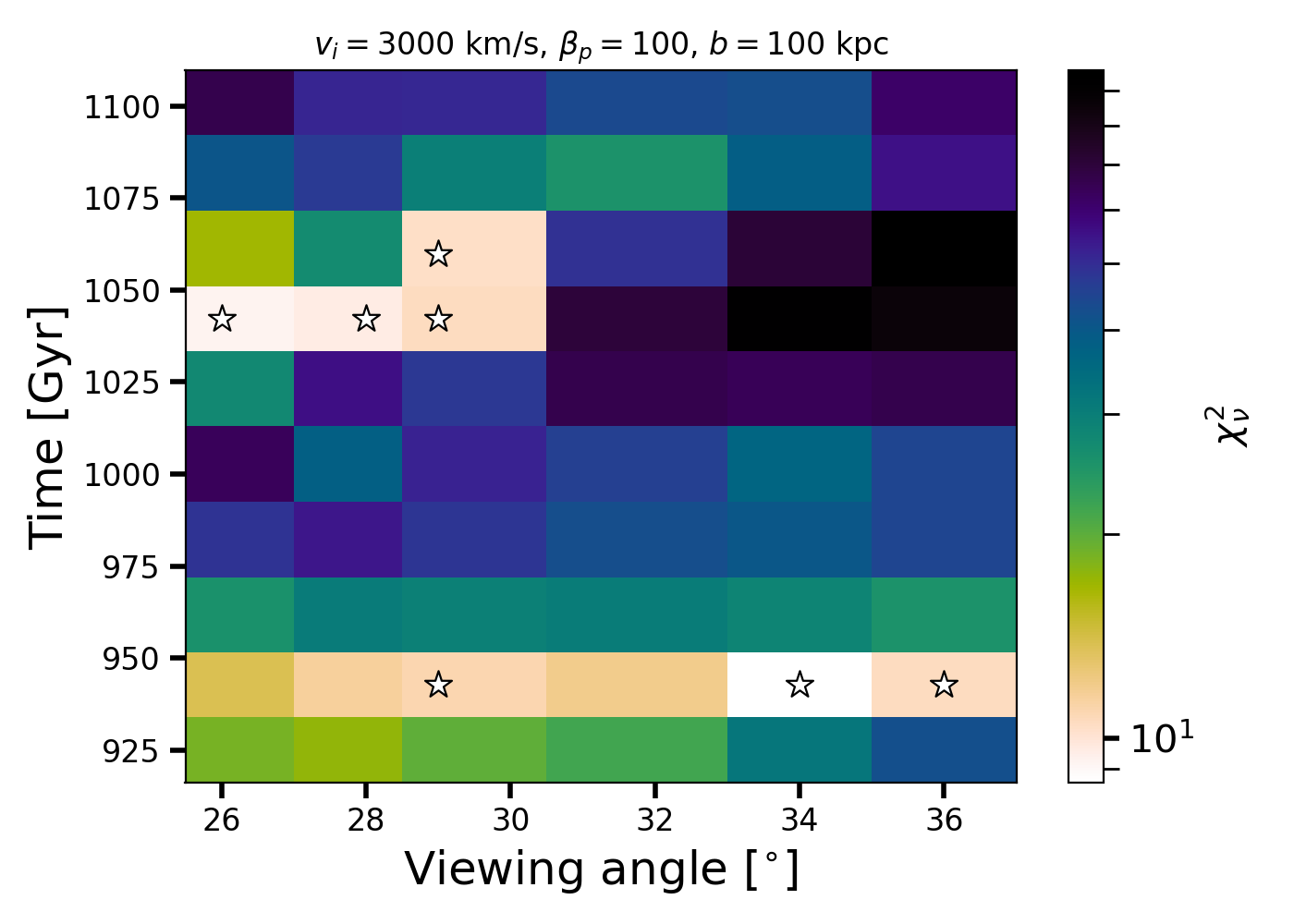} \\
    \includegraphics[width=0.41\textwidth]{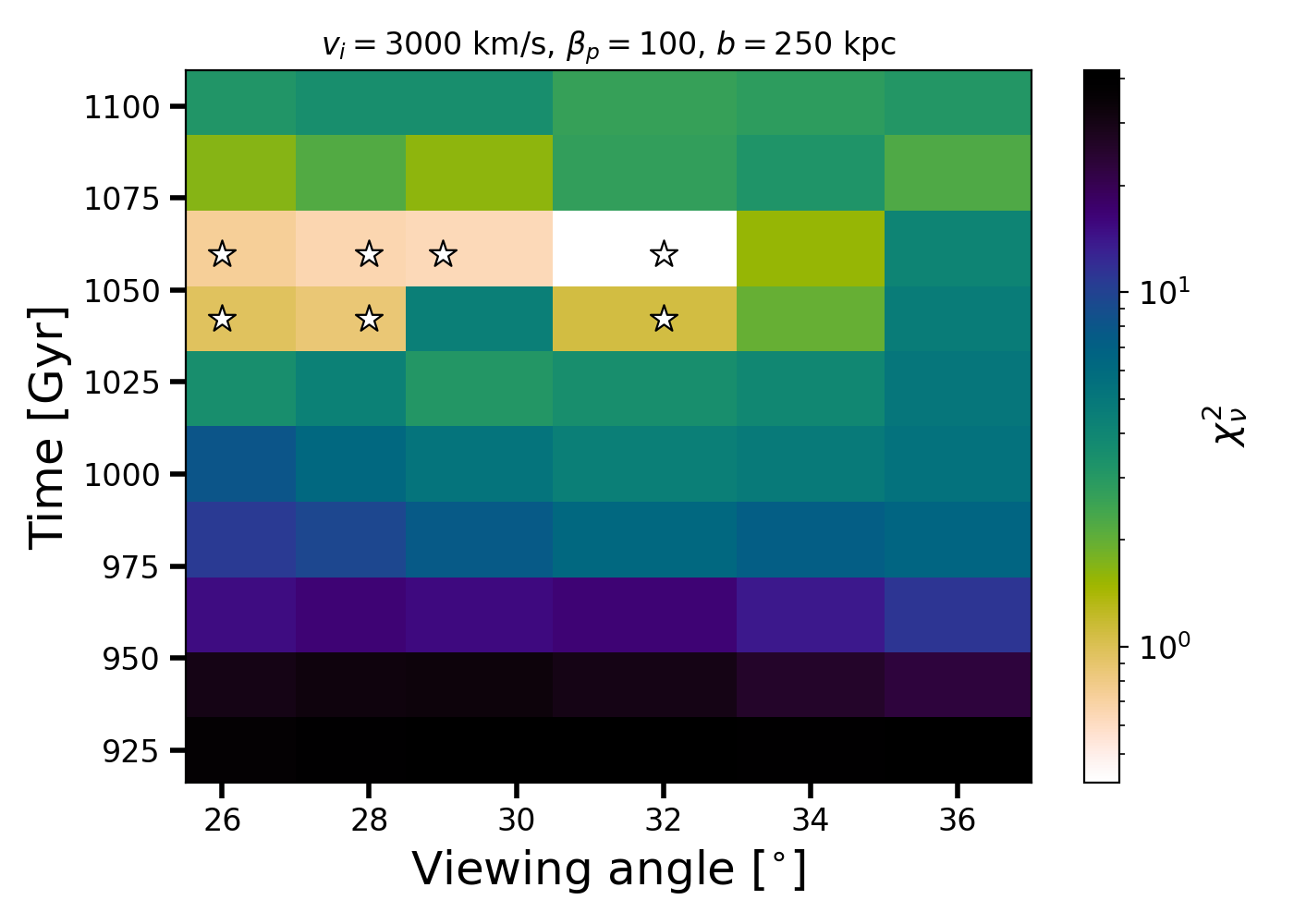}
    \includegraphics[width=0.41\textwidth]{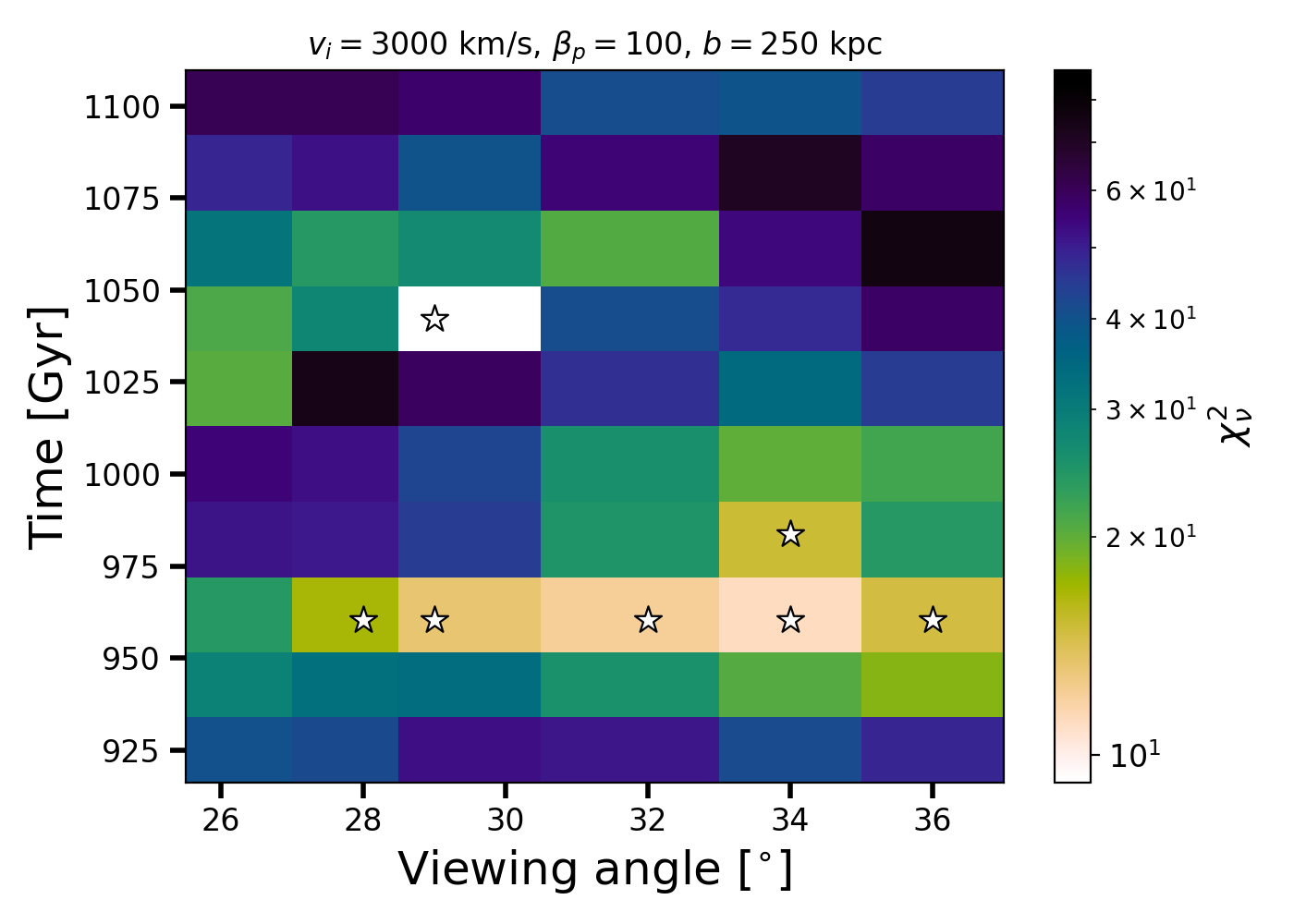} 
    \caption{Reduced $\chi^2_\nu$ values as a function of viewing angle and simulation epoch. Analysis with the radial profile from the LOFAR radio image at the highest resolution (taper None) is shown on the left column. On the right column we show the results from the LOFAR radio image at medium resolution (taper 8).}
    \label{fig:chi_obs00}
\end{figure*}

%%%%%%%%%%%%%%%%%%%%%%%%%%%
\section{Power to mass relation}
\label{app:radio_mass_relation}
%%%%%%%%%%%%%%%%%%%%%%%%%%%

\setcounter{figure}{0}
\renewcommand{\thefigure}{D.\arabic{figure}}

In Fig.~\ref{fig:power_relation}, we plot the radio power–mass relation derived from LOFAR observations at 150 MHz for both radio halos and radio relics. We show the radio power measured in this work with star symbols (see Tab.~\ref{tab:obs}). The same radio power is plotted for the two mass estimates reported in the literature for MACS~J0018.5+1626: the X-ray–derived mass of ${\rm M}_{500}=1.65 \times 10^{15}~{\rm M}{\odot}$ \citep[][]{2010MNRAS.406.1759M,2019ApJ...880...45S}, and the strong gravitational lensing–derived mass of ${\rm M}{500}=1.1 \times 10^{15}~{\rm M}{\odot}$ \citep[see Sec.~3.2 of][]{2024ApJ...968...74S}.
For radio halos, we show the observational relations reported in \citet{2023A&A...680A..30C}. Specifically, we use the fitting parameters listed in their Tab.~2 obtained with the Bivariate Correlated Errors and intrinsic Scatter (BCES) linear regression for radio halos only, as well as those derived using the Bayesian LInear Regression in Astronomy (LIRA) method, also restricted to radio halos. For radio relics, we use the best fit parameters reported in Tab.~3 of \citet{2023A&A...680A..31J} for data including and exluding candidate radio relics (cRR). We refer the reader to \citet{2023A&A...680A..30C} and \citet{2023A&A...680A..31J} for further details on these observational relations.

%%%%%%%%%%%%%%%%%%%%%%%%%%%
\section{Viewing angles beyond ICM-SHOX constraints}
\label{app:another_angle}
%%%%%%%%%%%%%%%%%%%%%%%%%%%

In Fig.~\ref{fig:ptp_squares2} we show the same plots as in 
Fig.~\ref{fig:ptp_squares}, but at a viewing angle of $60^{\circ}$. This angle lies beyond the current constraints of the ICM-SHOX pipeline for MACS J0018.5+1626 and, consistently with this, the two shocks show no significant spatial overlap, meaning that this projection 
would not reproduce the observations. Nevertheless, this case exemplifies how the viewing angle has a significant influence on the radio--X-ray point-to-point correlation. At $60^{\circ}$, the correlation shows larger scatter and the two-slope feature is less distinct. This is because the reduced projected surface area of the brightest emission shifts the dominant contribution towards dimmer regions (more aged CRe) producing a steeper slope.

\end{document}